# Trans-Neptunian objects as natural probes to the unknown solar system

Patryk Sofia Lykawka*

[1] Astronomy Group, Faculty of Social and Natural Sciences, Kinki University, Shinkamikosaka 228-3, Higashiosaka-shi, Osaka, 577-0813, Japan

*******

This paper has been accepted for publication in Monographs on Environment, Earth and Planets on 4 December 2012.

patryksan@gmail.com

http://sites.google.com/site/patryksofialykawka/

*******

*E-mail address: patryksan@gmail.com

## Abstract

Trans-Neptunian objects (TNOs) are icy/rocky bodies that move beyond the orbit of Neptune in a region known as the trans-Neptunian belt (or Edgeworth-Kuiper belt). TNOs are believed to be the remnants of a collisionally, dynamically and chemically evolved protoplanetary disk composed of billions of planetesimals, the building blocks from which the planets formed during the early solar system. Consequently, the study of the physical and dynamical properties of TNOs can reveal important clues about the properties of that disk, planet formation, and other evolutionary processes that likely occurred over the last 4.5 Gyr.

In contrast to the predictions of accretion models that feature protoplanetary disk planetesimals evolving on dynamically cold orbits (with both very small eccentricities, *e*, and inclinations, *i*), in reality TNOs exhibit surprisingly wide ranges of orbital eccentricities and inclinations, from nearly circular to very eccentric orbits (putting some objects at aphelia beyond 1000 AU!) and ranging up to ~50 deg of inclination with respect to the fundamental plane of the solar system. We can group TNOs into several distinct dynamical classes: (1) Resonant: TNOs currently locked in external Neptunian mean motion resonances; (2) Classical: non-resonant TNOs concentrated with semimajor axes in the range $37 < a <$ 45-50 AU on relatively stable orbits (which typically feature only minor orbital changes over time); (3) Scattered: TNOs on orbits that suffer(ed) notable gravitational perturbations by Neptune, yielding macroscopic orbital changes over time; (4) Detached: TNOs typically possessing perihelia, $q > 40$ AU, $a >$ 45-50 AU and orbits stable over the age of the solar system.

Several theoretical models have addressed the origin and orbital evolution of the main dynamical classes of TNOs, but none have successfully reproduced them all. In addition, none have explained several objects on peculiar orbits, or provided insightful predictions, without which a model cannot be tested properly against observations. Based on extensive simulations of planetesimal disks with the presence of the four giant planets and huge numbers of modeled planetesimals (reaching up to a million test particles or several thousand massive objects), I explore in detail the dynamics of the TNOs, in particular their (un)stable regions over timescales comparable to the age of the solar system, and the role of resonances across the entire trans-Neptunian region. I also propose that, along with the orbital history of the giant planets, the orbital evolution of primordial embryos (massive planetesimals comparable to Mars-Earth masses) can explain the fine orbital structure of the trans-Neptunian belt, the orbits of Jovian and Neptunian Trojans (objects moving about the L4/L5 Lagrange points of Jupiter and Neptune, respectively), and possibly the current orbits of the giant planets. Those primordial embryos were ultimately scattered by the giant planets, a process that stirred both the orbits of the giant planets and the primordial planetesimal disk to the levels observed at 40-50 AU. In particular, the main constraints provided by the trans-Neptunian belt are optimally satisfied if at least one such primordial embryo (planetoid) survived in the outskirts of the solar system. Therefore, a model with a hypothesized resident planetoid yields results that fit the identified main dynamical classes of TNOs, including those objects on unusual orbits within each class.

# 1 Introduction

Ancient people had, in general, more time to observe celestial motion than we do today. Indeed, the night sky has been fascinating mankind since the dawn of civilization. The apparent erratic motion of the planets in the sky over the year was one of the first intriguing facts noticed about our solar system. This observational constraint was very important, because it served as a crucial test for the development of solar system models, from Ptolemy's geocentric, to Copernicus' heliocentric, models. Copernican heliocentrism was published more than four centuries ago! Since then, our knowledge about the solar system has increased enormously through both observational discoveries and the development of more elaborate theories. In particular, large telescopes equipped with CCDs can probe more distant and fainter objects, fast computers with powerful codes are capable of revealing dynamical phenomena more accurately, and theoretical models and laboratory facilities allow us to reproduce diverse astrophysical environments (e.g., space weathering). All these techniques together have been revolutionizing the study of the solar system in recent decades. The advances in planetary science are evident, but even today the search for explanations for unsolved questions motivates modern planetary scientists, just as the apparent motion of the five known planets of antiquity intrigued ancient astronomers.

Modern theories for the formation of the solar system are essentially based on the idea that our system originated from a rotating cloud of gas and dust. Today, the solar system family includes distinct types of bodies, such as the Sun, the planets, dwarf planets, asteroids, satellites, comets, dust, among others. Nevertheless, all these objects originated from the same protoplanetary disk of gas and dust that existed billions of years (Gyr) ago. Based on the oldest collected meteorites on Earth, it is well known that the solar system is almost 4.6 Gyr old (e.g., Montmerle et al. 2006 and references therein). Therefore, the solar system represents the outcome of a complex structure that has been evolving chemically, collisionally, and dynamically over the last ~4.6 Gyr. More importantly, we should recall that the current characteristics of the solar system's members offer important clues on the solar system's history, including the initial conditions of the protoplanetary disk (e.g., composition, disk size, etc.) after the collapse of interstellar gas and dust. Therefore, the study of the members of the solar system provides crucial clues and a better understanding of the solar system itself.

Models of the solar system have several characteristics in common which can be summarized as follows. While the protosun was accreting matter from the gaseous/dusty protoplanetary disk, dust grains grew and started to sediment towards the disk midplane, growing up to cm-sizes in typically ten, or a hundred, thousand years (kyr). After that, these small grains grew further in a few million years (Myr) to km-sized planetesimals through mutual accreting collisions, or alternatively as large clumps in local instability niches (e.g., Morbidelli et al. 2009a). Shortly after the formation of such planetesimals, planetary-sized ones grew quickly (during the so-called runaway+oligarchic growth) leading to the formation of terrestrial planets and the core of the giant planets on ten-million-year timescales (Hayashi et al. 1985; Pollack et al. 1996; Montmerle et al. 2006). An important constraint on the formation of giant planets is that the disk gas probably dissipated in less than ~10 Myr, implying that Jupiter and Saturn acquired their massive atmospheres within this time span. A similar argument is valid for the existence of the (thinner) atmospheres of Uranus and Neptune, suggesting that the formation of these planets was accomplished a short time after the dissipation of the disk gas (e.g., Montmerle et al. 2006).

Although the scenario discussed above is somewhat simplistic, and many unsolved problems remain, the basic structure of the solar system was probably formed during the first 100 Myr after the birth of the Sun, so that all subsequent evolution in the solar system has taken place in the last ~4.5 Gyr. This scenario also predicts that the protoplanetary disk was composed of planetesimals in very cold orbital conditions, namely objects on near circular (eccentricities, $e$, ~ 0) and very low inclination, $i$, orbits. The distribution of mass in the disk is often described by its surface density, $\sigma$, usually following a power law with exponents in the range -3/2 ± 1/2, that describes how the mass distribution decreases with heliocentric distance, *R* (Hayashi et al. 1985; Morbidelli & Brown 2004; Jewitt 2008). This description is supported by estimates of the current surface density based on the masses and heliocentric distances of the planets, which extends continuously in the outer solar system starting at the orbit of Jupiter. The extrapolation of this distribution to distances beyond Neptune suggested the existence of several Earth masses ($M_\oplus$) of materials beyond 30 AU. Based on these considerations, researchers postulated the existence of a massive population beyond Neptune's orbit composed of a myriad of small icy trans-Neptunian objects (TNOs), to account for the extra mass in the disk. For instance, in the same year as the discovery of the ninth planet (currently classified as a dwarf planet[1]) (134340) Pluto by C. Tombaugh, Leonard (1930) speculated that Pluto could be the first member in the trans-Neptunian region. However, during that period, there was a search for a massive planet beyond Neptune (called "Planet X"), hypothesized to explain the "anomalies" in the motion of Uranus and Neptune (today known to be spurious; Standish 1993). Curiously, the search for Planet X led to the serendipitous discovery of Pluto orbiting at a semimajor axis, *a*, 39.4 AU, and, for this reason, astronomers thought Pluto was a massive planet, and not just one member of the expected large population of TNOs. However, subsequent observations gradually refined Pluto's estimated mass downwards (<0.15 $M_\oplus$; Duncombe et al. 1968; Delsanti & Jewitt 2006), so that Pluto was definitively too small to be Planet X, or to cause any important perturbation of Uranus or Neptune. With the discovery of Charon (Christy & Harrington 1978), the first satellite of Pluto, it was determined that its mass was only ~1/450 that of the Earth. In conclusion, Pluto was not the long-sought-after Planet X, nor was it truly considered a member of a trans-Neptunian reservoir, despite F. C. Leonard's correct "prediction".

If Pluto was too small to account for the outer solar system mass distribution, how was it possible to explain the abrupt discontinuity of the mass surface density in the outer solar system? There was no reason, *a priori*, to expect a truncation of the primordial planetesimal disk just beyond Neptune. This assertion led astronomers in the 1940-50s to consider TNOs as the remnants of the primordial planetesimal disk that did not accrete to form large planets because of insufficient surface density and/or too long timescales (Edgeworth 1949; Kuiper 1951)[2]. Edgeworth even speculated that TNOs would enter the inner solar system from time to time to be observed as comets. During that period, there were no telescopes equipped with CCDs, so that the task of finding the "first" TNO[3] was extremely difficult. On theoretical grounds, the predictions stated so far lacked scientific support. Thus, although the idea of a trans-Neptunian belt composed of a large number of icy bodies was attractive, it remained a matter of speculation.

In the 1970-90s, theoretical studies yielded promising results regarding the trans-Neptunian belt, especially those dealing with the origin of short period comets (SPCs). A well-known cometary reservoir at that time, the Oort cloud (Oort 1950), was not able to yield the correct flux of SPCs (Joss 1973). It turned out that only a flat disk of cometary bodies located beyond Neptune could explain the flux of SPCs (Fernandez 1980; Fernandez & Ip 1984). Following these previous works, Duncan et al. (1988) showed, through detailed numerical investigations, that the conjectured trans-Neptunian belt was an optimal candidate to be the source of SPCs, based on two main results. First, the inclination distribution of objects leaving the trans-Neptunian belt was dynamically preserved after entering into the inner solar system (modeled with $i < 30$ deg, roughly the same as that of the SPCs[4]). This ruled out the Oort cloud as a potential source, otherwise SPCs would not be confined to small inclinations. Second, the dynamical lifetime of a comet is much less than 4.6 Gyr, implying the source cometary reservoir should be stable over the age of the solar system in order to maintain a steady population of SPCs. Duncan et al. (1988) also suggested that there should exist an intermediate steady population of $\sim 10^5$ bodies on unstable orbits among the giant planets in order to supply the SPCs population. In fact, Duncan et al.'s prediction was fulfilled with the discovery of an unusual body orbiting between Saturn and Uranus, named Chiron (Kowal et al. 1979). Because of the dual nature of Chiron (Hartmann 1990), showing both cometary and asteroidal physical behaviour, this class of objects was referred to as the "Centaurs". The second Centaur, Pholus, was discovered only in 1992. In general, Centaurs have orbits that penetrate those of the giant planets, and consequently are unstable (Horner et al. 2003; Tiscareno & Malhotra 2003; di Sisto & Brunini 2007). Thus, the Centaurs represent observational evidence of the link between TNOs and SPCs, as predicted on theoretical grounds a few years before their discovery. Consequently, based on consistent theoretical predictions that were little by little confirmed by observations, it was just a matter of time before the "first" TNO was discovered. In short, the viewpoint that the Oort cloud and the trans-Neptunian belt were respective cometary reservoirs for long period comets and SPCs was confirmed, and this provided important new clues for the origin of the solar system (Mumma et al. 1993; Holman & Wisdom 1993; Duncan & Levison 1997; Horner et al. 2003; Fernandez et al. 2004; Emel'yanenko et al. 2005).

[1] According to the International Astronomical Union, a dwarf planet is an object that satisfies the following 3 conditions: (1) It orbits the Sun; (2) It is massive enough such its shape is controlled by gravitational forces rather than body material cohesion forces; (3) It has not cleared objects from the neighboring region of its own orbit. Pluto was reclassified from a planet to a dwarf planet in 2006.

[2] Although the trans-Neptunian belt is also commonly referred as the "Kuiper belt" or "Edgeworth-Kuiper belt", I opt to use the most general nomenclature in this paper.

[3] Today, we know Pluto was the first TNO discovered. However, at the time of Pluto's discovery (1930), Pluto was regarded as an isolated planet, not a member of a swarm of TNOs.

[4] Excluding "Halley-type" comets with very high inclinations (e.g., Morbidelli 2005).

In 1992, the "first" TNO ((15760) 1992 QB1) was discovered orbiting at $R \sim 41$ AU with an estimated diameter $D \sim 200$ km (assuming an albedo $p = 0.04$) (Jewitt & Luu 1993). Subsequent discoveries of TNOs (Williams 1997; Jewitt 1999) confirmed the existence of the trans-Neptunian belt and put Pluto into context as just one of the largest members of the belt (McKinnon & Mueller 1988; Stern 1998; Stern & Levison 2000; Davies 2001; Delsanti & Jewitt 2006). The number of discovered TNOs grew at an increasing rate during the subsequent 15 years. However, in recent years the rate has decreased, perhaps in the face of the new wide-area systematic surveys (Jewitt 2003; Jones et al. 2009; Petit et al. 2011). It is now widely accepted that TNOs constitute the remnants of the primordial planetesimal disk, and that they offer clues about the dynamical, collisional, chemical and thermal evolution of the solar system over billions of years (Davies 2001; Luu & Jewitt 2002; Delsanti & Jewitt 2006; Kenyon et al. 2008).

As of 1 April, 2012, more than 1200 TNOs with $D > 50$-100 km and approximately 100 Centaurs of size exceeding tens of kilometers have been discovered on a regular basis. Updated orbital elements of TNOs and Centaurs are available on public databases[5]. It is worth noting that the observed population of TNOs corresponds to only ~1-2% of the predicted population in the same size range (Luu & Jewitt 2002; Sheppard 2006). More than 700 TNOs possess orbital elements with uncertainties small enough to be considered "reliable" (i.e., observed during two or more oppositions). Figure 1 illustrates the orbital distribution of TNOs in *a-e* and *e-i* element space. While the currently known TNOs probe the large end of the size distribution ($D > 50$ km), the existence of several billions of cometary-sized TNOs has been predicted (Trujillo et al. 2001a; Bernstein et al. 2004). Even with the current limited sample, the currently known TNOs reveal a surprising and unexpectedly complex orbital structure. That is, most of the TNO population is currently evolving on excited "hot" orbits, with moderate-large eccentricities and/or high inclinations that reach almost 50 deg. Recalling the previous discussion on the nature of the primordial planetesimal disk, it is worth noting that the present orbital excitation seen in the trans-Neptunian region is in large contrast with the predictions of a dynamically cold swarm of primitive planetesimals (e.g., Morbidelli 2005). Regarding nomenclature, I will hereafter use the following definitions: *trans-Neptunian region* as the region with $a > 30$ AU. Any object in this region will be referred as a TNO. In the same region, I distinguish two main reservoirs, the *trans-Neptunian belt* (30 AU < $a$ < 1000 AU), and the *scattered disk* (perihelion, $q = a(1 - e) < 37$-40 AU), where TNOs with a great variety of orbits (including unusual ones) are found (Duncan & Levison 1997; Luu et al. 1997; Morbidelli et al. 2004). The *classical region* of the belt (or *classical belt*) is confined at 37 AU < $a$ < 45-50 AU and $q$ > 37-40 AU, and contain the bulk of TNOs. Thus, the classical region best represents the remnants of the primordial planetesimal disk in the outer solar system (Morbidelli & Brown 2004; Lykawka & Mukai 2007b; Gladman et al. 2008).

[5] The Asteroids Dynamic Site (AstDyS) – http://hamilton.dm.unipi.it
Minor Planet Center – http://cfa-www.harvard.edu/iau/TheIndex.html

The current trans-Neptunian region architecture is so complex that distinct dynamical populations have been unambiguously identified (Morbidelli & Brown 2004; Elliot et al. 2005; Lykawka & Mukai 2007b). In particular, apart from the unstable class of the Centaurs, there are four main classes of TNOs: classical, resonant, scattered, and detached, TNOs. Classical TNOs are non-resonant objects that orbit in the classical region of the trans-Neptunian belt. These particular TNOs represent the superposition of two different subclasses, the cold and hot populations, which are defined by classical bodies with $i$ < 5-10 deg and $i$ > 5-10 deg, respectively (Morbidelli & Brown 2004; Lykawka & Mukai 2005c; Chiang et al. 2007; Lykawka & Mukai 2007b; Gladman et al. 2008). This division is supported by studies of several properties of these bodies, including distributions of their colors, sizes, inclinations, luminosity functions, and dynamical origin/evolution (Brown 2001; Levison & Stern 2001; Doressoundiram et al. 2002; Hainaut & Delsanti 2002; Trujillo & Brown 2002; Gomes 2003b; Bernstein et al. 2004; Morbidelli & Brown 2004; Peixinho et al. 2004; Lykawka & Mukai 2008; Fraser et al. 2010; Benecchi et al. 2011). The great majority of classical TNOs are considered primordial and possess orbits stable over the age of the solar system. Resonant TNOs are currently trapped in resonances with Neptune[6]. Resonant populations so far observed include the Neptune Trojans at the 1:1 resonance and as distant as ~108 AU, at the 27:4 resonance. Some resonances seem to be more populated, such as the 3:2 ($a$ = 39.4 AU), 5:3 ($a$ = 42.3 AU), 7:4 ($a$ = 43.7 AU), 2:1 ($a$ = 47.8 AU), and 5:2 ($a$ = 55.4 AU) resonances (e.g., Lykawka & Mukai 2007b) (Fig. 1). Importantly, although the orbits of several resonant TNOs may seem to bring them quite close to Neptune's orbit, the nature of their host mean motion resonance acts to protect them from close encounters with that planet, and, as such, they move on orbits that are typically dynamically stable on Gyr timescales (Section 2.1). Scattered TNOs have experienced significant gravitational scattering by Neptune (Duncan & Levison 1997; Gladman et al. 2002; Morbidelli & Brown 2004; Lykawka & Mukai 2006; Lykawka & Mukai 2007c). Some of these bodies can be considered as "scattering" objects as a result of having orbits that are currently strongly interacting with Neptune. The largest and most massive TNO known, (136199) Eris (Bertoldi et al. 2006; Brown et al. 2006a; Brown & Schaller 2007), is likely a scattered TNO (Lykawka & Mukai 2007b; Gladman et al. 2008). Curiously, since several scattered TNOs are weakly interacting with Neptune or even temporarily trapped in resonances with the planet, it is the "scattering" component of this population that may, in fact, be the principle source of the Centaur population, which in turn are the main source of SPCs, as discussed above (Levison & Duncan 1997; Horner et al. 2003; Horner et al. 2004a; Horner et al. 2004b; Volk & Malhotra 2008; Bailey & Malhotra 2009). For simplicity, we call all such unstable objects the scattered TNOs. These bodies have no particular boundaries in semimajor axis, but usually possess perihelion distances close to the giant planet, $q$ < 37-40 AU. Finally, the detached TNOs are non-classical objects that do not encounter Neptune, so they appear to be "detached" from the solar system. Detailed studies have shown that these bodies evolve in very stable orbits, and without macroscopic changes, over the age of the solar system (Gladman et al. 2002; Lykawka & Mukai 2007b). Here, we consider TNOs with $a$ > 45-50 AU and $q$ > 40 AU as part of the detached population. See Lykawka & Mukai (2007b) for a detailed discussion of the identification of these TNOs. Figure 2 illustrates the orbits of TNOs with emphasis on the domains of the scattered and detached populations.

Trujillo et al. (2001a) estimated that the populations of classical and scattered TNOs are roughly comparable for the same size range, but that resonant TNOs should represent no more than 10% of the classical population. However, more recent estimates suggest a higher ratio of classical to scattered populations (Petit et al. 2011). In addition, the intrinsic fraction of resonant TNOs is likely to be higher because only the 3:2 and 2:1 resonant populations were taken into account by Trujillo et al. and other studies conducted before the identification of other resonant bodies (Lykawka & Mukai 2007b). In line with that expectation, recent surveys suggest that the resonant population should consist of ~15-20% of the population in the trans-Neptunian belt (Petit et al. 2011; Gladman et al. 2012). The discovery of the detached population might also reduce the intrinsic fraction of classical and scattered TNOs to the total population in the trans-Neptunian region. Indeed, because of strong observational biases against the discovery of objects evolving with large perihelia, the intrinsic population of detached TNOs probably surpasses that of scattered bodies (Gladman et al. 2002; Allen et al. 2006).

[6] Here, and henceforth, 'resonances' refers specifically to external mean motion resonances with Neptune, described by *r:s*, where *r* and *s* are integers that describe the ratio of the orbital periods of Neptune and the object.

How did the trans-Neptunian belt acquire such a complex excited orbital state composed of dynamically distinct populations? A straightforward explanation is through the gravitational perturbation by the planets over Gyr. In particular, continuous scattering by Neptune naturally results in the excited orbits of scattered TNOs. However, this does not hold for other classes of TNOs (Morbidelli & Brown 2004). Resonance dynamics also played an important role in TNO excitation over similarly long timescales. The origin of the resonant populations in the trans-Neptunian region is better understood by the sweeping mechanism of mean motion resonances. That is, driven by the conservation of energy and exchange of angular momentum between the disk planetesimals and the newly formed giant planets, the latter must have migrated from their birthplaces to current orbits within around 100 Myr (Fernandez & Ip 1984; Malhotra 1995; Hahn & Malhotra 1999; Gomes et al. 2004; Levison et al. 2007). During Neptune's outward migration, all of its exterior mean motion resonances moved in lockstep, sweeping the planetesimal disk, and it is likely that many planetesimals were captured until the end of migration. Thus, captured bodies that survived for billions of years would constitute the currently observed resonant populations. However, as the mean motion resonance capture process is not 100% efficient, possibly a significant fraction of planetesimals were not captured during the process. According to theory, the captured bodies are transported outwards by mean motion resonances with an increase in eccentricity (Peale 1976; Murray & Dermott 1999; Chiang et al. 2007). Depending on the characteristics of the resonance, the inclination may also become moderately excited. Today, the idea that the four giant planets migrated during the early solar system is well supported by several lines of evidence (Liou & Malhotra 1997; Ida et al. 2000b; Levison & Stewart 2001; Chiang & Jordan 2002; Gomes 2003a; Hahn & Malhotra 2005; Levison et al. 2007; Lykawka & Mukai 2008; Lykawka & Horner 2010). Resonant perturbations could also explain the origin of the detached TNOs (Fig. 2). However, although scattered objects can be temporarily detached from the gravitational domain of Neptune by the action of Neptunian resonances, this mechanism cannot account for the intrinsic total population of detached bodies (Gomes et al. 2005a; Chiang et al. 2007; Lykawka & Mukai 2007b). Lastly, it is worth noting that the solely gravitational and resonant effects of the giant planets after the end of migration, even if considered on Gyr timescales, cannot explain the orbital structure of classical TNOs.

In summary, the existence of distinct classes of TNOs reflects the evolution of the solar system through several different processes. Some operated long ago (e.g., planetary migration), whereas others are still active today, such as the gravitational sculpting and resonant perturbations by the planets (Duncan et al. 1995; Lykawka & Mukai 2005c). However, the action of such perturbations by the four giant planets may not fully explain the complex orbital distributions of the classical TNOs and the existence of a substantial population of detached TNOs, thus suggesting that other sculpting mechanisms would be necessary. Alternative mechanisms include passing stars, large (massive) planetesimals that existed in the past, giant molecular clouds, an unseen planet, and a temporarily eccentric Neptune (Kobayashi & Ida 2001; Brunini & Melita 2002; Morbidelli & Levison 2004; Kenyon & Bromley 2004b; Kobayashi et al. 2005; Gladman & Chan 2006; Morbidelli et al. 2008; Lykawka & Mukai 2008; Levison et al. 2008).

In addition to the unexpectedly excited orbital state in the trans-Neptunian region, the discovery of several large TNOs have revealed objects in the 1000-2000 km range, larger than many planetary satellites (Tholen & Buie 1997; Brown et al. 2005). In particular, wide area surveys have suggested the existence of other still unknown objects of similar size or even larger (Brown et al. 2005; Brown et al. 2006a; Delsanti & Jewitt 2006). This is in agreement with the expected largest size in the distribution: >1000 km (Trujillo et al. 2001b). The importance of large TNOs is that we can get some information that otherwise would be impossible with current telescope technology, such as detailed surface spectra, size, and albedo measurements (Delsanti & Jewitt 2006; Muller et al. 2009). Moreover, large TNOs carry an important record of the late stages of planet formation, because the former is believed to be the remnants of accretion (Pollack et al. 1996; Chiang et al. 2007).

Another important evolutionary process involves collisional evolution. Collisions are related to the build up of km-sized TNOs, the loss of the trans-Neptunian belt total mass (via collisional grinding), and transport mechanisms of small bodies across the solar system (e.g., collisions can impart orbital changes to target and impactor objects) (Farinella et al. 2000; Stern 2002; Kenyon et al. 2008). Based on the size distribution of TNOs, the total mass of the trans-Neptunian region was estimated to be around 0.1 $M_\oplus$. However, this value is most likely an upper limit and has uncertainties of at least a few (Bernstein et al. 2004; Chiang et al. 2007). This small total mass for the belt confirms the discontinuity of the mass surface density beyond Neptune, thus yielding values 2-3 orders of magnitude smaller than expected, as discussed earlier. Stern (1995) and Stern & Colwell (1997) showed that 100 km-sized TNOs are unable to form via accretion in the current outer solar system even on billion-year timescales, because too little material is currently available. According to these and other accretion models, TNOs were able to form under the condition that the primordial disk planetesimals were evolving on very cold orbits (Stern & Colwell 1997; Kenyon & Luu 1998; Kenyon & Luu 1999a; Kenyon & Bromley 2004a). In contrast, the trans-Neptunian belt is currently so dynamically hot that the encounter velocities between TNOs are large enough to favor disruptive collisions in the region. Consequently, tens-km-sized bodies (or smaller) are evolving currently in an erosive regime. This implies that 100-km, or larger, TNOs would represent primordial bodies, albeit with surfaces significantly reworked by collisions with other smaller TNOs (Stern 1995; Durda & Stern 2000). Likewise, smaller TNOs (<10 km or so) would consist of collisional fragments, implying that the great majority of SPCs would not be primordial (Farinella & Davis 1996; Choi et al. 2002). The problem of how to grow TNOs can be solved by assuming that the trans-Neptunian region carried more mass in the past. In fact, placing ~10-30 $M_\oplus$ in an annulus from 35 to 50 AU (at least 100 times more massive than now) allows the formation of TNOs with sizes of 100-1000 km in a time scale constrained by the formation of Neptune (Davis & Farinella 1997; Stern & Colwell 1997; Kenyon & Luu 1998; Kenyon & Luu 1999a; Kenyon & Bromley 2004a). The existence of large rubble piles and collisional families in the trans-Neptunian belt also suggests a more violent collisional ambient in the past, which would be reasonable if the belt was initially a much more massive primordial planetesimal disk (Jewitt & Sheppard 2002, Sheppard & Jewitt 2002; Brown et al. 2006b; Brown et al. 2007). Although a massive ancient trans-Neptunian belt can solve the problem of the growth of TNOs, another problem emerges: how can we explain the loss of >99% of its total mass, in order to meet current estimates? Some models suggest that this loss never occurred because the current trans-Neptunian space would have been populated by migrating TNOs from inner regions of the solar system (Morbidelli & Brown 2004; Levison et al. 2008). Other models suggest a mutual contribution from dynamical and collisional erosion over the Gyr (Kenyon et al. 2008; Lykawka & Mukai 2008). In summary, collisional evolution can provide valuable information on the size distribution, internal structure and strength, albedos, colors, rotation states, formation of multiple systems, compositional evolution, and other physical properties, all of which can provide important clues on the solar system evolution as a whole.

In this monograph, I summarize the results of several investigations conducted by myself and collaborators in trying to satisfy the main constraints of the trans-Neptunian region (these constraints are discussed in detail in Section 4). In general, I have investigated some basic physical properties of large TNOs, dynamical (orbital) evolution in the trans-Neptunian belt, the role of mean motion resonances, planetary migration and the presence of massive planetesimals on the origin of the TNO populations. In particular, I present in Section 9 an exhaustive investigation of the effects of a hypothetical resident trans-Neptunian planet on the main populations of TNOs. Thus, this outer planet is possibly one of the largest primitive massive planetesimals that formed during the final assembly of the giant planets, and that supposedly survived in the outskirts of the solar system. I will often refer to this hypothetical planet as (a massive) 'planetoid'.

I finish this introduction with some of many open issues that often motivate researchers and students interested in planetary sciences. Tentative (and likely incomplete) answers to some of these questions can be found in the various sections of this Work.

- What are the albedo and size distribution of TNOs? Are there any correlations between physical properties and orbital elements?
- How did the trans-Neptunian belt acquire its very small total mass?
- What is the role of (mean motion) resonances in the trans-Neptunian region?
- What can the planetary Trojan populations tell us about the evolution of their host planets? What happened to the primordial massive populations of Trojans?
- What is the origin of the distinct classes of TNOs and how do they evolve on Gyr timescales? How did some TNOs acquire detached orbits; in particular, objects such as 2000 CR105 and Sedna?
- What are the most plausible mechanisms for the primordial excitation of the trans-Neptunian belt?
- When, and under what conditions, did the four giant planets migrate through the planetesimal disk during the early solar system?
- What role did collisions play in the evolution of TNOs? How many collisional families formed in the trans-Neptunian belt?
- What is the nature of the trans-Neptunian belt's outer edge at about 50 AU? Does the outer edge represent the true edge of the primordial planetesimal disk, or was it the result of a truncation event in the past?
- Are there more planets in the solar system? Could one or more planets be detected in the near future?
- How could observations help us to discriminate the likelihood of models trying to explain the trans-Neptunian region structure?
- Is a comprehensive model capable of consistently unifying terrestrial and giant planet formation and the evolution of the inner and outer solar system?

This monograph is organized as follows. The first part is dedicated to describing the trans-Neptunian architecture and detailing the importance of resonances in this region (Sections 1-2). Then, I introduce the gamut of TNO populations, discuss their main characteristics and other peculiarities in detail, but with an emphasis on the dynamical and orbital aspects of these populations (Section 3). In the subsequent Sections 4-5, I present the main constraints from the outer solar system with special emphasis on solar system models that include the gravitational perturbation of at least one massive planetesimal. The methods and techniques employed in models developed by myself and collaborators are given in Section 6. In particular, the results are based on models that use detailed long-term integration of fictitious bodies representing primordial planetesimals. The main results of several other investigations are summarized in Sections 7-9. In particular, I discuss the dynamics of certain trans-Neptunian mean motion resonances and their role in the trans-Neptunian region. In addition, other topics include the long-term evolution of the main classes of TNOs, their origin, and interrelation of these populations. The dynamical effects of planetary migration and the existence of massive planetesimals in the outer solar system are also discussed in detail. Finally, conclusions, summaries and future perspectives are given in Sections 10 and 11.

## 2 Resonances in the trans-Neptunian region

The gravitational interaction of planets with a central massive body, like the Sun, gives birth to rich dynamics and secular evolution in the system. That is, according to the planetary disturbing theory, semimajor axes, eccentricities and inclinations of planets and other small bodies suffer periodic variations, which may involve both short and long periods. The longitude of ascending node (Ω) and argument of perihelion (ω) display precessions of a given object with periods of millions of years due to the perturbation of the massive members of the system. The sixth orbital element, the mean anomaly (Λ), describes the position of the body in its orbit for a specific epoch. Thus, the three angular orbital elements (Ω, ω and Λ) circulate through 0-360 deg (see Murray & Dermott 1999 for figures and more details). Because all six orbital elements vary with time, they are also known as osculating orbital elements and their evolution can be described through partial derivatives of the disturbing function according to Lagrange's equations (Brouwer & Clemence 1961; Murray & Dermott 1999). In addition to osculating elements, the proper elements represent the intrinsic characteristics of motion after eliminating periodic perturbations from their osculating orbital elements (Knezevic et al. 2003). Proper elements can be approximately obtained by high resolution orbital averaging using numerical techniques (Morbidelli & Nesvorny 1999). Proper elements are also useful for determining the mobility of bodies in element space represented by dynamical paths transversed by these objects (e.g., Lykawka & Mukai 2006). An approximate, but straightforward, way to obtain proper elements in real time during orbital integrations can be achieved by determining:

$$\langle l \rangle = \frac{1}{N} \sum_{t'=t}^{t'=t+W} l(t') \ , \qquad (1)$$

where $l$ represents any orbital element, $<l>$ is the proper element, $N$ is the number of output values of $l$ over a window of time $W$. This time span should be long enough to cancel out important periodic oscillations of the osculating orbital elements. Objects evolving only in regular periodic orbits (not experiencing resonant behavior or close encounters with a planet) will have proper elements that practically do not change with time (Morbidelli & Nesvorny 1999; Lykawka & Mukai 2006). Thus, their mobility will be confined to tiny regions of element space.

In a system with eight planets and billions of smaller bodies experiencing mutual gravitational and resonant perturbations, bodies that are able to survive on periodic orbits over long timescales must, in principle, have a convenient mutual spacing, otherwise gravitational scattering could become important, destabilizing the system via mutual planet-planet scattering events, or causing small bodies to acquire unstable orbits (e.g., TNOs approaching Neptune, Centaurs evolving in the outer solar system, etc.). The minimum distance from a given planet (or a massive planetesimal) as a criterion for stability can be given according to the planetary Hill radius, or the mutual Hill radius of two adjacent planets. The Hill radius describes the region where the planet's gravitational dominance is more important than that of the Sun or another massive body nearby (Ida & Lin 2004, and references therein). The Hill radius, $R_H$, and mutual Hill radius, $R_{mH}$, are given by

$$R_H = a_P \left( \frac{m_P}{3M} \right)^{\frac{1}{3}} \qquad (2)$$

and

$$R_{mH} = \left( \frac{a_{P1} + a_{P2}}{2} \right) \left( \frac{m_{P1} + m_{P2}}{3M} \right)^{\frac{1}{3}} , \qquad (3)$$

respectively (e.g., Chambers et al. 1996). In these equations, $m$ represents the mass of the planet(s), measured in terms of the Solar mass (natural units, such that $M = 1$), and the subscripts $P$ and $P1$, $P2$ refer to one or two planets. The first case is useful when we consider one planet and its perturbation on small bodies. This is commonly the case for Neptune and TNOs on planet-encountering orbits, or the perturbations that a Centaur experiences when interacting with any of the giant planets (Duncan & Levison 1997; Levison & Duncan 1997). The second case is more useful when we are dealing with the stability of two or more planets in the system. In any case, if a body evolves on an orbit whose approaches to a planet are at least a number of (mutual) Hill radii from it, then, in principle, it can be said that this object has a stable orbit. Likewise, if a body enters the Hill sphere of a planet (defined by its $R_H$), it can evolve under the control of this planet and experience strong gravitational interactions with the latter (e.g., gravitational scattering). Non-negligible gravitational scattering can also occur when the planet-body distances are smaller than a few times the (mutual) Hill radius.

The criterion for stability based on planetary Hill radii is not unique, however. Interestingly, it has been shown that planets, satellites and other minor bodies have a preference for commensurabilities due to dissipative processes in the solar system (e.g., gas drag, tidal effects, disk torques) (Dermott 1973). Resonant configurations are often a natural outcome after these dissipating forces are gone. Those commensurabilities include mean motions, precession frequencies, combination of frequencies and other configurations. Some of these configurations are able to support long-term stability. Commensurability of mean motions (or orbital periods) between two bodies is one of the necessary conditions for establishment of a mean motion resonance. For example, the mean motions of Jupiter and Saturn are almost in a 5:2 ratio (2:5 for their orbital periods, respectively), and similarly Saturn-Uranus (1:3) and Uranus-Neptune (1:2) are also found near mean motion commensurabilities[7] (Malhotra 1998). Many satellites of the giant planets also show orbit-orbit and spin-orbit resonant configurations. Prominent resonant structures do exist in the asteroid belt and the trans-Neptunian region (Nesvorny et al. 2002; Lykawka & Mukai 2007b). Resonant configurations have been found in extrasolar systems as well (Udry et al. 2007; Lissauer et al. 2011). I will discuss mean motion resonances in more detail in the following sections.

Secular resonances occur when the frequency of variation of the orbital elements of a small body becomes nearly equal that associated with a planet. The most relevant secular resonances are those in which the frequency of node precession or perihelion precession of the body is commensurable to the precession frequencies of the major planets (i.e., first order) (e.g., Knezevic et al. 1991). A nodal secular resonance can be represented by $\nu_{1x}$, and a longitude of perihelion (or apsidal) secular resonance by $\nu_x$, with $x$ representing the $x$th planet in order of heliocentric distance from the Sun. For example, The $\nu_{18}$ secular resonance appears when the rate of Neptune's nodal precession is the same that of the body, which essentially affects the orbital inclination of the latter. Likewise, $\nu_8$ is the secular resonance of longitude of perihelion associated with Neptune, which affects orbital eccentricities. These secular resonances can pump up eccentricities and inclinations of solar system bodies on a variety of timescales. In the trans-Neptunian belt, the most important secular resonances are present near 35-36 AU ($\nu_7$ and $\nu_{17}$) and between 40-42 AU ($\nu_7$, $\nu_{17}$ and $\nu_{18}$). These resonances typically excite the orbits of TNOs so that they start encountering Neptune, thus leading to instability by gravitational scattering events. Other associated secular resonances have been found to affect orbits of TNOs (Levison & Morbidelli 2003).

The Kozai mechanism (KM) is defined as the libration of the argument of perihelion rather than circulation (Kozai 1962; Wan & Huang 2007). An object experiencing the Kozai mechanism will tend to avoid the plane of the solar system when located at perihelion in its orbit, thus offering an alternative protection mechanism against close encounters with a planet. Small bodies suffering the perturbations of this secular resonance typically show inverse variations of eccentricity and inclination on Myr timescales. So the eccentricity increases when the inclination drops, and vice-versa (Gallardo 2006a; Wan & Huang 2007).

[7] That is, similarly to the Jupiter-Saturn pair, the ratio of orbital periods of the Saturn-Uranus and Uranus-Neptune pairs are close to 1:3 and 1:2, respectively.

Secondary resonances usually involve commensurabilities between the characteristic libration frequency of objects within a given mean motion resonance and other frequencies, such as the precession and circulation frequencies (Murray & Dermott 1999; Lecar et al. 2001; Kortenkamp et al. 2004). These resonances are usually weak when compared to the other resonances discussed above. However, secondary resonances have shown to be crucial in explaining the chaotic capture of small bodies into the Trojan clouds of both Jupiter and Neptune (Morbidelli et al. 2005; Lykawka et al. 2009).

Three-body resonances involve commensurabilities among three bodies, hence the name. Typically, they can be observed with the mean motions, as in the case of two planets and a minor body. These resonant structures are observed in the asteroid belt, the Jovian satellite system and within the Neptunian Trojan clouds (Murray et al. 1998; Murray & Dermott 1999; Zhou et al. 2009). The nature of a three-body mean motion resonance is typically chaotic (e.g., Nesvorny & Morbidelli 1998).

In summary, the dynamics of resonances can have profound consequences for the dynamical evolution of a system of planets and small body reservoirs. Thus, a minor body may experience stability or instability, migration paths or “sticking” in element space, and/or chaotic behaviour. As a matter of fact, resonance overlapping gives rise to large-scale chaos in the solar system (e.g., Lecar et al. 2001). Indeed, the overlap of secular resonances and mean motion resonances in the asteroid belt is the basis of the origin of most Kirkwood gaps (Moons & Morbidelli 1995). In order to measure how chaotic a system is, the Lyapunov time is a commonly-used tool. For two nearly equal initial orbits, it yields the exponential time scale for the separation of both orbits. Values less than 1 Myr usually indicate chaotic behaviour, which can lead to instability in timescales typically much less than the age of the solar system.

## 2.1 Mean motion resonances

When the mean motions of two bodies are related as a ratio of integers, both objects are said to be in a near mean motion resonance. Consider the case of Neptune and a TNO located in an external near mean motion resonance with the giant planet, with the subscript $N$ referring to Neptune. We can express this relation as

$$\frac{n}{n_N} = \frac{s}{r} \ , \qquad (4)$$

or using the orbital periods,

$$\frac{T}{T_N} = \frac{|r|}{|s|} \ , \qquad (5)$$

where $n$ is the mean motion, $T$ is the orbital period and $r$, $s$ are integers. Using Kepler’s third law, we can derive the position in the semimajor axis for the nominal location of a mean motion resonance as

$$a_{res} = a_N \left( \frac{r}{s} \right)^{\frac{2}{3}} , \quad (6)$$

where $a_N$ is the semimajor axis of Neptune, and parameter $r$ minus parameter $s$ defines the order of the resonance. In general, the order is intrinsically linked to resonance strength since it scales with the body's eccentricity as $e^{r-s}$ in the coefficient of resonant terms of the disturbing function (see Eqs. (8)-(11), and Gallardo 2006b, for more details). However, note that the resonance order is not necessarily always the dominating factor in determining the resonance strength. Several TNOs experiencing Neptune-encountering orbits are often temporarily captured by the planet's mean motion resonances. In this particular case of small bodies evolving on eccentric orbits, the resonance strength becomes a complex function of the semimajor axis and other resonant parameters. Indeed, the parameter $s$ dominates over the resonance order during resonant motion (Lykawka & Mukai 2007c).

Now, consider that the TNO is in conjunction with Neptune. As the time for the next conjunction will be given by $n_N t - nt = 2\pi$, it is easy to find that the time between successive conjunctions for the object at $a_{res}$ is

$$t_{conj} = T_N \frac{r}{r-s} . \quad (7)$$

In the case of a precessing object in an eccentric orbit, the considerations above are valid assuming mean motions ($n_N - \dot{\varpi}$ and $n - \dot{\varpi}$) in a reference frame rotating at the same rate as the perihelion line of the TNO. In this frame, the orbit of the TNO is stationary and, after further derivations, the resonant angle, or resonant argument, is described as

$$\phi = r\lambda - s\lambda_N - (r-s)\varpi , \quad (8)$$

where $\lambda = \Lambda + \varpi$ is the mean longitude and $\varpi = \Omega + \omega$ is the longitude of perihelion (e.g., Murray & Dermott 1999).

The behaviour of the resonant angle tells us whether the TNO is truly locked in a mean motion resonance. If the angle librates about a certain value, the body is said to be locked in mean motion resonance, otherwise the angle will circulate (the near mean motion resonance state). Therefore, a TNO in a mean motion resonance must satisfy two basic conditions: commensurability of mean motions and libration of the resonant angle. It is noteworthy that the librating state offers a protection mechanism against close encounters with Neptune, because when the TNO is at perihelion, the giant planet will always be at a significant distance from that point. Thus, the resonant angle can be understood as the angular distance between the TNO's perihelion line and the longitude of conjunction of the small body with the planet (Fig. 3). The resulting dynamics is such that a mean motion resonance tends to maximize the distance of approaches of the TNO with Neptune. A notable example in the trans-Neptunian belt is Pluto, which has $q \sim 29.7$ AU (i.e., "inside" the orbit of Neptune at 30.1 AU). However, thanks to the resonant protection mechanism, Pluto is always far away from the giant planet when near perihelion (Cohen & Hubbard 1965). The Kozai mechanism is also responsible for keeping Pluto far from the plane of the solar system during its perihelion approach, contributing to its long-term stability against the giant planet's gravitational influence (Williams & Benson 1971).

TNOs experiencing mean motion resonances with Neptune evolve through the libration of their resonant angle. Contrary to the circulation of the resonant angle for non-resonant TNOs, this angle varies periodically and symmetrically about a center of libration, usually 180 deg, for an object trapped in a mean motion resonance. However, symmetric libration around 0 deg is possible for objects possessing sufficiently large eccentricities. In addition, $r$:1 resonances can also exhibit asymmetric libration around centers of libration other than 180 deg (Gallardo 2006a; Gallardo 2006b). Finally, in the case of the 1:1 resonance, libration can occur around 60 or 300 deg on tadpoles about the L4 and L5 Lagrange points, or around 180 deg on horseshoe orbits (see also Section 2.1.2.1). The amplitude of the resonant angle (or libration amplitude), $A$, is another useful quantity in the study of resonant motion, and is defined in this monograph as the maximum distance in longitude from the libration center (or half the full width of the resonant angle). Smaller values of libration amplitudes indicate larger relative distances for Neptune-TNO encounters (Malhotra 1996; Murray & Dermott 1999). When the libration amplitude is too large, the resonant motion may cease and the resonant angle will begin to circulate. The libration amplitude is an excellent proxy for the degree of stability a resonant body can evolve. That is, smaller values of libration amplitudes indicate that the TNO will be deeply locked in the resonance, thus implying long-term stability (from several hundred Myr to the age of the solar system), because the mutual distances between that object and Neptune will remain much larger than the critical values of stability based on the Hill radius of the planet. Likewise, resonant TNOs evolving with larger values of the libration amplitude will tend to display a more unstable orbital behavior, eventually leaving the resonance after several Myr.

### 2.1.1 Physics of mean motion resonances

Let us consider a TNO which is evolving on a resonant orbit with Neptune. In this configuration, the conjunctions of the TNO and Neptune are expected to occur at the same longitude. If this longitude corresponds exactly to the perihelion or aphelion, the tangential force experienced by the TNO is cancelled out for periods before and after the conjunction. Therefore, the change of angular momentum and angular velocity is null. Four arbitrary positions of repetitive conjunctions are indicated by dashed lines at points A, B, C and D in Fig. 4. During the period when the small body is approaching the conjunction at point A, the disturbing tangential force $F_t$ (red arrow) is larger than that immediately after the conjunction, because the TNO and Neptune are in diverging orbits. Since the TNO is slowing down as it approaches the aphelion, its angular velocity is comparable to that of Neptune before reaching point A than after conjunction. Therefore, the small body spends more time experiencing a larger $F_t$ before the conjunction at point A, and less time experiencing a smaller $F_t$ after it. The result is a net loss of angular momentum and a respective gain of angular velocity. Thus, the next conjunction will tend to occur near the aphelion. Similarly, in the case of conjunction at point B the TNO will experience a net gain of angular momentum and a loss of angular velocity, hence the next conjunction will tend to happen towards the aphelion again. The same considerations can be applied to the conjunctions at points C and D and again we find the resonant TNO to be driven closer to the aphelion. In summary, the resonant angle of an object trapped into a mean motion resonance will librate between two extremes around the aphelion (Fig. 3). An object evolving in such a way is known as an aphelion librator. Even if the argument of perihelion varies secularly, the mechanism described above remains a good description of the dynamical motion (Peale 1976; Murray & Dermott 1999).

The disturbing function can be used to describe the resonant motion through Legendre-type series expansions to many orders. As the ratio of the mean motions of the TNO and Neptune are represented by a commensurability, the frequencies of a few of these series terms will approach zero and lead to large amplitude perturbations. Therefore, one term will usually be dominant and the other terms can be neglected (e.g., Murray & Dermott 1999). The general term of the disturbing function averaged to lowest order can be written as

$$\langle \Re \rangle = \frac{Gm}{a} \left\{ \alpha \Re_D + e^{|c_3|} e_N^{|c_4|} S^{|c_5|} S_N^{|c_6|} \left[ \alpha f_d(\alpha) + f_i(\alpha) \right] \cos\phi \right\}, \qquad (9)$$

where $S = \sin(i / 2)$ and $\alpha = a_N / a$. The expressions $f_d(\alpha)$ and $f_i(\alpha)$ are functions of Laplace coefficients and are related to the direct and indirect parts of the disturbing function (Murray & Dermott 1999; Gallardo 2006b). $\Re_D$ is the secular contribution of the direct part and $c_x$ ($x = 1\ldots6$) are integers for which the sum must equal zero (the D'Alembert rule). The general form of the resonance angle is then

$$\phi = c_1\lambda + c_2\lambda_N + c_3\varpi + c_4\varpi_N + c_5\Omega + c_6\Omega_N\ , \qquad (10)$$

or following the case of any ($r$-$s$)th order mean motion resonance described in the form $r{:}s$, where $c_1 = r$ and $c_2 = s$:

$$\phi = r\lambda + s\lambda_N + c_3\varpi + c_4\varpi_N + c_5\Omega + c_6\Omega_N\ . \qquad (11)$$

A number of different classes of resonances can exist depending on their particular combinations of perihelia and nodes, for a mean motion resonance described by Eq. 11 (Dermott et al. 1988). For example, consider the second order 3:1 mean motion resonance. Associated possible resonance classes include: $\phi_1 = 3\lambda - \lambda_N - 2\varpi$ ($e$-type), $\phi_2 = 3\lambda - \lambda_N - 2\Omega$ ($i$-type) and $\phi_3 = 3\lambda - \lambda_N - \varpi - \Omega$ ($ei$-type). Likewise, combinations including $\varpi_N$ or/and $\Omega_N$ lead to more resonance classes, but these can be considered less important compared with $e$-, $i$-, or $ei$-type resonances, because current Neptune's eccentricity and inclination are close to zero. It is worth noting that $e$-type resonances play a major role in the trans-Neptunian belt. As such, the mean motion resonances discussed in this work are of that type (unless specifically stated otherwise). In this case, the general resonant angle reduces to its simpler form, as given by Eq. 8, where the coefficients are $c_1 = r$, $c_2 = s$, $c_3 = (r\text{-}s)$; $c_4 = c_5 = c_6 = 0$.

Based on expansions of the disturbing function, it is also possible to determine the region of influence of a resonance by calculating its boundaries in element space (Murray & Dermott 1999). Thus, the resonance boundaries demark the width of the resonance within which libration can occur. The resonance width scales as $e^{(r-s)/2}$, and is therefore likely to become wider for larger eccentricities. Nevertheless, when the eccentricities become large enough that strong interactions with a planet occur, the resonance width tends to shrink with increasing eccentricities. The libration period can be derived analytically (Murray & Dermott 1999), and scales as $e^{-(r-s)/2}$ (i.e., inversely proportional to the resonance width). Libration periods of typical resonant TNOs are on the order of 10-100 kyr, but their precise period values depend on several factors and the nature of the resonance. In comparison, the libration periods of secular resonances are typically 100-1000 times longer.

Another way to visualize the resonant motion and the interrelations involving the resonant angle, libration amplitude and TNO-Neptune relative distances, is to follow the path of the resonant body in a frame rotating with Neptune. The path in a synodic period is an epicycle with a "cusp" during perihelion passages. These cusps result from different angular velocities relative to Neptune. In the rotating frame, the resonant TNO orbit has $s$-fold symmetry. Therefore, the relative position of Neptune when the body is at perihelion is spaced as (360/$s$) (Malhotra 1996; Murray & Dermott 1999). The following equation gives the longitude of Neptune measured from the perihelion line of the resonant TNO, when the latter is approaching perihelion:

$$\lambda_N = \varpi + \frac{360^\circ b + \phi}{s}\ , \qquad (12)$$

where *b* is an integer. For example, in the case of the resonant motion of Pluto in the 3:2 mean motion resonance, Eq. 12 yields 90 deg and 270 deg after setting $\varpi$ = 0 deg and $\phi$ = 180 deg (the most stable point). Therefore, objects trapped in the 3:2 and other mean motion resonances tend to concentrate near these preferred regions in space, which affects the likelihood of their discovery by standard surveys (e.g., Gladman et al. 2012).

Hereafter, for the sake of brevity and simplicity, external *e*-type mean motion resonances with Neptune represented by the *r:s* ratio will simply be referred to as 'resonances'.

### 2.1.2 Dynamics of particular resonances

The resonant structure of the trans-Neptunian belt has been explored in several works (Morbidelli et al. 1995; Malhotra 1996; Nesvorny & Roig 2000; Nesvorny & Roig 2001; Lykawka & Mukai 2007a) with particular attention being paid to the 3:2 and 2:1 resonances, currently the strongest first order resonances in the trans-Neptunian region. These, and other, resonances play a significant role in the trans-Neptunian belt, leading to diverse orbital evolutions of TNOs when compared to those not in resonances (non-resonant TNOs) (Malhotra 1998; Morbidelli 1997). This discussion also includes higher order resonances; in particular, those located in the scattered disk reservoir (Morbidelli et al. 1995; Murray & Holman 2001; Lykawka & Mukai 2007b). Each resonance has its own particularities and must be studied individually, so that it is not possible to know *a priori* the dynamical behaviour of a specific resonance (Morbidelli et al. 1995). Because the 3:2 and the 2:1 trans-Neptunian resonances have been studied in some detail in the literature, I will give special attention below to other resonances in the trans-Neptunian belt.

#### 2.1.2.1 The 1:1 resonance

The nominal location of the 1:1 resonance is $a_{res}$ = 30.1 AU. The orbital periods of TNOs in the 1:1 resonance (Trojan asteroids or, simply, Trojans) are nearly the same as that of Neptune. In short, Trojans share the same orbit as their host planet, with a resonant angle defined as $\phi = \lambda - \lambda_N$. According to the dynamics of a system of three bodies (i.e., the Sun, a planet (Neptune) and a small body (a TNO)), the regions of stationary solutions for small bodies when "feeling" the gravitational perturbation of the Sun and Neptune are centred at the Lagrange points (L*x*) (Dotto et al. 2008; Horner & Lykawka 2011). However, only the L4 and L5 Lagrange points, displaced 60 deg ahead (leading point) and behind (trailing point) the planet's orbit, respectively, are capable of hosting a significant number of Trojans on long-term stable orbits (Lykawka et al. 2011) (Fig. 5). Trojans evolving on tadpole orbits evolve in such a way that their resonant angle librates about either the L4 (at 60 deg) or the L5 points (at 300 deg), in the same rotating frame where Neptune is fixed at longitude zero. These orbits display "tadpole" paths about either the L4 or L5 Lagrange point in the same frame. Other orbits follow much larger paths moving around both the L4 and L5 points. These orbits display "horseshoe" paths, but objects occupying such orbits usually become unstable on timescales much shorter than the age of the solar system (Lykawka et al. 2011). In the framework of the 3-body problem, objects on horseshoe orbits are likely to suffer a subtle drift in the rotating frame that will lead to instability in relatively short timescales (e.g., Dermott & Murray 1999). However, simulations of N-body systems have shown that the dynamics of objects on horseshoe orbits is complex, so that dynamical lifetimes of the order of Gyr are also possible (Lykawka et al. 2011; Cuk et al. 2012).

Trojans on tadpole orbits evolving with libration amplitudes not larger than 30-35 deg tend to survive for Gyr, even for varied eccentricities and inclinations (Lykawka et al. 2009; Lykawka et al. 2010). However, several studies have shown that the dynamical structure of the 1:1 resonance is very complex, so maps of stability cannot unambiguously constrain the boundaries of the Trojan clouds in element space (Zhou et al. 2009; Zhou et al. 2011). Nevertheless, stable Neptunian Trojans are expected to concentrate on tadpole orbits with $e < 0.15$ and $i < 50$-60 deg. Libration periods of the Trojans are on the order of 10 kyr and the libration mechanism behavior discussed above is an acceptable model to explain the motion of these objects.

Although only eight members have been identified as Neptune Trojans, the intrinsic population is estimated to be as large as their Jovian counterparts, which is on the order of $10^6$ objects larger than 1 km in diameter (Sheppard & Trujillo 2006; Sheppard & Trujillo 2010a; Sheppard & Trujillo 2010b). In this way, the 1:1 resonance is likely one of the most populated in the trans-Neptunian region.

### 2.1.2.2 The 7:4 resonance in the classical region

The nominal location of the 7:4 resonance is $a_{res}$ = 43.7 AU. A body in the 7:4 resonance has a rational relation of periods so that $T = (7/4)T_N$. That is, in the time a 7:4 resonant TNO completes four orbits around the Sun, Neptune completes seven. The 7:4 resonance is a third order resonance with resonant angle defined as $\phi = 7\lambda - 4\lambda_N - 3\varpi$. In the case of a 7:4 resonant object, Eq. 12 yields 45, 135, 225 and 315 deg after setting $\varpi$ = 0 deg and $\phi$ = 180 deg. Therefore, objects trapped in the 7:4 resonance will tend to concentrate near these preferred regions in space. Lykawka & Mukai (2005a; 2006) performed detailed studies of the dynamics of the 7:4 resonance, where they determined the resonance's boundaries (width) in element space and the relation between libration periods and resonant amplitudes for a wide range of possible resonant orbits. In common with other resonant TNOs, the resonance width is quite narrow for smaller eccentricities. For example, at $e$ = 0.05 it is 0.05 AU and becomes larger than 0.2 AU only for $e > 0.12$. For larger eccentricities, the resonance region shrinks because bodies with large libration amplitudes can strongly interact and collide with Neptune. Thus, the maximum width of the resonance is set by objects with large libration amplitudes that do not approach too close to Neptune. The dependence of the resonance width on eccentricity is in agreement with the theoretical scaling, and other discussions in Section 2.1. Objects are expected to remain locked in resonance for Gyr if their libration amplitudes are less than about 150 deg. Moreover, theoretical resonant bodies with larger eccentricities have smaller libration amplitudes, and those with larger inclinations show a systematic increase in their amplitudes. Finally, the libration period as a function of libration amplitude for different eccentricities is illustrated in Fig. 7. In general, libration periods range within 10-100 kyr timescales and strongly depend on the object's eccentricity.

The 7:4 resonance is located within the classical region of the trans-Neptunian belt. This implies that several classical TNOs could be experiencing resonant dynamics over time. Indeed, Lykawka & Mukai (2005a) found evidence that the 7:4 resonance has been affecting the evolution of eccentricities and inclinations of classical TNOs located inside, and near, this resonance over Gyr timescales (Fig. 6). The importance of the 7:4 resonance gained weight with the identification of a substantial number of resonant TNOs currently locked in this resonance (Lykawka & Mukai 2007b).

### 2.1.3 Captures by the resonance sweeping mechanism

We have seen in the Introduction that the interaction of the newly-formed giant planets and the remaining planetesimal disk resulted in the migration of the four giant planets. That migration featured Jupiter moving a little bit inwards, and Saturn, Uranus and Neptune moving outwards, from their original birthplaces in the disk. According to the standard picture, Neptune would have migrated several AU to its current position (30.1 AU) in a period of tens of Myr. As discussed above, because the location of Neptunian resonances are associated with the planet's position, as Neptune migrated, all of its mean motion and secular resonances swept distinct regions of the disk of planetesimals. The first planetary migration studies usually assumed $a_{N0}$ = 23 AU to account for the orbital excitation of 3:2 resonant TNOs (Malhotra 1995; Gomes 1997; Gomes 2000). However, as models improved in complexity with the increasing number of observational constraints, giant planets placed in more compact orbital configurations are now preferred, implying that Neptune may have migrated from locations at ~15-18 AU, or experienced more complex orbital evolutions before its outward migration (Lykawka & Mukai 2008; Levison et al. 2008; Morbidelli et al. 2009b).

Planetary migration is usually implemented in computer codes by adding a small velocity kick along the velocity vector of the planet every time step. Following Hahn & Malhotra (2005), the process is described as follows:

$$\Delta v_x = \frac{\Delta a_x \Delta t}{2 a_x \tau} \exp^{-t/\tau} v_x \quad , \qquad (13)$$

where $v$ is the orbital velocity, $\tau$ is the migration timescale, $t$ is the time, $a_x$ is the instantaneous semimajor axis of the planet, and the index $x$ refers to the planets, where $x = J, S, U, N$ refer to Jupiter, Saturn, Uranus, and Neptune, respectively. $\Delta t$ is the timestep, and $\Delta a_x = \{\Delta a_J, \Delta a_S, \Delta a_U, \Delta a_N\}$ stands for the total radial displacement of the planet (in units of AU). Such exponential behavior was commonly observed in several self-consistent simulations, where the planets were embedded in massive planetesimal disks (Fernandez & Ip 1984; Hahn & Malhotra 1999; Gomes et al. 2004).

These additional velocity kicks result in a torque $\Pi_x = m_x a_x \Delta a_x / \Delta t$, causing the planet's semimajor axis to evolve at a rate

$$\dot{a}_x = 2 a_x \Pi_x / \Xi_x = \left(\Delta a_x / \tau\right) \exp^{-t/\tau} \quad , \qquad (14)$$

where $\Xi_x$ is the angular momentum of the planet.

When integrated, Eq. 14 gives the variation of the planet's semimajor axis,

$$a_x(t) = a_x(F) - \Delta a_x \exp^{-t/\tau} \quad , \qquad (15)$$

where $a_x(t)$ is the semimajor axis of the planet after time $t$, and $a_x(F)$ is the final (current day) value of the semimajor axis. Previous investigations suggest $\Delta a_x$ = {-(0.3-0.2), +(0.7-0.9), +(3-5), +(7-15)} for the four giant planets, $\tau$ = 1-10 Myr, and may represent a good simplification for the migration process (Malhotra 1995; Fernandez & Ip 1996; Friedland 2001; Chiang & Jordan 2002; Chiang et al. 2003; Gomes 2003a; Gomes et al. 2004; Hahn & Malhotra 2005; Levison et al. 2007). The migration timescale $\tau$ tells us how slow/fast the planets migrate from their initial locations to their current orbits. Due to the exponential behavior, after a period of about $5\tau$ the planets have essentially reached their current locations. In reality, given that the migration of Neptune would involve it perturbing and displacing a vast number of smaller bodies varying in size from dust grains to massive planetesimals, the true migration of the planet was probably somewhat stochastic (Hahn & Malhotra 1999; Murray-Clay & Chiang 2006).

In general, the orbital elements of all objects trapped by sweeping resonances evolve conserving Brouwer's integral, given by

$$B = a\left(\sqrt{1-e^2} - \frac{s}{r}\right)^2 \quad . \qquad (16)$$

In other words, because $B$ is conserved during migration (Hahn & Malhotra 2005), Eq. 16 can be used to predict the final semimajor axis and eccentricity of a captured TNO given its initial conditions, and vice-versa. For instance, Malhotra (1995) used this procedure in an attempt to explain the resonant and eccentric orbit of Pluto by backtracking the evolution of Pluto's current eccentricity to approximately zero (when Pluto was supposedly a typical planetesimal with $e = i \sim 0$ in the protoplanetary disk).

Another useful relation is the minimum eccentricity needed for adiabatic resonance capture:

$$e^{r-s} = \frac{\dot{a}_N M T_N}{8 K s m_N a_N} \ , \qquad (17)$$

where the rate at which Neptune is migrating is calculated as $\dot{a}_N = [30.1 - a_N(t)]/\tau$, $m_N$ is the mass of Neptune, and $K$ is a constant dependent on the specific resonances (Dermott et al. 1988; Murray & Dermott 1999; Hahn & Malhotra 2005). The orbital evolution of objects captured by sweeping resonances is usually dictated by the conservation of $B$, if they do not suffer scattering events by Neptune during its migration.

More recently, Lykawka & Mukai (2007a) used the above techniques to infer the orbital conditions of the primordial planetesimal disk during Neptune's migration and the origin of stable TNOs locked in more distant resonances, namely the 9:4 ($a$ = 51.7 AU), 5:2 ($a$ = 55.4 AU) and 8:3 ($a$ = 57.9 AU) resonances. The theory of the resonance capture mechanism, applied in a broad context, is also discussed in Mustill & Wyatt (2011).

### 2.1.4 The resonance sticking phenomenon

In principle, TNOs that acquire unstable Neptune-encountering orbits are expected to be ejected from the solar system or to collide with a planet or the Sun due to continuous gravitational scattering by the giant planets. However, apart from the scattering events by Neptune, these objects can also experience temporary resonance captures in several resonances across the trans-Neptunian belt. This phenomenon is called resonance sticking[8]. Therefore, in addition to the resonance sweeping mechanism, TNOs (in particular those on relatively unstable orbits due to encounters with a planet) are able to experience resonance captures any time in solar system history. In short, these TNOs tend to evolve chaotically through their dynamical histories by intermittent gravitational scattering by Neptune and temporary resonance captures with the same planet. The chaotic behaviour is believed to arise due to the action of high order secular and secondary resonances, beyond the scope of analytic methods (Morbidelli 1997). Therefore, numerical simulations are essential in order to grasp the details of the dynamics of TNOs evolving on such unstable orbits. An example is illustrated in Fig. 8.

Although this resonance sticking behavior was noted by Duncan & Levison (1997), dynamical surveys of scattered TNOs have focused exclusively on the gravitational scattering by Neptune (Gomes et al. 2008 and references therein). This picture changed with the detailed investigations performed by Lykawka & Mukai (2007c), which confirmed preliminary investigations (Lykawka & Mukai 2006) that showed that resonance sticking is a very common phenomenon in the solar system.

[8] It is worth noting that other solar system bodies can experience resonance sticking, such as, for example, the Centaurs (discussed in Section 4.6.1), and the asteroids (e.g., Gallardo 2006b). However, certain resonances in the asteroid belt are more prone to temporarily "stick" asteroids, while other resonances can lead to strong instability (the "Kirkwood gaps"). Such instability arises when resonances overlap with giant planets' secular resonances (see e.g., Nesvorny et al. 2002 for a detailed discussion on the dynamics of resonances in the asteroid belt).

In particular, Lykawka & Mukai (2007c) performed a detailed investigation of the dynamical evolution of 255 particles evolving as typical scattered TNOs that survived the full 4 Gyr of orbital integration[9]. Based on that work, in general scattered particles were on average captured in 88 distinct resonances, and spent ~38% of their lifetimes trapped in resonances. Temporary captures covered more than 600 distinct resonances. Resonance sticking proved to play an important role in determining the surprisingly high survival rate of scattered TNOs after billions of years, confirming earlier expectations (Malyshkin & Tremaine 1999). Even those objects that are eventually ejected from the solar system do so with ejection timescales enhanced by their previous temporary resonant capture evolutions (Lykawka & Mukai 2006; Lykawka & Mukai 2007c). The slow circulation of the argument of perihelion and the Kozai mechanism can also help to enhance the dynamical lifetimes of some scattered TNOs. Also, we should note that if the longitude of perihelion and the longitude of ascending node circulate much slower than those of the planets, we do not expect to find any strong (low order) secular resonances inside distant resonances beyond 50 AU (see also Fig. 2). In addition, resonant objects exhibited larger perihelia when compared to those not in resonances, and the $r$:1 and $r$:2 resonances played a major role in the whole evolution (Lykawka & Mukai 2004; Lykawka & Mukai 2006; Gallardo 2006a). Finally, resonance sticking is unimportant beyond about $a$ = 250 AU (Fig. 9).

The web of resonances that determined the multiple captures often followed captures in combined resonances of increasing parameter $s$ lying between two $r$:1 resonances, following the Farey sequence (Hardy & Wright 1988), and given by $(r_1 + r_2) / (s_1 + s_2)$, where the pair of arbitrary resonances is described by $r_1$:$s_1$ and $r_2$:$s_2$. Of course, this relationship cannot explain all jumps between the resonances, nor can it predict the next resonance capture, since the evolution is chaotic.

The likelihood of capture into a resonance, and the ability of that resonance to retain a captured object (i.e., timescale) is defined by the stickiness concept. Resonances of lower order (and lower $s$), or those located closer to the Sun, have larger stickiness, so TNOs moving on unstable orbits should preferentially be captured, and stay longer, in these resonances. In other words, resonances with the lowest argument $s$ and within the region at $a$ < 250 AU dominate the resonant evolution in the trans-Neptunian region. Finally, it is worth noting that the KM often operated for objects captured in $r$:1 resonances within the same region (Gomes et al. 2005; Gallardo 2006a; Gallardo 2006b; Lykawka & Mukai 2007c).

Resonance stickiness is intrinsically related to the strength of the resonance described by the strength function $SR(a,e,i,\omega) = \langle\Re\rangle - \Re_{min}$, where $\Re$ and $\Re_{min}$ represent the resonant disturbing function and its minimum value (see Gallardo 2006b, and Lykawka & Mukai 2007c, for details). Objects trapped in resonances with high $SR$ are expected to suffer stronger dynamical effects during their evolution. The close similarities between resonance stickiness and resonance strength imply that, the stronger the resonance, the longer the temporary captures and the higher the capture probabilities during resonance sticking. This is in agreement with the idea that scattered TNOs should preferentially be captured in stronger resonances, because the latter possess wider resonance widths and penetrate to lower eccentricities (see Eq. 9). The significant contribution of resonances at $a$ < 250 AU during resonance sticking reflects the presence of sufficiently strong resonances, which have lower $s$. The increase of perihelion is strongly related to resonances with higher resonance strength (stickiness) and correlated with resonance residence time. A detailed map of strength/stickiness of several resonances is illustrated in Fig. 9.

[9] These simulations are unlikely to have produced artificial resonance sticking, since temporary captures in resonances are not dependent on the minutiae of how the calculation is done during close encounters (Holman & Wisdom 1993; Duncan & Levison 1997).

# 3 Observing small bodies in the trans-Neptunian region

## 3.1 Observations and biases

The list of TNOs and other small bodies in the solar system is based on observations taken by ground and space telescopes. The orbits are available in public online databases (see Footnote 3). However, the available sample of TNOs with orbits accurate enough for follow-up observations, statistical analysis and dynamical models is currently very small (1-2% of the intrinsic population) and limited to large sizes, $D > 50$-100 km. This is not surprising, since such observations are challenging: TNOs are dark, relatively small and orbit at large distances from the Sun (Williams 1997; Davies 2001; Jewitt 2008). Usually, observations are concentrated near opposition, when a TNO has the highest relative sky motion due to parallax. This allows the heliocentric distance at discovery to be determined, and can therefore be used as a guide to detect TNOs, since the apparent motion of the candidate is usually 2-5 arcsec/h (e.g., Jewitt 1999), as can be determined by:

$$\dot{\theta} = \frac{148}{R+\sqrt{R}} , \qquad (18)$$

where $R$ is given in AU. In general, TNOs with observations at more than one opposition are said to have reliable orbital elements. That is, these orbits typically have observational arcs longer than 1 yr. Figures 1 and 2 illustrate the still poorly-characterized orbital distribution of the trans-Neptunian region.

Most surveys focus on ecliptic latitudes, $\beta$, near the ecliptic plane ($\beta \sim 0$), where more TNOs are expected to be found based on cosmogonical grounds (i.e., based on their origin in a primordial disk of planetesimals with very cold orbits, $e = i \sim 0$). This strategy discriminates against the discovery of higher inclination TNOs because they spend less time crossing the ecliptic plane. Those TNOs with larger eccentricities can spend even less time in this region. For larger ecliptic latitudes, it is impossible to detect objects such that $i < \beta$. Besides, other observational biases include the fact that discovered bodies have, in general, $i \sim \beta$ and objects with smaller perihelia are easier to detect, given the same detection limit based on brightness. Other biases include assumptions when computing the orbits to meet the criterion of the “simplest orbit” that fits the data. A few examples include: (1) A circular orbit ($e = 0$) if the object is not in a close encounter situation with Neptune; (2) A 3:2 resonant orbit if the fit implies a close encounter of the object with Neptune; (3) A “Väisälä” orbit with the object near perihelion for very poor circular orbit fits, if the object is not encountering Neptune. Several observed TNOs are biased toward such orbits (Bernstein & Khushalani 2000). Other biases, and observation limitations, are discussed in Jewitt & Luu (1995) and Horner & Evans (2002), while systematic biases in the distribution of TNOs are detailed in Jones et al. (2010). Lastly, surveys typically employ different observation strategies, telescopes, and analysis techniques, so biases and difficulties arise when trying to compare or combine observations carried out in distinct surveys (Fraser et al. 2010; Fuentes et al. 2011). For this reason, recent surveys have been carefully tracking their discovered objects in order to provide “bias-free” observational constraints (e.g., the CFEPS survey. See Petit et al. 2011, and references therein).

In addition, a large fraction of TNOs have orbital elements calculated from observations with less than one year arcs (only one opposition). Several of these TNOs are presumably “lost” because of large orbital uncertainties that preclude follow-up observations. After analyzing the discovery circumstances of a particular group of TNOs via inspection of the current values of their orbital elements and those assumed at discovery, the observational uncertainties from first time detection and current orbital elements are commonly very large for semimajor axes and eccentricities. In the case of inclinations and absolute magnitudes, the uncertainties are much smaller. The reason is that the inclination is easily determined from the proper motion in the sky when taking the ecliptic as reference, so even short arc observations can yield relatively accurate inclinations (Virtanen et al. 2003). In addition, the determination of inclination is not dependent on other orbital elements (Williams 1997). Concerning absolute magnitude distributions, since the distance of observation is well constrained, the estimates are quite good and have small uncertainties.

This simple analysis implies that TNOs with orbits determined from short arc observations are not suitable for dynamical classification (which depends mostly on semimajor axis and eccentricity), and are only useful for studies involving the distribution of inclinations, and/or absolute magnitudes.

## 3.2 The Cumulative Luminosity Function (CLF) and size distribution

By knowing the heliocentric distance, absolute magnitude and surface reflectivity (albedo) of a small body, it is possible to estimate its size. In fact, TNOs possess a variety of sizes, spanning a spectrum from dust to planetary-sized (dwarf planet) objects, which defines the size distribution of the population. Important information can be extracted from the size distribution of TNOs. For example, the distribution of mass, accretion conditions and collisional evolution in the trans-Neptunian belt (Fraser 2009). Normally the size distribution is derived from the Cumulative Luminosity Function (CLF). The CLF describes the cumulative number of TNOs per sky surface area for a given limiting magnitude. The CLF is denoted by

$$\Sigma(h) = 10^{C(h-h_0)} \quad , \qquad (19)$$

where $\Sigma(h)$ is the number of objects per square degree brighter than the apparent magnitude $h$, $C$ is a constant defining the slope of the curve and $h_0$ is a constant for which $\Sigma = 1$ deg$^{-2}$. In general, surveys have revealed small sky densities for brighter bodies (bigger TNOs) and larger sky densities for fainter ones (smaller TNOs). Steeper slopes $C$ indicate that more small bodies and less large objects are present in the distribution. For example, for such a surface density one would expect to find one 1000 km-sized TNO per 1000 deg$^2$, compared to about ten 100 km-sized bodies for 1 deg$^2$. Therefore, bright TNOs are more likely to be discovered in volume-limited surveys than those that are flux limited. Conversely, the opposite is true for smaller TNOs. According to dedicated observations, typical values are $C = 0.6 \sim 0.7$ and $h_0 \sim 23$ (Trujillo et al. 2001b; Elliot et al. 2005; Jones et al. 2006). However, note that because the trans-Neptunian populations consist of a mix of distinct main classes of TNOs with somewhat different physical/orbital properties (Section 4), it is not possible to derive a single CLF (and an associated single size distribution, as discussed below) to the entire population of TNOs. For instance, populations of TNOs moving on dynamically cold or hot orbits seem to possess statistically distinct CLFs (Bernstein et al. 2004; Fraser et al. 2008; Fraser et al. 2010).

The size distribution is typically derived from Eq. 19 under the assumptions of constant heliocentric distance and albedo (e.g., Elliot et al. 2005). The first assumption is reasonable, but the second is more problematic, since the distribution of TNO albedos displays a wide range of values, which also appear to depend on an object's size and dynamical class (Lykawka & Mukai 2005b; Stansberry et al. 2008; Muller et al. 2009). The size distribution can be expressed as

$$N(D)dD = \gamma D^{-j} dD \quad , \qquad (20)$$

or its associated cumulative size distribution:

$$N(>D) = \Gamma D^{-k} \quad , \qquad (21)$$

where $N$ is the number of objects, $\gamma$ and $\Gamma$ are constants and $j$ and $k$ represent the slopes of the distributions. The slope $j$ can be obtained from the CLF via $j = 5C + 1$. For smaller values of the slope $j$, more mass is concentrated in the larger bodies. From the observed CLF and derived size distributions, it is possible to estimate the total mass of particular TNO populations.

More recent surveys, including those that used the Hubble Space Telescope, found a dearth of small TNOs in the $25 < h < 29$ mag range; viz., discovery statistics of only ~3% of that expected from a single power law differential sky density for bright TNOs (Bernstein et al. 2004). This strongly suggests that a broken power law may be more appropriate to describe the size distribution of TNOs. That is, the size distribution of TNOs probably breaks at a size range around ~50-120 km, so smaller objects would follow a shallower distribution than their larger counterparts. The current distribution of TNOs with larger diameters is best described by $j = 4.8$, while the slope for smaller objects is unknown (Gladman et al. 2001; Fraser et al. 2008). Information on the size distribution of smaller TNOs relies on collisional evolution studies or upper limits derived from occultation surveys (e.g., Fraser 2009; Bianco et al. 2010). This is an important issue, since it relates to the total mass of distinct TNO populations and the origin and evolution of SPCs as small TNOs that penetrate the inner solar system after evolving over millions of years from trans-Neptunian distances (e.g., Jewitt 2002). In the near future, the observation of TNOs of cometary size (typically a few kilometers in diameter) will probably rely on stellar occultations (Cooray & Farmer 2003; Roques et al. 2003; Schlichting et al. 2009) or highly-sensitive survey programs (Larsen et al. 2007; Yoshida et al. 2011).

### 3.3 Identifying groups of TNOs in orbital space: Motivations and goals

- What are the main classes of TNOs, and what are their boundaries in orbital element space?
- What are the origin, evolution, and relative intrinsic fractions of distinct classes of TNOs?
- Which TNOs are currently in resonance with Neptune? Under which conditions are TNOs likely to evolve to Neptune-encountering orbits? (suffering gravitational scattering by the planet).
- What can peculiar TNOs (or TNO groupings) tell us about the solar system?

A dynamical classification system is useful because it can provide clues to the aforementioned questions, and consequently provide a better understanding of relevant evolutionary processes in the outer solar system. TNOs, properly classified, can provide new insights, or be used in statistical studies to investigate correlations of orbital elements with other parameters of interest, such as physical properties.

The identification of distinct classes of TNOs is a difficult task, since all TNOs appear to mix smoothly in element space so that no clear boundaries are known. During the first years of observations, even when the number of discovered TNOs was still very small, the existence of resonant populations in the 3:2 and 2:1 resonances was readily apparent based on clusterings of TNOs around the nominal semimajor axes of both resonances (Jewitt 1999). However, even if orbital uncertainties are very small, the osculating semimajor axis is not adequate for resonance identification, since it can vary by 0.5-1.5 AU, or even a few AU, after a short integration into the future for orbits in the classical region, and beyond, respectively. That is, the osculating semimajor axis at a certain epoch of a TNO near a resonance location cannot really tell us whether the body is in resonance or not. Instead, it is necessary to integrate the orbits of TNOs to verify if the resonant angle is librating or circulating (Section 2.1; Malhotra 1996; Murray & Dermott 1999).

Classical TNOs concentrate in a region of orbital element space up to about 50 AU. However, the existence of subpopulations and resonant families within the same region further complicates the task of identifying and characterizing this population. For instance, classical TNOs are traditionally considered to lie between an inner edge at 40-42 AU and an outer edge at 48-50 AU (Jewitt et al.

1998; Kuchner et al. 2002), but the inner and outer boundaries are difficult to establish. Moreover, the presence of resonances (e.g., 5:3, 7:4, 2:1, etc.) and possible close encounters with Neptune at perihelion distances $q$ < 37 AU, which may indicate the existence of scattered TNOs within the classical region, further complicate the situation (Kuchner et al. 2002; Lykawka & Mukai 2005c; Lykawka & Mukai 2007b; Volk & Malhotra 2011). Tentative classification schemes have been proposed as the number of observed objects increased beyond Neptune and among the giant planets (Horner et al. 2003; Tiscareno & Malhotra 2003; Morbidelli & Brown 2004; Elliot et al. 2005). It turned out that there are at least five main dynamical classes in the outer solar system: classical, resonant, scattered, and detached TNOs, and the Centaurs, consisting of non-resonant objects that cross the orbit of a giant planet. On the other hand, these classification systems provide quite limited resonance identification, use arbitrary thresholds to define orbital boundaries of non-resonant TNOs, and do not discuss the origin, evolution, nor the interrelation of identified classes. Nevertheless, classification schemes recently achieved a higher degree of sophistication and accuracy (Lykawka & Mukai 2007b; Gladman et al. 2008). In this paper, I focus on the scheme proposed by Lykawka & Mukai (2007b), which has several improvements over earlier work and accurate detailed features, namely secure identification of resonant TNOs in the entire trans-Neptunian region, determination of the boundaries for all non-resonant populations, secure identification of detached TNOs, identification of Kozai TNOs (resonant TNOs that experience the Kozai mechanism) and the determination of resonant properties (e.g., libration amplitudes) for all resonant populations.

The classical region of the trans-Neptunian belt is of special interest because the classical TNOs carry important clues about the early solar system, since they are supposed to be the best representatives of the primordial planetesimal disk. However, key questions include the following:

- What are the boundaries of the classical region?
- How can we distinguish between classical TNOs and other dynamical classes?
- What is the best threshold in inclinations to divide cold and hot subpopulations?
- Which parts of the classical region are stable over the age of the solar system?

The 40-42 AU subregion is also of interest because it is related to the inner boundary of the classical region, and TNOs in that region possess exclusively high inclinations ($i$ > 10 deg). Are these bodies hot classical TNOs? It is important, therefore, to carry out a proper identification and characterization of classical TNOs to better understand the origin and evolution of these objects and the early conditions of the primordial planetesimal disk from which these objects originated.

It is important to note, here, that the classical TNOs can be significantly affected, in the long-term, by resonances. For instance, TNOs on orbits near, or inside, the 7:4 resonance ($a$ = 43.7 AU) can experience non-negligible orbital evolution (Fig. 6). Indeed, this is one of the strongest resonances in the trans-Neptunian belt (Lykawka & Mukai 2005a; Gallardo 2006b). There are other strong resonances that could sculpt the orbital structure of classical TNOs over Gyr timescales, namely the 8:5 ($a$ = 41.2 AU), 5:3 ($a$ = 42.3 AU), 9:5 ($a$ = 44.5 AU), 11:6 ($a$ = 45.1 AU), 2:1 ($a$ = 47.8 AU), and other higher order resonances (Fig. 10). One such high order resonance is the 12:7 resonance ($a$ = 43.1 AU), which proved crucial in providing constraints on the dynamical history of the dwarf planet (136108) Haumea, and its associated collisional family (Brown et al. 2007; Lykawka et al. 2012). Therefore, the long-term effects of several resonances in the classical region are of great interest, as they may be inducing eccentricity and/or inclination changes over the age of the solar system (e.g., Volk & Malhotra 2011). Such orbital changes may be large enough to cause a classical TNO to change its subclass (i.e., cold to hot, after an increase of $i$) (Fig. 11).

In addition, there are several open questions about the role of distant resonances, defined as those resonances located beyond the 2:1 ($a$ > 47.8 AU) or simply as scattered disk resonances. For example:

- Could distant resonances harbor members that have survived there since the early solar system?
- What is the origin of resonant TNOs in the scattered disk? Can planetary migration give any clue

as to their origin? What can they tell us about the early solar system?
- Can the resonance sticking phenomenon alone account for the resonant structure in the scattered disk?
- Is the Kozai mechanism relevant for the evolution of TNOs located beyond the 2:1 resonance?

Strong resonances in the trans-Neptunian belt can capture most planetesimals in nearly circular and low-*i* orbits for typical migration timescales (Section 2.1.3; Malhotra 1995; Murray & Dermott 1999; Chiang et al. 2007). This scenario can reasonably account for the resonant populations out to the location of the 2:1 resonance, but fail for more distant resonances, such as the 5:2 (Chiang et al. 2003; Hahn & Malhotra 2005; Lykawka & Mukai 2008). The reason is that the probability of capture into these distant resonances (or resonances with order greater than, or equal to, two, in general) is close to zero for initially low-*e* planetesimals (e.g., Dermott et al. 1988). Lykawka & Mukai (2007a) went further, and investigated mechanisms that could reproduce not only TNOs in the 5:2 resonance, but also in the other occupied resonances beyond 47.8 AU, namely the 9:4, 7:3, 12:5, 8:3 and 3:1 resonances. These studies showed that resonant TNOs in distant resonances could have originated either from the migration of Neptune through a stirred planetesimal disk (via sweeping resonance capture) or from the evolution of typical scattered TNOs (via resonance sticking of temporary captures with significant residence time).

## 4 Trans-Neptunian populations and constraints for solar system models

The current orbital structure in the trans-Neptunian region is complex and consists of different families of TNOs evolving on orbits with wide ranges of eccentricities and inclinations. This is surprising, because in the absence of planet formation, remnants of a primordial planetesimal disk are expected to preserve its cold orbital conditions (Kenyon & Luu 1999b; Luu & Jewitt 2002; Kenyon & Bromley 2004a) (Figs. 1 and 2). Since all TNOs were once part of this disk, ~4.5 Gyr ago, the current existence of distinct dynamical classes in the trans-Neptunian belt suggests several mechanisms have been sculpting the outer solar system. Did the main classes of TNOs originate from similar or distinct regions of the disk? How have they evolved over long timescales? Can we identify their formation sites based on a dynamical classification of TNOs?

Dynamical stability studies can reveal in which regions of *a*-*e*-*i* element space TNOs are likely to survive, for the age of the solar system, without significant perturbations. Dynamical lifetimes and the characterization of certain unstable regions are also important to give an insight into the process of the delivery of SPCs into the inner solar system.

Several studies pointed to the following picture in the trans-Neptunian belt:
- Orbits among the giant planets are unstable with time scales of 10-100 kyr between Jupiter and Saturn and of 1-10 Myr for the Saturn-Uranus-Neptune regions (Holman & Wisdom 1993; Grazier et al. 1999; Kuchner et al. 2002; Tiscareno & Malhotra 2003; Horner et al. 2004a; Horner et al. 2004b; di Sisto & Brunini 2007).
- The region surrounding Neptune is rapidly cleared up to about 35 AU (Holman & Wisdom 1993; Malhotra 1995; Lykawka & Mukai 2005c). This can be understood in terms of the overlapping of first order mean motion resonances near Neptune resulting in chaos (e.g., Chirikov 1979).
- Considerable instability between 35-36 AU and 40-42 AU (at low inclinations) with timescales of 10-100 Myr (Holman & Wisdom 1993; Levison & Duncan 1993; Lykawka & Mukai 2005c). Both regions are associated with the overlapping of secular resonances featuring Uranus and Neptune respectively, causing instability (Knezevic et al. 1991; Morbidelli et al. 1995b).
- In general, objects with perihelion $q < 35$ AU are unstable, with exceptions for bodies in resonances and at $i < 35$ deg (Duncan et al. 1995; Kuchner et al. 2002; Lykawka & Mukai

2005c).

## 4.1 Classical TNOs

Classical TNOs orbit within the classical region of the trans-Neptunian belt, which extends out to about 50 AU. Inspecting the distribution of TNOs in Fig. 1, there seems to be a lack of objects beyond the 2:1 resonance. In addition, there is an absence of low-*i* TNOs in the 40-42 AU subregion. Therefore, TNOs that remained on dynamically cold orbits representing the primordial disk planetesimals would reside between about 42 and 45-50 AU. I follow the findings of Lykawka & Mukai (2007b) in this paper, who proposed a classification scheme for classical TNOs based on the direct comparison of long-term simulations (4 Gyr) of fictitious bodies with integration of observed classical objects, resulting in classical TNOs being those bodies spread between ~37 AU < $a$ < 45(50) AU with $q$ > 37 AU. However, the classical region is complex, and uncertainties arise because of a number of factors: (1) Stability is $i$-dependent; (2) Resonances affect stability and orbital evolution in their neighborhood; (3) The existence of the cold and hot populations (Duncan et al. 1995; Kuchner et al. 2002; Gomes 2003b; Chiang et al. 2007; Volk & Malhotra 2011).

It should be noted that although the classical region boundaries cover the range 37-45(50) AU and $q$ > 37 AU, TNOs trapped in resonances in this region are not normally considered members of the classical population, though we cannot know *a priori* which of these resonant bodies were originally classical or not. The perihelion boundary is tentative, but tells us the lower limit at which classical TNOs could be found. However, the bulk population of classical TNOs stable over the age of the solar system would be confined to $q$ > 39-40 AU (Duncan et al. 1995; Lykawka & Mukai 2005c; Lykawka & Mukai 2007b).

Other important questions are: what is the intrinsic outer boundary of the classical region? How can we differentiate between classical and detached TNOs? A hint comes from the unbiased radial structure of the trans-Neptunian belt. Strong evidence suggests an outer edge at about $R$ = 50 AU, with uncertainties of a few AU (Trujillo & Brown 2001; Morbidelli 2005). Assuming that the classical region is composed of objects formed in situ, the trans-Neptunian belt's outer edge could naturally set the outer boundary of the classical region. However, because observations are scarce, we cannot exclude the possibility of a slightly larger $a$-value. If this is the case, some classical bodies would possibly appear as detached TNOs just beyond the 2:1 resonance. Indeed, 2003 $UY_{291}$ ($a$ = 49.5 AU, $i$ = 3.5 deg, $q$ = 41.3 AU) could be one such object. In sum, there is no way to set an accurate division between classical and detached TNOs, based on current observations.

Concerning statistical predictions of the number of large classical TNOs, Trujillo et al. (2001b) have estimated that a few ~2000 km-sized TNOs and tens of ~1000 km-sized bodies would exist in the classical region. Earlier studies estimated the number of classical TNOs larger than 100 km to be around 38,000 for a size distribution slope $j$ = 4.0 (Trujillo et al. 2001a), or 47,000 bodies assuming $j$ = 4.2 (Trujillo et al. 2001b), with an approximate mass of 0.03 $M_{\oplus}$. Bernstein et al. (2004) also obtained similar total masses for the classical region. However, more recent dedicated surveys estimated the classical region would contain about 103,000~160,000 objects (Petit et al. 2011). Finally, the growth of large classical TNOs is supported by accretion studies (over 10-100 Myr timescales) (Kenyon & Luu 1998; Kenyon & Luu 1999a; Kenyon & Luu 1999b).

### 4.1.1 Cold and hot populations

The classical TNOs are strongly believed to be the superposition of two different dynamical populations, also known as the cold and hot classical TNOs. That is, the cold classical population would have formed locally in the planetesimal disk between about 35 AU and the original edge of the disk. The hot population would represent planetesimals that formed in the inner solar system (15-35 AU) and later entered the classical region (Gomes 2003b). The superposition of cold and hot populations is apparent only among the classical bodies. Thus, the cold classical TNOs could constitute the local population of remnant planetesimals left after planet formation in the outer solar system. Alternatively, cold classical TNOs could also have come from the inner solar system, transported by the sweeping 2:1 resonance (Levison & Morbidelli 2003; Levison et al. 2008). In both scenarios above, the cold and hot classical TNOs would presumably have different compositions as a result of their distinct birthplaces in the planetesimal disk. The orbital distribution of both populations is illustrated in Fig. 12.

The identification of two populations in the classical region is supported by the following observations.

(1) *Colors/Spectra*. The cold population contains predominantly redder objects, while the hot population has slightly red to very blue bodies. The majority of the redder classical TNOs are concentrated at approximately $i < 10$ deg, while a few objects with neutral colours are found at larger inclinations (Tegler & Romanishin 2000; Doressoundiram et al. 2001; Doressoundiram et al. 2002). Indeed, a correlation between inclination and color was found with a tendency for neutral/bluer classical TNOs towards higher inclinations (Trujillo & Brown 2002; McBride et al. 2003; Benecchi et al. 2011). These correlations were also confirmed by statistical analysis of inclinations with spectra (Fornasier et al. 2009).

Contrary to this idea, one could expect the color difference between both populations to be caused by the inclination itself, as more energetic collisions could be capable of blueing the more excited high-inclination TNOs. In this case, we would expect a correlation between colors and other orbital elements related to impacts, such as semimajor axis and eccentricity, as well as a correlation between colors and inclination inside any class or subpopulation of TNOs. These hypothetical correlations or trends have not yet been observed. Therefore, distinct distributions of colors for both subpopulations and the existence of the mentioned correlation are together strong evidence that the cold and hot classical TNOs are compositionally different, suggesting different dynamical classes (Morbidelli & Brown 2004).

(2) *Size distribution*. The largest cold classical TNOs are predominantly smaller than hot ones. In fact, the hot population contains the majority of the largest TNOs. These features are likely not caused by observational biases (Levison & Stern 2001; Lykawka & Mukai 2005b). Apparent correlations between colors and sizes with inclinations have been found with high confidence levels, implying different compositions and/or birthplaces in the planetesimal disk (Tegler & Romanishin 2000; Levison & Stern 2001; Trujillo & Brown 2002; Doressoundiram et al. 2002; Hainaut & Delsanti 2002; Doressoundiram 2003; McBride et al. 2003; Fraser et al. 2010). Noteworthy, these properties and correlations with inclinations are not caused by impact effects (Morbidelli & Brown 2004).

Moreover, Bernstein et al. (2004) found evidence of different size distributions for both a “classical” population ($i < 5$ deg) and an “excited” population ($i > 5$ deg), which are analogs to the cold and hot populations discussed here. Their total masses were estimated at 0.01 $M_{\oplus}$ for the cold population, and a few times that value for the hot counterpart. These results were confirmed by more recent dedicated surveys, which indicate that the cold population has a steeper size distribution than the hot counterpart (Fraser et al. 2010). From an analysis of the absolute magnitude distribution of classical TNOs, Lykawka & Mukai (2005b) confirmed that cold and hot classical TNOs present a different concentration of large bodies. In addition, more massive classical bodies are anomalously present at $a < 43.5$ AU, a result statistically significant (>99.8%), and apparently not caused by observational biases.

(3) *Inclination distribution*. The inclination distribution of classical TNOs seems to be best fit by a sum of two Gaussian distributions, thus indicating that at least two dynamically distinct classes of objects populate the same region (Brown 2001). Morbidelli & Brown (2004) revisited the inclination distribution finding that approximately 60% of the classical TNOs belong to the hot population, based on an $i$ = 4 deg threshold to divide the cold and hot populations. Analysis of the DES observational programme confirmed Brown (2001)'s results and found that, in general, 80% of TNOs are in the high inclination group (Gulbis et al. 2010). Distinct color, size, and inclination distributions can be interpreted as evidence that the cold and hot populations represent distinct populations.

(4) *Dynamics*. During planetary migration, large amounts of planetesimals were scattered by the giant planets. In this process, less than 1% of them would have entered and been captured in the classical region. These objects could have inclination distributions up to about 30 deg, and moderate eccentricities. Coming from inner regions of the planetesimal disk, those invading bodies (representative of hot classical TNOs) would be, on average, larger than the local objects (representative of cold classical TNOs) in the classical region because more material was available for accretion in these inner regions. With different origins, cold and hot populations would show incompatible colors indicating different compositions (Gomes 2003a; Levison et al. 2008; Lykawka & Mukai 2008). Therefore, this would naturally explain the correlation and trends involving inclinations and physical properties, as discussed above.

There are several phenomena that can affect the orbital distribution of classical TNOs, which, in turn, could affect any correlations/trends in the classical region, and alter their implications. First, interactions and/or temporary captures in local resonances could promote some cold TNOs to the hot population as a result of inclination excitation. This excitation is in general small ($i$ < 10 deg), but large enough to allow TNOs to enter the hot population (Kuchner et al. 2002; Lykawka & Mukai 2005a; Lykawka & Mukai 2005b; Volk & Malhotra 2011). Secondly, a few TNOs currently locked in weaker resonances may, in reality, be classical TNOs trapped temporarily by these resonances. Resonances also cause important changes in the distributions of eccentricities and inclinations, survivability, and the number density of objects situated near those resonances, over Gyr (Lykawka & Mukai 2005c). In conclusion, the commonly-adopted arbitrary inclination threshold of 4-5 deg to divide classical TNOs into cold and hot populations may be leading to misclassification of a fraction of both populations.

On the other hand, inspection of the colors and inclinations of TNOs, the sculpting effects of several resonances in the classical region (from the 7:5 to 2:1 resonances; See Fig. 12), collisional evolution and the likely gravitational perturbation of massive planetesimals suggest, instead, that a higher value, of ~10 deg, may be more appropriate to satisfy the physical and dynamical constraints of both subpopulations (Lykawka & Mukai 2007b). This implies some mixing in the classical region at $i$ < 10 deg. In support of this conclusion, the spectral slopes of classical TNOs do not allow the choice of any obvious $i$-threshold (Chiang et al. 2007). Moreover, recent statistical analysis of the colors and orbits of classical TNOs suggest that a threshold at ~12 deg better discriminate both cold and hot populations (Peixinho et al. 2008). Correlations of spectral slopes and inclinations also favor a 12 deg threshold (Fornasier et al. 2009). Finally, some mixing of inclinations is expected in the entire classical region, which is caused by the action of several resonances (including those of high order) located in that region and chaotic diffusion (Volk & Malhotra 2011). Given these uncertainties, instead of a fixed inclination threshold to divide cold and hot populations, it would be more realistic to consider that these TNOs concentrate at $i$ < 5 deg and $i$ > 10 deg. Indeed, a more firm classification of hot classical TNOs is obtained only at $i$ > 10 deg, because some mixing of cold and hot populations is expected in the 5-10 deg range. For practical purposes, here I follow the tendency of recent work and adopt a threshold at 10 deg henceforth.

### 4.1.2 Classical region dynamics

After evolving a large ensemble of particles representing classical TNOs, Lykawka & Mukai (2005c) found that the great majority (>99%) of classical objects did not suffer any significant radial change due to the perturbation of the giant planets. Significant orbital changes in eccentricities and inclinations were seen mainly for bodies evolving near, or inside, the main resonances of the classical region. In particular, a significant fraction of objects near the 5:3, 7:4, and 2:1 resonances suffered changes in their inclinations large enough to change their dynamical subclass (i.e., cold to hot, or vice-versa).

The main resonances in the classical region (4:3, 5:3, 7:4, 9:5, 11:6, 2:1) have an important role in determining the orbital structure of classical TNOs. In fact, many unstable regions arise near the location of those resonances, driven by resonant excitation in eccentricities or chaotic diffusion out of the resonance. Escapees from both low and high-*i* classical components are possible. Although this is just a minor contribution, it implies that the classical region can contribute to the population of scattered TNOs, which, in turn, sustain the populations of Centaurs and SPCs. Several of the above results have been confirmed and extended by more recent work (e.g., Volk & Malhotra 2011).

Classical TNOs possess an unexpected excitation in their eccentricities and inclinations. In particular, although cold classical TNOs are concentrated in very low inclination orbits, they surprisingly possess moderate-large eccentricities. Such excited orbits cannot be attributed to gravitational perturbations by Neptune (assuming its current orbit) or to mutual planetesimal gravitational stirring (Morbidelli 2005). There is also an apparent lack of TNOs on low eccentricity orbits beyond about 45 AU, which is probably not due to observational biases (Morbidelli & Brown 2004; Morbidelli 2005). Moreover, the long-term evolution of objects in the outer skirts of the trans-Neptunian belt is stable at low eccentricities, so the absence of low eccentricity TNOs beyond about 45 AU is unexpected (Holman & Wisdom 1993; Duncan et al. 1995; Kuchner et al. 2002; Lykawka & Mukai 2005c).

### 4.1.3 The outer edge of the trans-Neptunian belt

Despite the observational capability to detect sufficiently large TNOs on near circular orbits beyond 45 AU, several observations support the existence of an outer edge, which is characterized by the absence of near-circular TNOs beyond about 48 AU and the dearth of these objects at $R$ ~ 45-50 AU (Gladman et al. 1998; Jewitt et al. 1998; Allen et al. 2001; Gladman et al. 2001; Trujillo et al. 2001a; Bernstein et al. 2004; Morbidelli & Brown 2004; Larsen et al. 2007).

Integrations of the orbits of fictitious objects in near-circular orbits around 45-55 AU have shown that this region is stable over the age of the solar system, and that the 2:1 resonance is unable to produce the observed edge (Duncan et al. 1995; Brunini 2002; Lykawka & Mukai 2005c). Alternative explanations for the edge, such as cold thin disks composed of small objects, extreme variations of maximum sizes/albedos, surface properties with heliocentric distance, peculiar eccentricity distributions beyond 50 AU, and other effects been ruled out (Allen et al. 2001; Trujillo & Brown 2001; Trujillo et al. 2001a; Allen et al. 2002). Possible explanations for the edge include perturbations by massive planetesimals, passing stars, or UV photoevaporation of the protoplanetary disk (Petit et al. 1999; Ida et al. 2000a; Brunini & Melita 2002; Adams et al. 2004); inward radial migration of 1 km-sized planetesimals during the early solar system (Weidenschilling 2003); and the outward transportation of bodies by the 2:1 resonance (Levison & Morbidelli 2003).

### 4.1.4 The 40-42 AU subregion

A strong instability is known to affect the 40-42 AU portion of the classical region at low inclinations (Knezevic 1991). However, it was only recently that detailed studies using numerical simulations have explored this intriguing part of the classical region (Lykawka & Mukai 2005c). The results of simulations of classical objects in the 40-42 AU region indicates that this population survives, in general, with $q > 35$ AU and $i > 10$ deg, while a strong depletion of objects with $i < 10$ deg is observed up to ~42-42.5 AU (Fig. 10). More importantly, the outcome in this region reflects the initial conditions of the system. That is, currently observed TNOs in this region did not originate from lower inclination orbits. Therefore, TNOs in the 40-42 AU subregion did not suffer inclination changes over Gyr, implying that they are genuine hot classical TNOs. The results also suggest that cold classical TNOs might be found trapped in the 8:5 resonance, although no member of this resonance has yet been found.

## 4.2 Resonant TNOs

TNOs have been found trapped in the following resonances: 1:1 (Neptune Trojans), 5:4, 4:3, 11:8, 3:2, 18:11, 5:3, 12:7, 19:11, 7:4, 9:5, 11:6, 2:1, 19:9, 9:4, 16:7, 7:3, 12:5, 5:2, 8:3, 3:1, 7:2, 11:3, 4:1, 11:2, and 27:4 ($a$ = 107.5 AU) (Lykawka & Mukai 2007b). Apparently, these resonant TNOs represent ~1/3 of the entire trans-Neptunian population, but unbiased estimates currently indicate a 15-20% fraction (Petit et al. 2011; Gladman et al. 2012). As discussed in Sections 2.1.3 and 2.1.4, this configuration is believed to be the outcome of either the sweeping mechanism (early solar system) or resonance sticking (active throughout solar system history). Strong/sticky resonances in the trans-Neptunian region include the 1:1, first order (5:4, 4:3, 3:2 and 2:1 resonances), the 5:3, 7:4, and the $r$:1 or $r$:2 resonances located at 50 AU < $a$ < 250 AU (Section 2.1.4). Figures 1 and 12 illustrate some resonances of interest. It is interesting to note that there are no strong resonances located at ~44.5-47.5 AU. Therefore, although there are TNOs on excited orbits in that region, resonances are unlikely to have originated such orbital excitation. Another remarkable characteristic of resonant TNOs is their ability to avoid close encounters with Neptune thanks to the libration protection mechanism. Indeed, some TNOs in the 3:2 and 2:1 resonances “cross” the orbit of Neptune with perihelion passages $q < 30.1$ AU. Resonant populations that are stable over the age of the solar system are found throughout the entire trans-Neptunian region, including prominent populations in more distant resonances, such as the 5:2 resonance (Chiang et al. 2003; Hahn & Malhotra 2005; Lykawka & Mukai 2007b). In addition, the origin of long-term resonant bodies beyond 50 AU strongly suggests that the ancient trans-Neptunian belt was excited in eccentricities and inclinations before the Neptunian resonances swept across the belt (Lykawka & Mukai 2007a; Lykawka & Mukai 2008).

The resonances that appear to be most populated in the trans-Neptunian belt are currently the 3:2, 5:3, 7:4, 2:1 and 5:2 resonances. The 3:2 resonance is apparently the most populated among all resonances. However, this is a result of strong observational biases (Jewitt et al. 1998) such as: (1) Smaller semimajor axes than other TNOs (overestimation); (2) Easily observed when near perihelion, an epoch at which they are concentrated in particular longitudes away from Neptune (as dictated by Eq. 12); (3) Kozai TNOs are mostly located far from the ecliptic when near perihelion. Noting that most observations concentrate near the ecliptic, this would produce an underestimation of this particular group of 3:2 resonant bodies. Similarly to the 3:2 resonants, the members of the 2:1 resonance are also evolving in apparently quite stable orbits. Members of both 5:3 and 7:4 resonances seem to occupy the disaggregated stable regions found by previous works (Melita & Brunini 2000; Lykawka & Mukai 2005a). Roughly ~15% of the total trans-Neptunian resonant population is found in distant resonances beyond the 2:1 resonance. However, this number is certainly a lower limit, because of strong observational biases against the discovery of TNOs in such distant resonances. Apparently, the most populated of these resonances are the 9:4, 7:3, and 5:2 (Lykawka & Mukai 2007b).

Detailed results showing the resonant properties of TNOs locked in resonances and full lists of resonant TNOs can be found in Lykawka & Mukai (2007b) and Gladman et al. (2008). A more updated analysis of resonant populations and their intrinsic fractions in the trans-Neptunian region can be found in Gladman et al. (2012)[10]. The orbital elements of several resonant TNOs and their dependence on libration amplitudes are illustrated in Fig. 13. Finally, it is important to note that models must be capable of explaining the observed resonant populations, their intrinsic populations, orbital distributions and other properties.

#### 4.2.1 The 1:1 resonant population: Neptunian Trojans

There are currently eight observed Trojans librating about the L4 (six bodies) and L5 (two bodies) Lagrange points (Table 1). The discovery of Neptunian Trojans is difficult, so that the intrinsic population is severely biased. Indeed, it has been estimated that the Neptunian Trojan population would be comparable to the Jovian Trojan population, or even surpass it over the same size range (Sheppard & Trujillo 2006; Sheppard & Trujillo 2010b). Seven of the known Trojans appear to be primordial[11], but two of them (2001 QR322 and 2008 LC18) are likely representatives of a decaying population of captured Neptunian Trojan objects (Section 7.3. See also Horner & Lykawka 2010b; Zhou et al. 2011; Horner et al. 2012).

The intrinsic population in the Neptunian Trojan clouds also occupies wide ranges of inclination (up to 28 deg) and moderate eccentricities (< 0.2). The broad inclination distribution is surprising because standard models of planet and Trojan formation predict the existence of Trojans essentially on cold dynamical orbits, consistent with a flat disk of material that formed the planets themselves (Dotto et al. 2008, and references therein). On the other hand, recent models have found that the most promising way to explain the Neptunian Trojan orbital structure is the chaotic capture mechanism. That is, Neptune probably acquired its Trojan populations via the capture of planetesimals scattered by the four giant planets during planetary migration in the early solar system. The migration allowed chaotic capture to operate during particular resonance crossings of Neptune with Uranus and/or Saturn (Nesvorny & Vokrouhlicky 2009; Lykawka & Horner 2010). Lykawka et al. (2009) suggested that an optimal way to form the Neptune Trojans would be if Neptune migrated over large distances and over timescales of a few tens of Myr.

The efficiency for capture into the Neptunian Trojan clouds was found to be only on the order of 0.1%. However, because the planetesimal disk through which Neptune migrated likely contained tens of Earth masses, a low capture efficiency would still imply that a substantial population was captured as Trojans by the end of planetary migration. After evolving the captured Trojan populations obtained at the end of migration, Lykawka et al. (2011) found that only ~1-5% of that population would be able to survive until today. This implies that the current Trojan population may be a very small remnant of the original population. This also means that a large fraction of these captured Trojans left their resonant state, thus evolving to unstable orbits typical of scattered TNOs or Centaurs.

The long-term dynamical evolution of hypothetical Trojans captured in this manner, and that of five of the 8 currently known Trojans is essentially quasi-static, implying that their orbits should reflect the primordial conditions when these objects become trapped in the 1:1 resonance in the first place.

[10] I became aware of this paper during the revision of this Work. Although I leave an in-depth analysis of that paper to future work, I strongly encourage readers to see that paper for a detailed discussion on the resonant structure beyond Neptune.

[11] We recently found that the newest member of the Neptunian Trojan clouds, 2004 KV18, has a very unstable orbit. This object is probably a temporarily-captured Trojan that originated in the scattered disk. See Horner & Lykawka (2012) for more details.

This strengthens the importance of studying Trojan populations to better understand the early orbital conditions in the planetesimal disk and dynamical evolution and formation of the host planet (Dotto et al. 2008; Lykawka et al. 2009).

### 4.2.2 The 3:2 resonant population

The 3:2 resonance is a first order resonance, and one of the strongest resonances in the trans-Neptunian belt. The 3:2 resonance has been widely explored in the academic literature (Morbidelli et al. 1995; Nesvorny & Roig 2000; Melita & Brunini 2000; Chiang & Jordan 2002; Wiegert et al. 2003; Tiscareno & Malhotra 2009). Most resonant members lie well inside the stable region, at an amplitude of resonant angles $A < 100$ deg for $e < 0.25$ and at smaller amplitudes for larger eccentricities (as indicated by the circles in panel **a** of Fig. 13), suggesting that they are primordial objects residing in the resonance over Gyr timescales. The remaining bodies may have evolved to larger libration amplitudes after billions of years of evolution, or may have been captured recently into the resonance via resonance sticking. The diffusion of bodies can lead to escape from the resonance, thus contributing to the injection of new objects into non-resonant orbits at low eccentricities or unstable ones, thus contributing to the scattered TNO population (Morbidelli 1997; Nesvorny & Roig 2000). Pluto's gravitational perturbation may also help unstable members to leave the resonance via chaotic diffusion over Gyr (Nesvorny et al. 2000; Tiscareno & Malhotra 2009). Finally, results of long-term dynamical evolution of particles within the 3:2 resonance also suggest that the currently observed resonant population represents only ~27% of that in the early solar system (Tiscareno & Malhotra 2009).

Secular resonances play an important role inside the 3:2 resonance region; in particular, the Kozai mechanism and the $\nu_{18}$ secular resonance. There is an apparent ratio of ~20-30% of 3:2 resonant TNOs experiencing the Kozai mechanism (Gomes 2000; Wan & Huang 2001; Chiang & Jordan 2002), where the uncertainties arise because some TNOs display only interactions and/or temporary Kozai mechanism orbits. Lastly, Kozai TNOs display large anti-correlated oscillations in their osculating eccentricities and inclinations (Section 2).

The current distribution of libration amplitudes seems to favor sweeping resonance capture at migration timescales of ~10 Myr (Chiang & Jordan 2002). However, it seems that a better match may be obtained if the 3:2 resonance swept a region of the planetesimal disk with hot initial conditions (eccentricities and inclinations with lower limits up to 0.3 and 25 deg) (Wiegert et al. 2003). Alternatively, the 3:2 resonant population may have been captured from scattered objects during the early solar system (e.g., Levison et al. 2008).

### 4.2.3 The 2:1 resonant population

Similarly to the 3:2 resonance, the 2:1 resonance is also one of the strongest resonances in the trans-Neptunian belt, so we can expect it to be one of the most populated in the trans-Neptunian belt. This resonance has been explored by many researchers (Melita & Brunini 2000; Nesvorny & Roig 2001; Chiang & Jordan 2002; Murray-Clay & Chiang 2005). In addition to the usual symmetric libration of the resonant angle about 180 deg or 0 deg, objects in the 2:1 resonance can also present asymmetric librations about 90 deg or 270 deg (Gallardo 2006b). Lykawka & Mukai (2007b) found that about 1/3 of 2:1 resonant particles exhibited symmetric librations and evolved on orbits within 33.9 AU $< q <$ 44.5 AU ($0.07 < e < 0.29$), and mostly $i < 15$ deg after 4 Gyr. In general, TNOs residing in the 2:1 resonance are well inside the stable regions previously discussed, with ~70-85% showing asymmetric librations (80% leading Neptune in orbital longitude; 20% trailing), and ~15-30% librating symmetrically about 180 deg (see also Fig. 13). Chaotic diffusion of objects within the 2:1 resonance suggests that the currently observed resonant population represents ~15% of that in the early solar system (Tiscareno & Malhotra 2009).

TNOs in the 2:1 resonance can also experience the Kozai mechanism (Nesvorny & Roig 2001), with perhaps an intrinsic fraction of ~10-30% of the total population of 2:1 resonants displaying such behaviour (Chiang & Jordan 2002; Lykawka & Mukai 2007b). Similarly to the 3:2 resonants, 2:1 resonant TNOs in Kozai mechanism evolve with large *e*-*i* oscillations for periods of a few Myr. Indeed, all TNOs inside the 2:1 resonance with the highest eccentricities (>0.34) experience the Kozai mechanism.

The spatial distribution of 2:1 resonant TNOs provides an important constraint for the migration history of Neptune. Indeed, the excess of 2:1 resonant TNOs moving on asymmetric orbits about 90 deg supports migration timescales exceeding 10 Myr (Murray-Clay & Chiang 2005).

Captured bodies from the scattered disk are able to remain in the 2:1 resonance for 4 Gyr (Duncan & Levison 1997; Lykawka & Mukai 2006a). Therefore, an unknown fraction of the currently observed 2:1 resonants likely did originate from unstable orbits (typical of scattered TNOs), rather than having a local origin in the planetesimal disk that was swept by the resonance during Neptune's migration. Lykawka & Mukai (2006a) found that these captured 2:1 resonant objects experienced moderate-large Kozai *e*- and *i*-oscillations.

#### 4.2.3.1 Comparing the 2:1 and 3:2 resonant populations

The intrinsic ratio of 2:1 to 3:2 resonant TNOs, and the resonant properties and spatial distributions of resonant TNOs, in general offer important constraints for migration models (Hahn & Malhotra 1999; Ida et al. 2000b; Melita & Brunini 2000; Friedland 2001; Zhou et al. 2002; Chiang & Jordan 2002; Luu & Jewitt 2002; Murray-Clay & Chiang 2005). Earlier observations suggested that it was ~3 times easier to detect 3:2 resonants than 2:1 resonant objects (Jewitt et al. 1998), so the apparent ratio between 3:2 and 2:1 resonants is highly biased. We can estimate the true ratio between members of both resonances by determining a simple observational bias correction factor. Consider a flux proportional to $R^{-4}$, and account for bias against discovery for eccentric orbits as $(Q / q)^{3/2}$ (Chiang et al. 2007), where $Q$ is the aphelion. Approximate $R$ as the average distance, giving

$$d = a\left(1 + \frac{\bar{e}^2}{2}\right) , \qquad (22)$$

substitute $a$ for $a_{res}$ and use the median $e$ to calculate $d$, $Q$ and $q$ for the 3:2 and 2:1 resonant populations, respectively. Using this method, Lykawka & Mukai (2007b) found the intrinsic fraction of 3:2 to 2:1 resonant TNOs to be ~ 2.8 using a correction factor of ~0.39, in agreement with values found in the literature (Nesvorny & Roig 2001; Trujillo et al. 2001a; Chiang & Jordan 2002). Nevertheless, due to various uncertainties, the true population of 3:2 resonant TNOs should range between three and four times that of their counterparts in the 2:1 resonance according to dedicated surveys (Petit et al. 2011; Gladman et al. 2012). It is worth noting that chaotic diffusion operating in both the 3:2 and 2:1 resonances likely caused the ratio of objects within these resonances to almost double over the age of the solar system (Tiscareno & Malhotra 2009). This suggests that that ratio was probably 1.5-2 in the early solar system, presumably at the end of planetary migration. Nevertheless, it should be noted that if the 2:1 resonant population was significantly more massive in the past, the estimations above become more uncertain (Levison & Morbidelli 2003; Tsukamoto 2011).

### 4.2.4 The 5:4, 4:3 and 5:3 resonant populations

Lykawka & Mukai (2007b) and Gladman et al. (2008) identified several TNOs locked in the 5:4, 4:3 and 5:3 resonances. Long-term evolution inside these resonances indicate that 5:4 and 4:3 resonant particles can survive for 4 Gyr on orbits with $0.03 < e < 0.15$ and $i < 11$ deg, and $0.03 < e < 0.18$ and $i < 20$ deg, respectively. These boundaries match those for observed TNOs and stability dynamical maps (Nesvorny & Roig 2001). Since 5:4 resonant TNOs show relatively low libration amplitudes, these objects may well be primordial relics of the early solar system. Indeed, most 5:3 resonant objects concentrate in orbits with $0.09 < e < 0.27$ and $i < 20$ deg. Again, the 5:3 resonant TNOs typically inhabit the stable regions, but seem to occupy the resonance disaggregated regions (Melita & Brunini 2000). The Kozai mechanism is also possible in the 5:3 resonance (Lykawka & Mukai 2007b).

### 4.2.5 The 7:4 resonant population

The 7:4 resonance is one of the most populated in the trans-Neptunian region, rivaling the 2:1 resonance in intrinsic number, if we account for observational biases (i.e., using the rough method above. See also Gladman et al. 2012). Since this resonance has been poorly explored, and given its important role in the classical region, I have summarized the main results of Lykawka & Mukai (2005a) and Lykawka & Mukai (2006) below. First, the 7:4 resonance has stable regions aggregated with many chaotic islands. The most stable 7:4 resonant bodies concentrate on orbits at $0.05 < e < 0.2$ and $i < 8$ deg and $0.25 < e < 0.3$ and $i < 5$ deg (Fig. 11). Thus, several 7:4 resonant TNOs could have been evolving in the resonance since the early solar system. However, no such highly-eccentric resonant TNOs have been detected in the latter region (Fig. 13). Concerning the influence of the inclination, strong unstable islands are present at $i > 10$ deg, although small stable niches were also found. Except in the aforementioned region at $0.25 < e < 0.3$, almost all stable objects displayed significant changes in eccentricities and inclinations (Fig. 6). In this case, 7:4 resonant TNOs are likely to experience high mobility in *e*-*i* element space and even leave the resonance after reaching a critical $e \sim 0.2$, an outcome caused by chaotic alternation between circulation and libration of the resonant angle. In summary, the 7:4 resonance is weakly chaotic, meaning that many bodies with orbits near, or within, this resonance should suffer significant eccentricity and/or inclination evolution after billions of years.

The Kozai mechanism is also present inside the 7:4 resonance (Morbidelli et al. 1995), in particular for members with high inclinations. Indeed, it is possible that 30-40% of the 7:4 resonant population are evolving on Kozai orbits (Lykawka & Mukai 2005a). However, these objects, in particular, can behave quite irregularly by displaying alternation of libration centers as well as extensive periods of temporary circulation.

Simulations have also shown that the fate of escapee TNOs from the 7:4 resonance is highly varied, with particles experiencing re-capture in other Neptunian resonances (including captures on Myr and Gyr timescales), inward orbital diffusion towards the inner solar system (thus becoming Centaurs or even SPCs), collision with a planet, and other dynamical histories involving temporary resonance captures, and gravitational scattering by the giant planets (Lykawka & Mukai 2006).

### 4.2.6 The 9:4, 5:2, 8:3 and other distant resonant populations

After the discovery of several TNOs in the 4:3, 3:2, 5:3 and 2:1 resonances, it was thought that these resonances were the most important in the trans-Neptunian belt, implicitly suggesting that the resonant structure of the trans-Neptunian region was limited to the 2:1 resonance (Nesvorny & Roig 2000; Luu & Jewitt 2002; Nesvorny & Roig 2001; Morbidelli & Brown 2004). Nevertheless, this picture started to change with the first identification of 7:3 and 5:2 resonant TNOs, located at *a* = 53.0 AU and *a* = 55.4 AU, respectively (Chiang et al. 2003; Elliot et al. 2005).

Despite the strong observational biases that afflict the discovery of TNOs on distant orbits, a significant number of TNOs have now been found locked in several resonances beyond 50 AU (Lykawka & Mukai 2007a). In particular, one of the strongest resonances in the trans-Neptunian region, the 5:2 resonance, seems unsurprisingly to be the most occupied of these distant resonances (Gallardo 2006b; Lykawka & Mukai 2007c). However, the intrinsic 5:2 resonant population intriguingly rivals that in the 3:2 resonance! (Gladman et al. 2012). Resonance occupancy also includes very high order, and still more distant, resonances, such as the 16:7, 11:2, and 27:4 resonances. Therefore, the resonant structure in the trans-Neptunian region continues far beyond the 2:1 resonance, which should be taken into account by both future theoretical models of the solar system and attempts to achieve a better characterization by future observational surveys. In particular, we should expect significant resonant populations in the 9:4, 7:3, 12:5, 8:3, 3:1 and many other *r*:1 or *r*:2 resonances beyond 50 AU, a prediction that appears to have been confirmed by recent surveys (Gladman et al. 2012). Figure 13 illustrates resonant properties of four distant resonances, namely the 9:4, 7:3, 5:2 and 8:3. Lastly, objects in such resonances can also experience the Kozai mechanism (Lykawka & Mukai (2007a).

### 4.2.7 Minor resonant populations

A few TNOs orbit in the 11:8, 18:11, 12:7, 19:11, 9:5, and 11:6 resonances. It is surprising that such weak resonances can harbor TNOs, because they possess quite small stickiness and widths in element space. Lykawka & Mukai (2007a) determined a stable region at $e < 0.185$ and $i < 15$ deg in the 9:5 resonance, which is located in the classical region (Fig. 12). Finally, objects trapped in the 18:11, 12:7 and 19:11 resonances showed irregular behavior, alternating between *e*-type and *i*-type resonant configurations (Murray & Dermott 1999). It is worth noting that the 12:7 resonance played an important role in the dynamics of Haumea and its associated family. This suggests that other similarly high order resonances are worth detailed dynamical investigations in the future (Section 3.3). This is also evidence that even minor (weak) resonances can provide useful information and new clues on the dynamics of the trans-Neptunian region and its members (e.g., Volk & Malhotra 2011).

## 4.3 Scattering vs. scattered TNOs

Beginning with the discovery of 1996 TL66 (Luu et al. 1997), subsequent observations confirmed a new class of objects with perihelia close enough to Neptune to interact strongly with that planet and suffer gravitational scattering over long timescales. Hence, these TNOs are often called scattered TNOs (or scattered disk objects). The majority of scattered TNOs are thought to be the relics of a much larger primordial population, rather than being sustained by TNOs coming from unstable regions of the trans-Neptunian belt (Morbidelli 2005). However, the orbital distribution of scattered TNOs is highly biased, since objects near perihelion have a greater probability of being found by the surveys. If scattered TNOs originated from the disk located around Neptune's location (~25-35 AU), then the present population would represent only ~1% of the original population that existed on Neptune crossing orbits in the past (Duncan & Levison 1997; Morbidelli & Brown 2004; Lykawka & Mukai 2007c). Moreover, scattered TNOs are also probably the main source of the Centaur and SPC populations (Duncan & Levison 1997; Emel'yanenko et al. 2005).

The scattered TNOs are typically classified as objects with $a > 50$ AU. However, different thresholds for the perihelion were adopted among different studies in the literature. Here, I consider scattered TNOs as those bodies that suffer close encounters with Neptune at least once during their dynamical lifetime, based on orbital integrations of currently known TNOs and particles from several numerical investigations over the last 4-5Gyr (Lykawka & Mukai 2007b). Therefore, the essential orbital element of interest is the perihelion distance. Indeed, objects typically suffer frequent scattering when they possess q < 37 AU, independent of other orbital elements, such as semimajor axes or eccentricities. This is supported by the orbital evolution of non-resonant bodies in the classical region over 4 Gyr, which indicates that TNOs with $q$ < 37 AU will suffer strong scattering by Neptune at least once during their lifetimes, even for those undergoing temporary resonance captures (limited to residence timescales < 3 Gyr).

One problem with the $q$ < 37 AU boundary is that resonances can often drive scattered TNOs to large perihelia so that they stop strongly interacting with Neptune (i.e., the gravitational scattering ceases) (Gladman et al. 2002; Hahn & Malhotra 2005; Gomes et al. 2005, Lykawka & Mukai 2006; Lykawka & Mukai 2007c). In addition, the trans-Neptunian region is very complex and full of chaotic islands, implying that thresholds in perihelion do not necessarily hold for any particular place in element space (Torbett 1989; Gladman et al. 2008). It is important to recall that not only scattered TNOs inhabit the scattered disk, but also other dynamical classes, such as the resonant (between the 9:4 and 27:4 resonances) and detached TNOs. The perihelion boundary of 37 AU offers a guide for the likelihood that a TNO may have been scattered at some point during its lifetime, but only costly long-term orbital integrations can ultimately answer this question. Gladman et al. (2008) proposed calling objects suffering scattering within 10 Myr as scattering TNOs (instead of scattered). However, note that in this case the population of commonly known “scattered disk” objects would shrink because only those showing truly unstable orbits on short timescales would count as scattered TNOs.

In this Work, I consider all TNOs with $q$ < 37 AU as scattered TNOs, for simplicity. Among this population, “scattering objects” would be experiencing gravitational scattering currently or “recently” (within the last 10 Myr, as proposed in Gladman et al. 2008). This subpopulation would be directly contributing to the Centaur population, and consequently SPCs as well (e.g., Duncan & Levison 1997). The remaining scattered TNOs would represent objects on unstable orbits on longer timescales (e.g., 10 Myr or longer), so they would not be detached but could instead display either slow chaotic orbital evolution or temporary capture in resonances. Indeed, resonance sticking is a fundamental mechanism governing the orbital evolution of all these bodies in the scattered disk.

A caveat with the classification adopted here is that it neglects the existence of scattered TNOs in the range 37 AU < $q$ < 40 AU, whose identification is tricky because detached TNOs also share similar orbits. These boundaries seem to delimit an intermediate region between Neptune scattering ($q$ < 37 AU) and no scattering by the giant planet ($q$ > 40 AU) over Gyr timescales. That implies there is no clear division between scattered and detached TNOs, so that they will appear to be mixed in this intermediate region.

Inspecting Figs. 1 and 2, common characteristics of scattered TNOs include a significant range in semimajor axes, large eccentricities and low to high inclinations. Note also that the Centaurs share similar orbital elements, so that, unsurprisingly, Centaurs and scattering TNOs do not have precise dynamical boundaries.

Earlier studies estimated the number of scattered TNOs larger than 100 km to be around 31,000 objects with an approximate mass of 0.05 $M_{\oplus}$, assuming a size distribution with slope $j$ = 4 (Trujillo et al. 2000). This value is in agreement with the upper limits of 35,000-250,000 objects, as determined by an analysis of more recent surveys (Parker & Kavelaars 2010, and references therein). However, note that Petit et al. (2011) estimated the scattering population within the scattered disk to be only 2,000~10,000 objects. Trujillo et al. (2000) also proposed extending the size distribution of scattered TNOs to sizes as small as 1 km, obtaining an estimation of about $4{\cdot}10^9$ objects. However, this is likely an overestimate because smaller TNOs appear intrinsically less common than what a single power law would predict (see Section 3 for details). Given the above uncertainties, more observations and theoretical studies of objects in the scattered disk are warranted.

### 4.4 Detached TNOs

At first, detached TNOs were not recognized as a distinct class of their own, but rather an extension of the scattered population (Gladman et al. 2002). Although scattered TNOs are expected to show large variations in semimajor axis and eccentricity, their perihelia is not supposed to change beyond the 40 AU limit over the age of the solar system (Gladman et al. 2002). However, the first three members of this "extended" scattered disk class of TNOs never suffered approaches with Neptune over 5 Gyr of orbital evolution (Emel'yanenko et al. 2002; Gladman et al. 2002).

In this Work, detached TNOs are bodies that never suffer close encounters with Neptune, so these objects appear to be "detached" from the solar system. In particular, this is evident in the case of 2004 XR190 ($a$ = 57.7 AU; $q$ = 51.6 AU; $i$ = 46.6 deg), (145480) 2005 TB190 ($a$ = 76.4 AU; $q$ = 46.2 AU; $i$ = 26.4 deg), (148209) 2000 CR105 ($a$ = 223.7 AU; $q$ = 44.0 AU; $i$ = 22.8 deg), 2004 VN112 ($a$ = 349.2 AU; $q$ = 47.3 AU; $i$ = 25.5 deg), and (90377) Sedna ($a$ = 535.5 AU; $q$ = 76.4 AU; $i$ = 11.9 deg), and low-$i$ detached TNOs located near the classical region (Figs. 1 and 2. See also Fig. 14). Of all the detached TNOs identified in Lykawka & Mukai (2007b), only one member is in resonance (8:3 resonance). Using this criterion, detached TNOs can be classified as non-resonant objects with $q >$ 40AU, with the caveats that it is virtually impossible to distinguish detached TNOs in the classical region (i.e., at 45-50 AU), and the fact that resonances can contribute to the formation of detached TNOs (Gomes et al. 2005; Lykawka & Mukai 2006; Lykawka & Mukai 2007c).

The apparent fraction of detached TNOs beyond the 2:1 resonance is approximately 10%, but the intrinsic fraction must be much larger than that, when one accounts for the observational biases that discriminate against their discovery. In fact, the population of detached TNOs likely surpasses that of scattered TNOs (Gladman et al. 2002; Allen et al. 2006; Petit et al. 2011). It is also important to note that the $r$:1, $r$:2 and $r$:3 resonances could lift the perihelion of temporarily captured objects beyond 40 AU for $a$ < 250, 200 and 170 AU, respectively. This was observed in several previous studies (Lykawka & Mukai 2004; Fernandez et al. 2004; Gomes et al. 2005; Gallardo 2006a; Lykawka & Mukai 2007c). This mechanism predicts that approximately ~5-25% of all TNOs beyond the 2:1 resonance would evolve onto detached orbits. However, this is far too little to account for the intrinsic population of detached TNOs, which is at least 50% of the population in the scattered disk (Allen et al. 2006). Moreover, detached bodies produced by resonances have, in general, high inclinations ($i >$ 25-30 deg), a property largely unseen in the observational data (except for a few outliers, but these are not trapped in resonances) (Gomes et al. 2005; Gallardo 2006a; Lykawka & Mukai 2006). In short, it is intriguing that the sole presence of the giant planets may not explain the origin of detached TNOs (Gomes et al. 2008).

### 4.5 Very high-$i$ TNOs

There are currently three TNOs known whose inclinations are higher than 40 deg: 2004 DG77 ($i$ = 47.8 deg), Eris ($i$ = 43.8 deg), and 2004 XR190 ($i$ = 46.6 deg). Allen et al. (2006) and Lykawka & Mukai (2007b) found that 2004 XR190 has only a 4-5% chance of being a 8:3 resonant object excited by the Kozai mechanism, although this object may have achieved its orbit by the influence of that resonance during planetary migration (Gomes 2011). Very high-$i$ TNOs and other extreme objects are illustrated in Fig. 14.

Inclinations higher than 40 deg cannot easily be obtained, according to theory and current formation models of the trans-Neptunian region. The conservation of the Tisserand parameter relative to Neptune is an useful tool in this analysis, and is given by

$$P_N = \frac{a_N}{a} + 2\sqrt{\frac{a}{a_N}\left(1-e^2\right)}\cos(i) \ , \qquad (23)$$

where the orbital elements refer to the TNO, and the subscript $N$ stands for Neptune. Although this parameter is a constant of motion in the restricted three-body problem (Sun-Neptune-particle), provided the TNO does not encounter any other planets except Neptune, the Tisserand parameter is quite well conserved during its dynamical evolution in an N-body system, remaining nearly constant, and thus explaining why the bulk of the unbiased $i$-distribution of scattered TNOs is found at <40 deg (Morbidelli et al. 2004). Simulations confirm that: ~99% of TNOs experiencing gravitational scattering by Neptune over 4-5 Gyr have $i$ < 40 deg (Duncan & Levison 1997; Gomes 2003a; Lykawka & Mukai 2007c).

The discovery of very high-$i$ TNOs has been extremely difficult in most surveys, because these objects spend a very small fraction of their time near the ecliptic (see also Section 3.1). When probing a region near the ecliptic, the probability of discovering TNOs with $i$ = 40 deg is approximately 20-50 times smaller than for objects at $i$ ~ 0 deg (Trujillo et al. 2001; Jones et al. 2005). This implies that the intrinsic fraction of very high-$i$ TNOs should be roughly ~10-25%, after considering the apparent fraction (about 0.5%). Therefore, taking into account the results of the simulations discussed above, there appears to exist an, as yet, unseen large population of minor bodies with higher inclinations ($i$ > 40 deg) that cannot be produced by standard scenarios, based on resonant or gravitational perturbations by Neptune. The Kozai mechanism is one potential mechanism that could help to produce these bodies (e.g., Gomes et al. 2005).

## 4.6 Other populations

### 4.6.1 Centaurs

I consider non-resonant objects with 6.3 AU < $q$ < 27.8 AU to be Centaurs. That is, objects on unstable orbits experiencing close encounters with at least one giant planet. For this reason, Centaurs have quite short dynamical lifetimes (a few Myr) (Holman & Wisdom 1993; Duncan & Levison 1997; Levison & Duncan 1997; Tiscareno & Malhotra 2003; Horner et al. 2004; Emel'yanenko et al. 2005; di Sisto & Brunini 2007). However, four Centaurs possess exceptional high inclinations >70 deg (Fig. 14). Such peculiar bodies could be compatible with an origin in the Oort cloud (Levison et al. 2001; Brasser et al. 2012). It is also interesting to note that Centaurs, similarly to scattered TNOs, can also experience intermittent gravitational scattering by the giant planets and temporary captures in resonances with those planets (i.e., the resonance sticking phenomenon!) (e.g., Bailey & Malhotra 2009).

Centaurs are believed to be in transitional orbits filling the dynamical link between TNOs on unstable orbits and SPCs. Therefore, to account for a steady population of SPCs, it is assumed that the Centaurs also consist of a population regularly fed by trans-Neptunian populations. There are about 100 Centaurs larger than 100 km, and an estimated 10 million bodies larger than 1 km (Sheppard et al. 2000).

### 4.6.2 Short period comets

Comets with shorter orbital periods have strong links to origins in the trans-Neptunian belt, since they share a similar inclination distribution typically confined to inclinations less than 30-40 degrees. These SPCs are also typically classified into Jupiter family comets ($2 < P_J < 3$; i.e., controlled by Jupiter) and Halley-type comets ($P_N < 2$; i.e., beyond the control of Jupiter). Another important group of comets are the long period comets (LPC) ($P_N < 2$), which possess an isotropic inclination distribution. LPCs are thought to come from the Oort cloud (Oort 1950; Brasser et al. 2006). Here, I focus on the general SPC population and their origin in the trans-Neptunian region.

The link between SPCs and TNOs has been well established since the 1980s (Fernandez 1980). On the other hand, it is much less clear where these comets are coming from. Specific regions of the trans-Neptunian belt can contribute to supplying objects to the current SPCs population, such as 3:2 resonant TNOs associated with unstable regions in that resonance (Morbidelli 1997; Ip & Fernandez 1997; Yu & Tremaine 1999; Nesvorny et al. 2000; Wan & Huang 2001; di Sisto et al. 2010), scattered TNOs via chaotic layers within the scattered disk (Torbett 1989; Holman & Wisdom 1993; Duncan & Levison 1997), the long-term decaying Trojans of Jupiter and Neptune (Lykawka & Horner 2010), or classical TNOs leaking out via chaotic diffusion (Kuchner et al. 2002). Because the Centaurs share similar Tisserand parameters as the "scattering" TNOs, it is not possible to distinguish their origin based on the dynamics of SPCs alone (Duncan & Levison 1997). Ultimately, the scattered disk is probably the main source of SPCs, but it is not clear which of the aforementioned sources is the major one feeding the "scattering" TNOs (Duncan & Levison 1997; Gomes et al. 2008). Comparisons of the size distributions of SPCs with those populations in the scattered disk, classical region, and Jovian Trojan clouds add further evidence that the scattered TNOs cannot be the sole source of SPCs (Fraser et al. 2010).

Recent studies found evidence that the Trojan populations of Jupiter and Neptune are supplying a significant fraction, or even the majority, of the SPC population! (Horner & Lykawka 2010a; Horner & Lykawka 2010c). An illustrative case showing this dynamical path is shown in Fig. 15. Note that, since we observe an apparent steady population of SPCs, the main source, whether the scattered disk or resonance sites (e.g., the Trojans), must provide a diffusion timescale comparable to the age of the solar system. Regions which are depopulated too fast are not able to supply the modern influx of SPCs. Finally, dynamical studies point to long-term stability for the classical TNOs, so that, in principle, they should not be contributing significantly to the Centaur population (Kuchner et al. 2002; Lykawka & Mukai 2005c).

### 4.6.3 Haumea and collisional families in the trans-Neptunian belt

Collisional families in the belt can provide evidence of the importance of collisions between TNOs during the early stages of the evolution of the solar system (Chiang 2002; Brown et al. 2007). The first family identified consists of the dwarf planet (136108) Haumea, locked in the 12:7 resonance ($a \sim 43$ AU), and at least nine other ~100 km-sized TNOs located around $a$ = 42-44.5 AU. The nine family members concentrate in orbital elements at $a$ = 42-44.5 AU, $q$ = 37-39 AU and $i$ = 24-29 deg, and share the peculiar C-depleted, and $H_2O$ ice rich, surface spectra exhibited by Haumea. The long-term orbital evolution of Haumea's family, over 4 Gyr reveals that, for a plausible set of initial assigned ejection velocities, the family fragments probably spread over a range of several AU in semimajor axis, almost 0.1 in eccentricities and a few degrees in inclinations, providing predictions for new family members that can be addressed by future observations (Lykawka et al. 2012). These particular results also imply that collisionally-generated populations of small bodies in the trans-Neptunian belt can populate all four main populations of TNOs (Fig. 16). In addition, the orbital diffusion of the stable theoretical family fragments over 4 Gyr is extremely small. Therefore, the observed orbital distribution of Haumea's family can be used to draw conclusions about the nature of the collision that created the family (Fig. 17).

In conclusion, a better characterization of the Haumean family and the identification of new collisional families in the trans-Neptunian belt will provide valuable new clues on the origin and evolution of the entire belt! (e.g., Marcus et al. 2011).

### 4.7 The current low total mass of the trans-Neptunian belt

As discussed in the introduction, estimates of the total current mass in the trans-Neptunian belt are of order of 0.1 $M_\oplus$, which is at most 1% of that required for TNOs to grow locally to their current sizes in the planetesimal disk in reasonable timescales (a few tens of Myr) (Jewitt et al. 1998; Gladman et al. 2001; Bernstein et al. 2004; Chiang et al. 2007). It is unclear whether TNOs, in particular the cold classical TNOs, formed locally in the disk at ~35-50+ AU or were transported from inner regions of the disk (< 35 AU) billions of years ago. However, there is evidence that both cold and hot classical TNOs experienced collisional evolution (Fuentes et al. 2010).

### 4.8 Summary of constraints for theoretical modeling

In summary, a successful model for the trans-Neptunian region must explain: (1) The excitation of the primordial planetesimal disk, as evinced by orbital distribution of the trans-Neptunian populations discussed in earlier sections; (2) The existence of distinct classes of TNOs and their orbital structure, physical properties, and intrinsic ratios; (3) The creation of the trans-Neptunian outer edge, characterized by a dearth of TNOs at 45-50 AU and the absence of near-circular TNOs beyond 48 AU; (4) The 2-3 orders of magnitude mass discontinuity beyond Neptune's orbit, given the very small total mass of the trans-Neptunian belt.

## 5 The physical properties of TNOs

The orbital structure of TNOs provides several important clues to the origin and evolution of the solar system. In addition, the physical properties of TNOs also provide fundamental constraints that should be taken into account together with their dynamical characteristics. Ideally, we would like to obtain distributions of sizes, albedos, colors, spectra, compositions, and lightcurves, among others. Since TNOs are distant and relatively dark objects, the sample of objects with determined physical properties is still small[12]. However, in particular, the number of objects with known colors and spectra has been increasing thanks to dedicated surveys (Luu & Jewitt 2002; Delsanti & Jewitt 2006; Delsanti et al. 2006; Cruikshank et al. 2007; Fornasier et al. 2009; Perna et al. 2010; Merlin et al. 2010; Barucci et al. 2011). A recent review on the surface compositions of TNOs can also be found in Brown (2012). Here, I focus on albedos and sizes, which are the fundamental properties with best available statistics (Stansberry et al. 2008).

### 5.1 Large TNOs

Our knowledge about the size distribution of TNOs is still observationally limited to TNOs with sizes larger than about 100 km. Among them, many large-scale TNOs (500-2000 km) have been observed. An updated list of the largest TNOs discovered so far, including the trans-Neptunian dwarf planets, can be found in Stansberry et al. (2008). The potential importance of large TNOs can be summarized as:

[12] However, programmes dedicated to unveiling the physical properties of TNOs (e.g., "TNOs are Cool" and ALMA) are starting to change this picture (Muller et al. 2010; Moullet et al. 2011; Muller et al. 2012).

- According to accretion models, TNOs with $D > 100$ km must be primordial, reflecting the end of the accretion phase. Therefore, knowing the distribution of large TNOs can help us to find the largest diameter produced during accretion, since a cut off is expected at large size. In addition, large TNOs can also yield fruitful information about the original distribution of mass in the primordial planetesimal disk that formed the solar system (Kenyon & Luu 1998; Kenyon et al. 2008).
- With current observational technology, as large TNOs are brighter, it is possible to more easily obtain spectra in order to probe compositions and structure.
- Knowledge about the color and albedo distributions of large TNOs can help us to constrain collisional models and to understand surface dynamics (e.g., processes responsible for the dispersion in colors) (Stern 2002; Thebault & Doressoundiram 2003; Kenyon et al. 2008).
- Having a significant sample of large TNOs, dynamical models can be constrained (e.g., hot classical TNOs origin), correlations with orbital elements can be proved or disproved, and the dependence on mass can be better understood considering different populations.

## 5.2 Albedos and sizes

Knowledge of the sizes (diameters) and albedos of TNOs is of fundamental importance for a better understanding of their distributions among the trans-Neptunian populations, further development of models that infer TNOs compositions via spectral measurements, surface and interior dynamics (e.g., presence of atmospheres, space weathering, etc.), and to constrain the orbital evolution of all TNOs by investigating correlations of orbital elements with size or albedo.

A better understanding of the size distribution is important to obtain an insight into the formation of planetesimals and planets. In addition, knowledge about the initial size distribution is essential for studies of collisional evolution in the trans-Neptunian region (Fraser 2009). Finally, as far as I am aware, the effects of albedo variation on the size distribution of TNOs have been explored very poorly in the literature. It is thus important to provide more accurate estimates of the diameter and albedos of TNOs for which no direct measurements are currently available or possible. Indeed, it is intriguing that large TNOs possess higher albedos, in particular in the case of dwarf planet TNOs (as of August 2012, Pluto, Eris, Haumea and Makemake).

Albedos and sizes are normally obtained from simultaneous measurements of reflected light and that emitted in infrared by the TNO of interest (Jewitt & Sheppard 2002; Jewitt 2008). Such measurements permit the calculation of albedos and diameters using well-established formulae (Russell 1916),

$$p\left(\frac{D}{2}\right)\Psi = 2.25 \cdot 10^{16} R^2 E^2 10^{0.4(h_* - h)} \quad , \qquad (24)$$

where $p$ is the geometric albedo, $D$ is the object diameter (km), $\Psi$ is the phase function, $R$ is the heliocentric distance (AU), $E$ is the geocentric distance (AU), and $h$ is the apparent magnitude of the object ($h_*$ represents that of the Sun). $\Psi = 1$ and $E = R - 1$ are assumed when the body is at opposition. The apparent magnitude of the Sun $h_*$ is -27.1, -26.74 and -26.07 in red, visible and blue wavelengths respectively. The apparent magnitude, $h$, is related to the absolute magnitude $H$, distances and the phase angle as $h = H + 5\log_{10}(R \cdot E) + \psi$, where $\psi$ is a parameter of the phase function. For the sake of simplicity, I will assume that the last term is negligible for the purposes of this Work.

Measurements of size/albedo are available for Centaurs and TNOs. However, the albedos of Centaurs are presumably distorted because of contamination by cometary activity. Compared to TNOs,

Centaurs feature planet crossing orbits leading to enhanced processing due to periodic proximity to the Sun and the giant planets. It is possible that an unknown fraction of TNOs could be experiencing cometary activity, depending on their volatile composition, internal structure and other factors (Hainaut et al. 2000).

Even using the largest telescopes in the world, it is hard to measure diameters and albedos of TNOs (Altenhoff et al. 2004; Brown & Trujillo 2004; Brown et al. 2006). Since the albedo distribution of TNOs is unknown, a constant albedo $p = 0.04$ was widely assumed for TNOs in early studies. This assumption was inspired by the well-supported dynamical link between SPCs and TNOs, and the fact that cometary nuclei are quite dark, with $p = 0.02\text{-}0.05$ (Jewitt et al. 2001; Schulz 2002). Furthermore, very low albedos are expected from space weathering studies (e.g., Gil-Hutton 2002). However, large TNOs possess higher albedos on average, roughly proportional to their sizes. Smaller TNOs, and those TNOs surrounded by satellites, also show a wide distribution of albedos (Noll et al. 2008). These observational facts strongly imply that the long-assumed $p = 0.04$ cannot hold for the entire population, or even particular populations, of TNOs. Intrinsic activity, heterogeneous surface composition and other factors could contribute to an increase of albedos above 0.04. In addition, TNOs with $D > 100$ km could have suffered differentiation by the concomitant action of solar irradiation and internal radiogenic heating (de Sanctis et al. 2001; Choi et al. 2002; Merk & Prialnik 2006), thus causing important modifications to their volatile distribution, albedos and other surface features.

The albedos of large TNOs are significantly higher than all other trans-Neptunian populations. Interestingly, recently-observed planetary-sized TNOs, namely Eris and Haumea (Brown et al. 2006b; Rabinowitz et al. 2006), also possess very high albedos: 0.8 and 0.6, respectively. Large TNOs are probably intrinsically more reflective as a result of mechanisms dominant on these bodies, such as differentiated internal structure, the presence of large amounts of ices (near, or on, the surface), intrinsic activity (possibly linked to atmospheres and/or cryovolcanism), distinct collisional evolution, and temporary atmospheres, among others (Jewitt & Luu 1998; Jewitt 1999; Jewitt et al. 2001). The gravity of large TNOs would favor the existence of tenuous atmospheres and associated icy frosts on their surfaces, which would help to increase their albedos. In particular, this is feasible from two points of view. First, large bodies have high escape velocities (because they are more massive), implying common chemical species in the outer solar system (e.g., $CH_4$, CO or $N_2$,) would not easily escape from their surfaces via sublimation and/or collisional ejecta; Second, since these icy bodies must have formed in between, or beyond, the orbits of the giant planets, they should possess huge internal reservoirs of ices, which would sustain hypothetical atmospheres/icy frosts over the age of the solar system (Yelle & Elliot 1997; Schaller & Brown 2007).

Assuming that a large TNO can sustain a (temporary) atmosphere consisting of $CH_4$, CO or $N_2$, and considering bodies with mean densities of 0.5-1.0 $gcm^{-3}$, Lykawka & Mukai (2005b) obtained critical sizes as a function of heliocentric distance taking into account the molecule velocities and the body’s escape velocity. In conclusion, currently all large TNOs (>700-800 km) would be capable of sustaining thin atmospheres and associated icy frosts composed of $CH_4$, CO or $N_2$, even for body bulk densities as low as 0.5 $gcm^{-3}$, suggesting surface rejuvenation (caused by intrinsic activity) could explain the observed higher albedos.

### 5.2.1 Influence of albedos on the size distribution and the CLF

The largest TNOs possess albedos greater than the long-assumed $p = 0.04$, increasing as a function of size (Lykawka & Mukai 2005b; Grundy et al. 2005). Lykawka & Mukai (2005b) derived two empirical relations that can provide helpful estimations of diameters and albedos as a function of absolute magnitude:

$$D(H) = fg^{H} \quad , \tag{25}$$

where *f* and *g* are parameters that inform the range of sizes and the rate of albedo changes over those sizes.

$$p(H) = \frac{U \cdot 10^{0.4(h_{*}-H)}}{f^{2} g^{2H}} \quad , \tag{26}$$

where $U = 9 \cdot 10^{16}$ and $h_{*} = -26.74$ (visual apparent magnitude of the Sun).

Both equations are valid for $H < 5.5$, and the fitting parameters are $f = \left(2.05^{+0.10}_{-0.15}\right) \cdot 10^{3}$ and $g = \left(8.00^{+0.15}_{-0.30}\right) \cdot 10^{-1}$. In general, the estimated results obtained by using these equations yield diameters/albedos roughly compatible with the measured values (i.e., within, or not too far from, the observed error bars). Thus, these equations can be used to estimate the diameter or albedo of objects without observational measurements[13].

An albedo distribution behaving as described by Eq. 26 will strongly affect the number of large TNOs observed in the sky. That is, since these bodies are intrinsically brighter than other TNOs, they will be more likely to be found by surveys. Thus, for a given absolute threshold ($H = 5.5$), this will translate into shallower slopes of the CLF near the bright end, implying changes to the TNO sky densities. Therefore, discovery would be favored implying sky densities 50-100 times larger for dwarf planets ($m \sim 15$ mag), and 2-10 times for 1000 km-class TNOs ($m \sim 19.5$ mag).

In addition, since albedo scales as $p^{-3/2}$ with mass, (large) TNOs with systematic higher albedos as a function of size will reduce the total mass estimated (up to 50% less) in the trans-Neptunian region, in particular for the dominant population of TNOs with $D > 100$ km (Bernstein et al. 2004; Chiang et al. 2007).

## 6 General methods

In this section, I summarize a series of studies and methods related to the various results discussed in Sections 7 and 8, with emphasis on the origin and evolution of several small-body populations in the trans-Neptunian belt. In general, these results were compared to observations in an attempt to better understand the primordial evolution of the belt, the origin of its main populations and the implications for the giant planets and outer solar system evolution. In particular, these results are based on extensive computer simulations following timescales of 4-5 Gyr (the initial conditions and other details are listed below and discussed in Sections 7 and 8), analytical equations to better understand the simulation results (most of which are presented in Sections 2 and 4), and performed Kolmogorov-Smirnov (K-S) statistical tests (Press et al. 1992) using orbital data and physical properties of certain TNO populations to search for correlations.

Specifically, the topics of theoretical investigations include the long-term orbital behavior of particles representing primordial TNOs. Based on that behavior, the dynamical boundaries in terms of orbital elements ($a$, $e$, $i$) were obtained for the classical region. Other studies include the long-term stability of objects evolving on orbits in the 1:1, 5:4, 4:3, 5:3, 9:5, 2:1, 9:4, 7:3, 12:5, 5:2, 8:3, 3:1 resonances.

[13] However, these expressions are valid only for TNOs, not for other solar system small bodies.

In particular, it was possible to determine the fraction of long-term and transient members within these resonances. These long-term dynamical studies also allowed to better understand the dynamical states (e.g., stable vs. unstable, resonant or not, etc.) of the observed outer solar system bodies with long-arc orbits (TNOs and Centaurs), and the evolution of TNOs on typical scattered disk and detached orbits.

More recently, other investigations include the dynamical evolution of representative primordial trans-Neptunian belts under the gravitational influence of the four giant planets with and without the presence of massive planetesimals in the system. Other details and implications about these particular systems are discussed in the upcoming sections. All these investigations consisted of large-scale simulations, typically including several thousand particles with initial orbits covering the orbital elements of interest, and at least the presence of the four giant planets with either current or pre-migration orbits (i.e., in more compact orbital configurations than now). The planetesimal disks consisted of objects moving on trans-Neptunian orbits, either evolving on dynamically cold $e = i \sim 0$ or hot orbital regimes. Minor bodies in the modeled systems suffered only perturbations from massive bodies (e.g., the Sun, the planets or other massive objects within the system). Disk planetesimals were treated as massive or massless bodies, depending on the purpose of the investigation. The integrations completely accounted for collisions and close encounters with the planets and the Sun, and particles that collided with a massive body were removed from the integrations. The mass of the terrestrial planets was added to the Sun.

Although the giant planets were placed on their current orbits in several of the aforementioned simulations, for cases where planetary migration was included in the model, they were also initially considered in pre-migration compact orbital configurations (within ~17-20 AU). Several configurations were tested, where Jupiter, Saturn, Uranus, and Neptune were typically placed initially at 5.4, 8.7, ~13-16, and 17~23 AU, respectively. Planetary migration operated in the model as described in Section 2.1.3. In general, the giant planets migrated over several AU on timescales described by $\tau$ = 1, 5, or 10 Myr, bracketing the typical migration speeds determined in studies of the energy and angular momentum exchange of planets and disk planetesimals (e.g., Levison et al. 2007). Planetary migration was executed for durations equivalent to 5-10 times the typical timescale above, when the planets acquired their current orbits at the end of that migration.

Several of the integrations were conducted using the EVORB package (Brunini & Melita 2002; Fernandez et al. 2002). This integrator is based on a second-order symplectic method, essentially the same technique used in the MERCURY integrator (Chambers 1999), which provides reliable integrations over the age of the solar system even for timesteps as large as 1-2 yr (Wisdom & Holman 1991; Duncan et al. 1998; Chambers 1999; Horner et al. 2003). In the simulations presented in this review, the time steps ranged between 0.25-1 yr, with preference for smaller values when modeling planetary migration. Gravitational scattering by the planets was followed by the more accurate Bulirsch–Stoer algorithm, which is automatically used in the integrators above. To securely identify resonant objects, I created and used RESTICK (RESonance STICKing), a code capable of identifying objects in resonant motion in any part of the trans-Neptunian region. The RESTICK code permits automatic resonance identification, the calculation of libration amplitudes, and other resonant properties, including asymmetric librations ($r$:1 resonances) and resonance capture durations. The code was also tested exhaustively, providing reliable and accurate results (see Lykawka & Mukai 2007c for more details).

In the simulations that included one or more massive planetesimals (i.e., embryos or planetoids), in addition to the giant planets (which were initially in compact orbital configurations mimicking the system prior to planetary migration), these bodies were also modeled as massive perturbers in a self-consistent way in the calculations. The planetoids were placed in typical Neptune-scattered (i.e., with perihelion close to Neptune's semimajor axis), distant (~40-160 AU) and inclined, orbits (10-40 deg), which are the expected outcomes from gravitational scattering. In most cases, the masses of the planetoids ranged from 0.1 to 1.0 $M_{\oplus}$, representing the massive embryos that likely formed during late stages of planet formation. This range of masses falls well within the upper limits provided by several theoretical and observational constraints, namely 1-3 $M_{\oplus}$ inside 60-70 AU (Hogg et al. 1991; Melita et al. 2004; Gaudi & Bloom 2005; Parisi & Del Valle 2011). Finally, I conducted a large number of simulations to investigate the outcomes of a model containing the giant planets, and such a planetoid, before, during and after migration, as described in Sections 8 and 9.

# 7 Unveiling new clues and constraints for models of the architecture of the trans-Neptunian region

## 7.1 The structure of the classical region

The dynamical evolution of objects placed in the classical region over the age of the solar system seems to suggest a differentiated evolution between its inner and outer portions. In the inner classical region ($a$ < 45 AU), stable orbits were found for objects with $q$ > 39-40 AU. In addition, there is significant sculpting by the 5:3, 7:4, 9:5 and 11:6 resonances over Gyr timescales. In the outer classical region ($a$ > 45 AU), stable orbits are common, especially for $q$ > 40 AU and low-$i$, but perturbations near the 2:1 resonance are important, affecting stabilities and the number density distributions of classical bodies located near this resonance. In summary, the inner ($a$ < 45 AU) and outer ($a$ > 45 AU) subregions show a dependence on inclinations and the sculpting of resonances between the 5:3 and 2:1 resonances.

Although the role of the 2:1 resonance was remarkable, its erosive effect cannot explain the lack of classical TNOs beyond about 45 AU, nor can it explain the predominance of eccentric orbits there. Although orbits with low eccentricity at $a$ = 45-50 AU are stable over the age of the solar system, only classical TNOs with moderate-large eccentricity have been found beyond ~45 AU. In short, the existence of an outer edge at $a$ ~ 48 AU cannot be explained solely by the perturbation of the 2:1 resonance and of the giant planets in the current solar system.

How can we explain the mismatch between model results and observations in the outer classical region? Another perturbative mechanism apart from the four giant planets seems to be needed in order to explain the unexpected excitation of this region and the existence of the trans-Neptunian belt’s outer edge. In this Work, I focus on the gravitational perturbation of a massive planetesimal (a planetoid) that existed during the early solar system as a potential mechanism. For a given orbital configuration of the planetoid, its perturbation would become important in gravitationally sculpting the orbits of classical TNOs. Besides, the planetoid should be massive enough to efficiently perturb the outer classical region in reasonable timescales (i.e., <100 Myr), and also truncate the trans-Neptunian belt. One possibility is that the planetoid excited the eccentricities of TNOs beyond 45 AU, which would explain the paucity of low-eccentricity TNOs in that region and perhaps the trans-Neptunian belt’s outer edge. However, such an excitation should not last too long, otherwise the inner classical subregions at $a$ < 45 AU would be depleted too heavily. Likewise, for a certain timescale, the planetoid should be located neither too close nor too far away for its perturbations to be effective on classical TNOs.

An alternative explanation for the edge is that it would reflect the original size of the primordial planetesimal disk, although this may only marginally agree with observations of other protoplanetary disks (e.g., Jewitt et al. 2009).

## 7.2 Origins and dynamical interrelation of the main classes of TNOs

The classification schemes proposed by Lykawka & Mukai (2007b) and Gladman et al. (2008) are currently the most comprehensive. They provide several clues concerning the origin and evolution of the five major dynamical classes of outer solar system bodies (Centaurs + four classes of TNOs). In particular, the existence of classical, resonant, scattered and detached TNOs suggests several dynamical mechanisms have operated during the history of the solar system. At least two of them have been active until today: the gravitational sculpting applied by the giant planets, and the resonant dynamics associated with them. I briefly comment on the origin of the main classes of TNOs below.

*Classical TNOs*. The origin of cold classical TNOs is still under debate: inner trans-Neptunian belt ($a$ < 30-35 AU) vs. local formation (35 AU < $a$ < 48 AU) (Gomes 2003; Morbidelli 2004; Batygin et al. 2011). The hot classical population is believed to have originated in the inner primordial planetesimal disk (at about 25 AU < $a$ < 30 AU) (Gomes 2003).

*Resonant TNOs*. Most of this population probably was acquired by resonance sweeping through the (cold or excited) planetesimal disk at $a$ ~ 20-35 AU during Neptune's migration (e.g., Hahn & Malhotra 2005; Levison et al. 2008). However, in general, it is difficult to constrain whether these bodies formed locally or originated in other regions of the disk before being swept out by the resonances. It is also possible that resonant TNOs in distant resonances beyond 50 AU originated by capture at locations beyond 35 AU; in particular, the $a$ ~ 35-55 AU region (Lykawka & Mukai 2007a).

*Scattered TNOs*. These bodies represent ~1% of a larger population that has been scattered by Neptune over the age of the solar system (Duncan & Levison 1997; Morbidelli et al. 2004; Lykawka & Mukai 2007c). Consequently, scattered TNOs probably formed near Neptune's current location (25 AU < $a$ < 35 AU) and along the path traversed by the giant planet during its outward migration at roughly 15 AU < $a$ < 30 AU. As discussed in Section 9, the region beyond 35 AU may have also contributed to the scattered population (Lykawka & Mukai 2008).

*Detached TNOs*. These bodies could be originally scattered TNOs or local planetesimals of the primordial planetesimal disk. In either case, an extra perturbation is necessary. For instance, a temporarily scattered or a resident massive planetesimal can perturb the orbits of scattered bodies in such a way that their perihelia are easily lifted beyond the threshold of 40 AU that commonly defines the detached population reservoir (e.g., Gladman & Chan 2006). Thus, this process could produce a prominent population of detached objects.

*Very high-i TNOs*. This subclass is intriguing because no known model can produce a large population of TNOs with $i$ > 40 deg. As for detached TNOs, an extra perturbation seems needed to account for this population too.

Considering the various dynamical mechanisms relevant to the solar system, Neptune's gravitational perturbation can explain the scattered TNOs. Planetary migration is required to explain the hot classical and resonant populations. Finally, an extra perturbation is needed to account for the existence of detached TNOs and very high-$i$ TNOs. The same mechanism responsible for that perturbation may also explain the eccentricity excitation of classical TNOs and the formation of long-term resonant TNOs in distant resonances beyond 50 AU.

### 7.2.1 Resonant TNOs in the scattered disk

There is evidence for the superposition of resonant TNOs stable over the age of the solar system and counterparts on temporarily resonant orbits across the trans-Neptunian region (e.g., with timescales < 1 Gyr). As discussed in Section 4.3, the evolution of TNOs in the scattered disk can be well described by scattering by Neptune and resonance sticking with that planet. In this sense, the role of distant resonances beyond 50 AU in the high survival rate of scattered TNOs over the age of the solar system is fundamental, otherwise we should observe at least an order of magnitude fewer scattered TNOs if they evolved solely by orbital diffusion (Malyshkin & Tremaine 1999). In addition, these resonances can significantly change the object's perihelion, thus making scattered TNOs evolve temporarily as detached bodies ($q$ > 40 AU), a phenomenon enhanced if the TNO also experiences Kozai mechanism while inside a (mean motion) resonance.

The probability of capture into resonance depends on the initial orbital elements of the planetesimals, the migration speed of the giant planets, and the strength of the resonance (Dermott et al. 1988; Hahn & Malhotra 2005). Therefore, during planetary migration, the capture of TNOs by various sweeping resonances was not a 100% efficient process. For typical migration parameters ($\tau$ = 1-10 Myr), capture to the 9:4, 7:3, 5:2, 8:3, 3:1 and other high order resonances is possible only if the disk planetesimals had initially moderately or highly eccentric orbits. Focusing on the observed long-term resonant TNOs beyond 50 AU, and recalling that their evolution follows the conservation of Brouwer's integral (Eqs. 16 and 17), it turns out that only particles with initial $a$ > 39-40 AU and $e$ > 0.07-0.09 could have evolved to orbits that resemble the observed 5:2 resonant TNOs after migration. Similarly, to reproduce the 9:4 and 8:3 resonant TNOs, objects with initial $a$ > 42-43 AU and $e$ > 0.11-0.12 are required before migration. Notice that this is valid for a very wide range of migration parameters and initial conditions for Neptune, because Brouwer's integral does not depend on such parameters.

By comparing the resonant properties of objects in resonance obtained by resonance sweeping and temporarily captured from scattered orbits with those of observed 9:4, 5:2, and 8:3 resonants, it is possible to infer the origin of those resonant TNOs in terms of the scattered disk or an excited local planetesimal disk. Based on the results of Lykawka & Mukai (2007a), particles captured from the perturbed disk acquire a broad range of libration amplitudes, in accordance with those determined for resonant TNOs, while particles captured from the scattered disk resulted in resonants with very large libration amplitudes. Second, only the resonance sweeping model could easily reproduce the $e$ and $i$ of currently observed 9:4, 5:2, and 8:3 resonant TNOs. Finally, the efficiency of long-term capture in these resonances from the scattered disk is extremely small. Thus, if scattered objects were the source of the above-mentioned resonance populations, it would require far too massive a disk with planetesimals on initially Neptune crossing orbits, to account for the estimated total masses in each of the resonances. Thus, this would be in serious conflict with solar system formation models (Luu & Jewitt 2002; Morbidelli 2005; Chiang et al. 2007; but see Levison et al. 2008 for an alternative view).

The combination of distribution of libration amplitudes, orbital elements and resonant dynamical timescales of observed distant resonant populations suggest the resonant TNOs in the 9:4, 5:2, and 8:3 resonances with low libration amplitudes originated from an planetesimal disk with broad eccentricity distributions, or from typically scattered orbits, (i.e., current scattered TNOs). If true, the existence of several long-term resonant TNOs implies that they are observational evidence that the trans-Neptunian belt was excited by some mechanism, before the onset of Neptune's migration (Fig. 18).

### 7.3 The Trojan populations of the four giant planets

Of the four giant planets in the solar system, only Jupiter and Neptune are currently known to possess swarms of Trojan objects, and these populations may easily outnumber that of asteroids (Sheppard & Trujillo 2010b). There is evidence that the Jovian and Neptunian Trojans were mostly (or even entirely) captured from the planetesimal disk during planetary migration (Morbidelli et al. 2005; Lykawka et al. 2009; Nesvorny & Vokrouhlicky 2009; Lykawka & Horner 2010) (Figs. 19 and 20). In addition, Uranus and Saturn should also have captured large populations of Trojans via the same mechanism. Therefore, all four giant planets were able to capture and retain a significant population of Trojan objects captured from the disk by the end of planetary migration (see Table 2). The bulk of captured objects are, to some extent, dynamically-unstable, and therefore these objects tend to provide an important ongoing source of new objects moving on dynamically unstable orbits among the giant planets (e.g., the Centaurs and their daughter subpopulation, SPCs) (Lykawka & Horner 2010; Horner & Lykawka 2010c).

To what degree have the primordial populations of Jovian and Neptunian Trojans been dynamically depleted over the age of the solar system? Approximately 75% and 95-99% of the Jovian and Neptunian Trojan populations, and the entire Saturnian and Uranian Trojan populations, were lost over the age of the solar system (since the end of planetary migration). Lykawka & Horner (2010) estimated that the lost Trojans of Jupiter and Saturn probably contained 3-10 times the current mass of observed Jovian Trojans, while the lost Trojan populations of Uranus and Neptune amounted to tens, or even hundreds, of times that mass! This implies that the Trojan populations have been providing an important source of objects on unstable orbits throughout the entire history of the solar system. In addition to sourcing the Centaurs and SPCs, a fraction of these lost Trojans will have left scars after impacts on planets or the satellites of the giant planets (Horner & Lykawka 2010a).

### 7.4 Probing the dynamical signatures of early solar system massive planetesimals

In general, planet formation models are based on the accretion of planetesimals according to the main stages of runaway growth and oligarchic growth (Kenyon 2002; Rafikov 2003; Goldreich et al. 2004b; Schlichting & Sari 2011). Several massive planetesimals (or planetoids) are expected to have existed in the disk during the late stages of planet formation. After that, terrestrial planets and the cores of the giant planets presumably formed by the collisions of such planetoids (Pollack et al. 1996; Kenyon & Luu 1999; Goldreich et al. 2004a; Kenyon & Bromley 2004a; Rafikov 2004). In this way, the newly-formed giant planets cleaned their neighborhoods by scattering a remnant massive disk of planetesimals, including possibly a few, or tens of, planetoids (Fernandez & Ip 1984, 1996; Pollack et al. 1996, Jewitt 1999; Petit et al. 1999). In the end, scattered planetesimals may suffer: (1) Ejection from the solar system; (2) Collision with a planet, a satellite, or the Sun; or (3) Placement in distant reservoirs, such as the scattered disk or the Oort cloud (Oort 1950; Brasser et al. 2006; Kaib & Quinn 2008).

However, one may wonder what evidence would support the existence of a substantial population of massive planetesimals in the past. In fact, a number of characteristics of the present solar system represent such evidence: the formation of the Pluto, Eris, Haumea, and Orcus satellite systems (Brown et al. 2006b; Brown et al. 2010; Canup 2011), the high tilts of Uranus and Neptune (e.g., Brunini et al. 2002), the retrograde orbit of Triton (e.g., Agnor & Hamilton 2006), and the discovery of dwarf planets at trans-Neptunian distances. In particular, the origin of the multiple systems above, and the tilts of Uranus and Neptune, are better understood as a result of giant impacts of massive planetesimals (Stern 1991; Stern 1992; Stern 1998; Brunini & Melita 2002; Brunini et al. 2002; Canup 2005; Stern et al. 2006; Weaver et al. 2006; Barkume et al. 2006; Brown et al. 2006b; Lee et al. 2007; Parisi & Del Valle 2011). Moreover, the Neptunian satellites Triton and Nereid may have been also captured from the trans-Neptunian region (Brown 2000; Schaefer & Schaefer 2000; Luu & Jewitt 2002; Agnor & Hamilton 2006). Finally, upper limits of 1.4~4 $M_\oplus$ for individual massive planetesimals formed beyond Saturn at the end of giant planet formation were found based on constraints from the Neptunian system (Parisi & Del Valle 2011). In short, giant impact or close encounter events of massive planetesimals seem to represent a natural and consistent way to explain the evidence above. If true, such planetesimals must have existed in large numbers to allow such events (Stern 1991; Brown 2002).

The results of Lykawka & Mukai (2007a) suggest that the trans-Neptunian belt had a substantial population of planetesimals with eccentricities and inclinations ranging from near zero (up to ~45-50 AU) to the excited values illustrated in Fig. 18, prior to planetary migration, and that the radius of the planetesimal disk was at least 45-50 AU. This hypothesis is supported by the current distribution of classical TNOs (Fig. 21). This suggests that the primordial cold planetesimal disk suffered an excitation from outside before planetary migration, which ultimately led to the orbital excitation as observed in the outer classical region (Fig. 18). Here, I propose the perturbation of a planetoid (representing one of the massive planetesimals) as a potential mechanism to do the job.

For a given orbital configuration, the planetoid should be massive enough to efficiently perturb the outer regions of the trans-Neptunian belt, and truncate it at 48 AU. The planetoid must have crossed the classical region within a relatively short time, perhaps on the order of Myr, to avoid disrupting it. Thus, the planetoid must be inclined (>10 deg) and relatively distant (> 60-70 AU). Such a hypothetical planetoid was possibly one of the large planetesimals scattered by Neptune in the past, during the late stages of planet formation. On the other hand, since gravitational scattering occurs at fixed spatial positions, the perihelion of the outer planet would not depart much from the orbit of Neptune (Gladman et al. 2002). In addition, the interaction of this massive object with the background disk of planetesimals could lead to a near circular, and very low inclination, orbit via dynamical friction (e.g., Del Popolo et al. 1999), which would be in conflict with observational constraints (Morbidelli et al. 2002). One way to avoid this uncomfortable situation is to consider the action of resonances.

The strongest and dominant resonances in the scattered disk are those of the $r$:1 type located at $a <$ 250 AU. These resonances can induce large perihelion increase for objects evolving inside these resonances (Gallardo 2006a; Gallardo 2006b; Lykawka & Mukai 2007c). Because this mechanism is mass-independent, small TNOs, dwarf planets and other more massive objects could be captured in any of these resonances. The capture probability in $r$:1 resonances is also higher than in other resonances beyond 50 AU (Lykawka & Mukai 2007c). In short, after experiencing gravitational scattering events with the giant planets, one of the several planetoids present in the early solar system probably interacted with a Neptunian $r$:1 resonance, thus avoiding scattering evolution with Neptune.

# 8 The influence of a massive planetesimal (planetoid) in the trans-Neptunian region

As noted in Section 7.4, in addition to the giant planets, a large number of massive planetesimals formed simultaneously during late stages of planet formation. Since these planetesimals strongly gravitationally interact with each other and with the newly-formed planets, such planetesimals were scattered into inner and outer regions of the planetesimal disk. Could any of such objects acquire a stable orbit?

A massive trans-Neptunian planetoid could exist in the present solar system, provided that it is currently in a distant and inclined orbit to have avoided detection by any of the various surveys for outer solar system objects. It should be noted that this hypothesis is not completely new. A few modern works have investigated the influence of such planetoids on the orbital distribution of TNOs (Morbidelli & Valsecchi 1997; Brunini & Melita 2002; Melita et al. 2004). The planetoid proposed in the model described in the following sections would represent a single survivor from the population of scattered planetesimals. Before introducing this particular model, I first revisit two main models of interest: the massive planetesimal, and resonance sweeping, models.

## 8.1 The massive planetesimal model

The first variant of this scenario considers the temporary existence of massive planetesimals in the trans-Neptunian belt (Morbidelli & Valsecchi 1997; Petit et al. 1999; Gladman & Chan 2006), whilst the second suggests the existence of one undiscovered planetoid orbiting at trans-Neptunian distances (Matese & Whitmire 1986; Harrington 1988; Maran et al. 1997; Brunini & Melita 2002; Melita & Williams 2003; Melita et al. 2004).

Neptune-scattered massive planetesimals could have sculpted the primordial trans-Neptunian belt, leading to orbital excitation and dynamical depletion of a substantial population of TNOs that acquired unstable orbits during that time (Morbidelli & Valsecchi 1997; Petit et al. 1999). The perturbation of this massive planetesimal could also have led to the creation of the detached population (Gladman & Chan 2006). More recent studies using modern techniques have explored the prospects of a Mars-like planet at about 60 AU to explain the excitation in the classical region, the trans-Neptunian outer edge and the formation of the detached population (Brunini & Melita 2002; Melita & Williams 2003; Melita et al. 2004). In short, the massive planetesimal model is appealing because it could explain consistently several features in the trans-Neptunian region, as mentioned above.

Nevertheless, the orbital structure of the classical region obtained in the aforementioned studies is incompatible with that of the classical TNOs. Also, Melita et al. (2004) also found that a resident planetoid with such an orbit is unable to simultaneously create the trans-Neptunian outer edge, excite the orbits of classical TNOs, and explain the prominent resonant populations in the entire trans-Neptunian region. In addition, the massive planetesimals, as modeled in those works, would destroy the stability of primordial resonant populations; in particular, the stable 3:2, 2:1 and 5:2 resonant populations. In the case of the distant resident planetoids, those postulated in early studies can be ruled out based on constraints from the motion of planetary orbits/spacecraft and observations (Matese & Whitmire 1986; Harrington 1988; Hogg et al. 1991; Maran et al. 1997; Morbidelli et al. 2002; Gaudi & Bloom 2005). Moreover, the proposed planetoids in more recent work would be located at 50-60 AU on a near circular, and low inclination, orbit. Such a scenario would be in conflict with observational constraints. In other words, a surviving Mars (or less massive) embryo at 50-60 AU, on a low inclination orbit should have already been discovered if it existed (Morbidelli et al. 2002).

## 8.2 The resonance sweeping model

Today, the resonance sweeping model is considered a crucial mechanism to explain TNOs trapped in several resonances, and other properties of the solar system as a whole (Malhotra 1995; Fernandez & Ip 1996; Ida et al. 2000b; ; Chiang & Jordan 2002; Chiang et al. 2003; Gomes 2003b; Gomes et al. 2004; Hahn & Malhotra 2005; Murray-Clay & Chiang 2005; ; Morbidelli et al. 2008; ; Thommes et al. 2008; Minton & Malhotra 2011) (see also Section 2.1.3 for details).

In the standard resonance sweeping model, in which the Neptunian resonances sweep through a disk of planetesimals featuring cold orbital conditions, resonant TNOs are mostly reproduced within 48 AU. However, in such a scenario the stable resonant populations beyond 50 AU are not reproduced. In addition, the model cannot reproduce the excited and peculiar orbits of the classical and detached populations. Another problem is that the production efficiency of hot classical and detached objects is too low, and likely in conflict with observations (Gomes 2003a; Gomes 2003b; Chiang et al. 2007). Hahn & Malhotra (2005) performed simulations of planetary migration over both cold and hot planetesimal disks. In these variants of the model, resonant populations in the scattered disk can be well reproduced (see also Lykawka & Mukai 2007a). Nevertheless, the assumption of an initially-excited planetesimal disk is not justified in those models, and those models fail to reproduce the detached population.

In general, the resonance sweeping models published so far (cited above) are unable to explain the trans-Neptunian outer edge and the low total mass of the belt. Moreover, independent of model parameters and published variant scenarios, the obtained detached populations are too small to account for that which we observe today.

## 8.3 The Planetoid-Resonance hybrid model

### 8.3.1 Non-migrating planetoids

The parameter space explored in massive planetesimal, and in resonance sweeping, models was not exhaustive. In this way, what is the feasibility of a model containing a planetoid in a distant trans-Neptunian orbit? I have investigated in detail the influence of such planetoids, varying their initial orbital and mass parameters, in an attempt to reproduce the excitation of the classical region and form the trans-Neptunian outer edge. The giant planets were modeled at their current orbits in these systems. The typical time span of the simulations was 4 Gyr. All the main results presented below are based on the findings of Lykawka & Mukai (2008).

A single planetoid is able to disrupt the local cold planetesimal disk by removing objects with $e < 0.1$ at 48 AU, particularly for planetoids with 50 AU < $q_P$ < 56 AU, $i_P$ < 20-25 deg and mass of several tenths of $M_\oplus$, where, henceforth, the subscript P refers to the planetoid. Planetoids with smaller perihelia resulted in too strong an excitation in the classical region, whereas those with larger perihelia and/or higher inclinations were unable to produce a trans-Neptunian outer edge. In the successful runs, the formation of the outer edge occurred on timescales of at least 2 Gyr for planetoids with $a_P$ = 60-140 AU. Also, less massive planetoids ($m_P$ = 0.01-0.05 $M_\oplus$), placed in the 2:1 resonance, excited planetesimals in narrow regions of the trans-Neptunian belt to large eccentricities, and to inclinations of up to 20 deg in a few hundred Myr. However, 2:1-resonant planetoids either led to too little perturbation in the classical region or too much depletion of the inner parts of that region.

In conclusion, non-migrating planetoids are unable to reproduce the observed orbital distribution in the classical region. When the classical region is sufficiently perturbed by the planetoid, the obtained distribution is incompatible with observations. Therefore, non-migrating planetoids cannot form the trans-Neptunian belt outer edge and reproduce the orbits of TNOs in the classical region. Other drawbacks of non-migrating planetoids are presented in Section 7.4.

### 8.3.2 Migrating planetoids

Lykawka & Mukai (2008) investigated systems where the giant planets and the planetoid studied experienced migration through the planetesimal disk. The main results of that study are summarized below. The simulations spanned typically 100-200 Myr. The giant planets were placed in more compact configurations than their present architecture, in line with migration models (Section 2.1.3), and the planetoid had initial perihelion not greater than 10 AU beyond Neptune's initial semimajor axis, prior to migration. Thus, the planetoid represented a primordial Neptune-scattered body in the simulations. In agreement with that, the planetoid's initial inclination was set at $i_P$ = 10-30 deg, and its semimajor axis (> 40 AU) and mass (0.1-1 $M_\oplus$) were considered free parameters, but limited to reasonable ranges shown within parentheses. In these simulations, along with the giant planets, the planetoid was forced to migrate outwards obeying a predefined radial displacement according to the position of a distant and strong *r*:1 resonance with Neptune. Planetary migration was modeled as described in Section 2.1.3. The giant planets and the planetoid migrated according to Eq. 15 using $\tau$ = 5-15 Myr. In particular, Neptune started at 23.1, 20, and 17.1-18.1 AU in the sets of initial orbital configurations for the giant planets. By the end of the simulations, planetoids had perihelia around 50-80 AU. A word of caution is needed, because this procedure was an idealization of a resonance capture followed by interactions with the planetesimal disk and the Kozai mechanism. In self-consistent simulations, a similar behaviour was achieved only in less massive planetesimal disks, and the evolution of the planetoid was stochastic before experiencing captures into *r*:1 resonances (Lykawka & Mukai 2008). Further investigations are warranted in the future.

A single migrating planetoid was typically successful in the simultaneous formation of resonant populations at $a$ < 50 AU, orbital excitation in the classical region, and the disruption of the disk beyond about 48 AU, by the end of migration. Fast migrations (i.e., using smaller $\tau$) of the planetoid yielded too little excitation in the disk, whereas incursions with $q_P$ < 50 AU caused too much perturbation in the region.

A migrating planetoid represents an improvement over fixed planetoids (Section 8.3.1). However, inclinations of objects in the classical region experienced little excitation, resulting in a clear lack of hot classical objects (those objects with $i$ > 10 deg). That is, although hot classical particles were obtained in most cases, the efficiency was as low as < 0.2%, and their obtained eccentricities were, in general, higher than those of hot classical TNOs. Another problem is that none of the runs produced any significant resonant populations in the 9:4, 5:2, and 8:3 distant resonances. This occurred because the capture probability was too low due to the small eccentricities of disk planetesimals as those resonances crossed the belt. In short, the excitation in the planetesimal disk prior to planetary migration was insufficient (Lykawka & Mukai 2007a).

Yeh & Chang (2009) proposed a model in which, in addition to the four giant planets, a fifth planet of mass 0.1-2.0 $M_\oplus$ migrated while trapped inside the 3:2 resonance with Neptune. In summary, that model produced results that could satisfy several of the main constraints in the trans-Neptunian belt. In particular, their main results include producing: (1) The high inclination component of 3:2 resonant TNOs; (2) The orbital excitation in the classical region; (3) The small ratio of the 3:2 resonant population compared to the classical population; (4) The excess of 3:2 resonant TNOs compared to their 2:1 counterparts; and (5) The bulk of the 5:2 resonant population.

In summary, there are two major problems for the migrating planetoids scenarios. First, the excitation of the trans-Neptunian belt occurs too late when distant resonances have already swept the disk. Second, the resulting excitation in inclinations is not high enough inside 50 AU, even in the case of $M_P > 0.5\,M_\oplus$. The obtained eccentricities usually span wide ranges in qualitative agreement with that observed in the classical region. However, depending on the mass and/or the timescales for the perturbation of the planetoid, TNOs in the classical region may acquire too excited orbits. Other potential problems include the production of peculiar groupings of TNOs not currently seen (e.g., TNOs on near-circular orbits beyond 50 AU).

### 8.3.3 Survival of resonant TNOs with the presence of resident trans-Neptunian planets

The migrating planetoids in successful runs were able to satisfy a relatively-large number of constraints. However, because the perihelia acquired by these planetoids were within 50-60 AU, how this would affect the long-term stability of eccentric resonant TNOs with orbits reaching such distances? (see Figs. 1 and 22). Indeed, such TNOs may be dislodged from their host resonance by gravitational interactions with the planetoid, in particular during the latter perihelion passages.

Objects that leave resonances will no longer be protected from close encounters with Neptune by the libration mechanism (Peale 1976; Malhotra 1998; Murray & Dermott 1999). Thus, these objects will be ultimately scattered by the giant planet becoming scattered TNOs (for $q$ < 37 AU), or local TNOs on fossilized orbits near their parent resonance location, $a_{res}$ (for $q$ > 37 AU).

Lykawka & Mukai (2008) investigated in detail the survival of the stable resonant TNOs located in the 3:2, 2:1, 5:2 and other resonances with the presence of hypothetical resident planetoids. Unsurprisingly, all resonant populations tested were strongly perturbed by the planetoids with $q$ = 50-60 AU. In particular, the 3:2 resonant population was depleted at ~30-90% levels, and 3:2 bodies with $e$ > 0.25-0.3 often left entirely the resonance. Similarly, the 2:1 resonant population was completely disrupted in almost all runs, and yielded resonant survivors with $e$ < 0.26. Therefore, in general, planetoids with $q_P$ = 50-60 AU and $i_P$ < 40 deg led to the destruction of the 2:1 resonant population at high eccentricities in less than 1 Gyr, and significant depletion of the 3:2, 5:3, and 7:4 resonances over 4 Gyr. In short, a fraction of resonant objects can survive 4 Gyr if the following condition is satisfied: $q_P > Q$ + 2-3 AU for $m_P$ = 0.1-0.5 $M_\oplus$, where $Q$ is the aphelion distance.

By considering the results above, the observed 3:2 and 2:1 resonant populations could survive over the age of the solar system only if $q_P$ > 53 AU and $q_P$ > 67 AU, respectively. Although a planetoid with perihelion around 50-60 AU would be able to create the trans-Neptunian belt outer edge and excite the classical region (Section 8.3.2), it would also destroy the structure of the 2:1 resonant population (Fig. 22). This implies that only planetoids that acquired $q_P$ > 70-80 AU after migration could allow the survival of resonant populations at $a$ < 50 AU in a way compatible with observations. These results, also rule out resident trans-Neptunian planets proposed in several earlier studies because they generally have 50 AU < $q_P$ < 60 AU (Matese & Whitmire 1986; Harrington 1988; Brunini & Melita 2002; Melita & Williams 2003; Melita et al. 2004).

The survival of the 5:2 resonant population offers another important constraint for massive planetesimal models (Section 8.1). In general, the 5:2 resonant population can survive for 4 Gyr if the planetoid's perihelion is about the same, or slightly greater than the aphelia experienced by the resonant members. This result is valid for $a_P$ = 60-140 AU, $i_P$ = 10-40 deg and 0.3-1.0 $M_\oplus$. This implies that any resident planetoid must have $q_P$ > 80 AU to guarantee the survival of a fraction of the resonance population with orbital and resonance properties compatible with the stable 5:2 resonant TNOs (Fig. 23). Therefore, these results also rule out the existence of any hypothetical planet at $a$ ~70-80 AU that could explain Sedna's orbit (Brown et al. 2004).

# 9 The origin and evolution of the trans-Neptunian belt with the presence of a massive planetoid

In this section, I summarize the main results of the scenario modeled by Lykawka & Mukai (2008) and discuss the implications based on these results and the new constraints discussed in Section 7. In short, this scenario suggests that after a planetoid was scattered by Neptune and the other giant planets, it managed to temporarily excite the primordial planetesimal disk to provide an early excitation to the disk that could account for the formation of 9:4, 5:2, and 8:3 resonant TNOs. Later, it would have been transported outwards by resonant interactions and scattering with Neptune. Thus, this scenario follows the idea of a hypothetical resident planetoid in the scattered disk, and the time $t$ = 0 for the model is set at the late stages of giant planet formation, a few tens of Myr after the birth of the solar system (e.g., Montmerle et al. 2006).

## 9.1 Summary and preliminary results

The model is divided into three main stages:

I. *Pre-migration excitation of the planetesimal disk*. The giant planets are assumed to have formed in a more compact configuration than they currently occupy, in line with several solar system models (Section 8). In particular, Neptune would have formed within 17-20 AU. Before planetary migration, following a typical scattered orbit, the hypothetical planetoid should be orbiting in an eccentric moderately-inclined orbit ($i_P$ = 10-15 deg) with perihelion, $q_P$, a few AU larger than Neptune's semimajor axis. With such an orbit, the planetoid would have temporarily perturbed the primordial cold planetesimal disk. Based on the most successful results, and incorporating other constraints as discussed in previous sections, planetoids with $a_{P0}$ = 50-80 AU and $m_P$ = 0.3-1.0 $M_\oplus$ were adopted. The planetesimal disks were represented by a few thousand small-mass particles occupying cold orbits in both uniform and $R^{-1}$ disk distributions. Such initial conditions reflect the "optimal" disks found earlier (Section 8.3.2). None of the planets experience migration at this stage. This is consistent with a late start of planetary migration (Gomes et al. 2005). This stage probably lasted for several tens of Myr.

The stirring of a scattered planetoid satisfied the required eccentricity excitation for the formation of distant resonant TNOs beyond 50 AU (Fig. 24). In addition, the perturbed planetesimal disks presented orbital distributions very similar to the non-resonant observed populations in the 40-50 AU region (Compare Fig. 24 with Fig. 21). A representative case is also illustrated in panel **a** of Fig. 25.

II. *Planetary migration*. The excited disks obtained at the end of stage I were taken as initial conditions for the planetary migration simulations (see panel **a** of Fig. 25). Along with the migration of the giant planets, the planetoid is assumed to be transported to large distances via gravitational scattering by the giant planets and through resonance interactions with a strong Neptunian *r*:1 resonance. The best candidate resonances to trap the planetoid would range from 6:1 to 14:1 resonances, which are located within the preferred region of $a$ = 50-80 AU if Neptune experienced orbits within 15-24 AU prior to migration (see also Section 2.1.4).

Recalling the results of Section 8.3.3, in this scenario, two mechanisms would have the potential to satisfy the constraint $q_P > 80$ AU: (1) Dynamical friction within the planetesimal disk, and (2) The Kozai mechanism. Here, given the strength/stickiness of *r*:1 resonances and the common action of KM within these resonances, the planetoid probably exhibited decreased eccentricity (increased perihelion) at the expense of increased inclination while interacting with one of the strong *r*:1 resonances mentioned above. This effect obeys the relation $\sqrt{1-e_P^2}\cos(i_P)$ as a consequence of vertical angular momentum conservation. This stage probably lasted for approximately 100 Myr.

In the main simulations, the planetoid acquired approximately $q_P$ = 80-85 AU and $i_P$ = 30-45 deg in timescales of ~100-200 Myr, by the end of migration (see panel **b** of Fig. 25). The candidate *r*:1 resonances mentioned earlier would translate to semimajor axes of 100-175 AU for the planetoid at the present day. Based on capture probabilities in the 5:2, 3:1 and 4:1 resonances and the comparison of strength/stickiness of the *r*:1 resonances of interest, the capture probability of initially-scattered bodies in the 6:1, 7:1, ..., 14:1 resonances can be estimated to be roughly 0.5-3% (smaller for farther resonances) (Gallardo 2006b; Lykawka & Mukai 2007a; Lykawka & Mukai 2007c). Moreover, large populations of massive planetesimals were present and were scattered by the giant planets during the early solar system, so resonance capture of a single planetoid is plausible.

III. *Long-term sculpting by the planets*. This stage represents the long-term evolution of the system consisting of the giant planets, the planetoid and the remaining disk planetesimals over 4 Gyr. Several of the planetary migration runs were extended to 4 Gyr to investigate the final system outcomes (see panel **c** of Fig. 25). Following the reasoning of stage II, the trans-Neptunian planetoid would have acquired final orbits near one of the strong *r*:1 resonances mentioned above, namely within $a$ = 100-175 AU.

In the most successful simulations, planetoids of 0.4 and 0.5 $M_{\oplus}$ were transported outwards following the location of the 6:1 and 9:1 resonances, respectively. In the long-term stage, the obtained disk planetesimals at the end of planetary migration were cloned out to 54 AU, obtaining an excited planetesimal disk composed of several thousand particles (up to 17,000). Finally, these systems were evolved to 4 Gyr.

In summary, this model considers some basic and well-supported aspects of solar system history, such as the scattering of massive planetesimals by the newly-formed giant planets, planetary migration, and the dynamics of resonance capture. These are essentially the main features of both the massive planetesimal and resonance sweeping models (Sections 8.1 and 8.2).

## 9.2 Classical region

First, objects were removed up to about the current location of the 3:2 resonance through Neptune's gravitational scattering and the overlapping of its resonances within 5 AU from the planet. In particular, the overlapping of resonances in this region gives rise to strong instability. Another mechanism was the capture of local planetesimals by sweeping resonances (e.g., 5:4, 4:3, 3:2) that swept through the primordial disk to ~40 AU. Thus, the lack of TNOs in the region up to about 39 AU is evidence of the outward migration of Neptune.

For both simulations and observations, I considered cold ($i \leq 10$ deg) and hot ($i > 10$ deg) classical populations as non-resonant particles with $a < 50$ AU and $q > 37$ AU. In agreement with the results presented in Sections 8.1 and 8.3, the cold population was strongly perturbed during the pre-migration stage by the planetoid. In particular, the final eccentricities of classical bodies were remarkably similar to the observed values, while the cold nature of their orbits was well preserved ($e < 0.1$ and $i <$ 5-10 deg) (Fig. 25). Furthermore, planetesimals that were initially on very cold orbits were able to acquire $i$ = 5-15 deg. Taking these results together with the fact that resonances in the classical region can also excite inclinations within, or near, such resonances, a significant part of the observed cold population was likely promoted to the hot one, especially in the 5-15 deg range of inclinations. These results suggest that the cold population formed in situ at approximately $a > 37$ AU and were able to populate the element space up to ~15 deg. This also suggests that a fraction of the hot classical TNOs formed by the excitation of local planetesimals to the moderately-inclined range of 5-15 deg, whereas the remaining fraction entered the classical region from inner regions of the disk to form the component with higher inclinations (15-35 deg). Therefore, the hot classical population would have originated from both the local planetesimal disk (~35-50 AU) and the inner solar system (~15-35 AU).

One important implication is that the distinct physical properties of the cold and hot classical populations would become more evident among those objects with low inclinations and those with $i >$ 10-15 deg. This would also mean that both populations could be used as representatives of the primordial planetesimals that formed at 15-35 and 35-50 AU, respectively. Today, unfortunately, details of the initial locations of disk planetesimals are virtually lost. However, in support of this scenario, the distribution of inclinations combined with the spectral slopes of classical TNOs shows much clearer differences in these distributions for classical TNOs with $i >$ 10-15 deg than the typically-adopted 5 deg inclination threshold of cold and hot classical TNOs (Chiang et al. 2007). Finally, these results also suggest that the cold and hot populations underwent significant different dynamical histories, which likely affected their accretion evolution during late stages of planetesimal formation (Section 7.4). This is in line with analysis of CLFs and size distributions of classical TNOs (Fraser et al. 2010; Fuentes et al. 2010).

The above scenario could also explain the intrinsic fraction of cold and hot populations because an important fraction of cold classical objects was promoted to the hot population. The ratios of cold to hot classical bodies from the simulations resulted in comparable values to those observed, despite an initial preference for a large population in the cold component. This is compatible with the intrinsic ratio estimates based on observations (e.g., Morbidelli & Brown 2004), and avoids the problem of overpopulation of cold to hot bodies, as seen in previous studies (e.g., Gomes 2003b).

The planetoid's perturbation could explain the general orbital structure of the classical region, even including the lack of objects on low-eccentricity orbits beyond 45 AU. The obtained excited orbital distribution is a consequence of the perturbation from the planetoid during the pre-migration stage (Figs. 24 and 25). The time span necessary to reproduce the observed excitation was about 20-100 Myr for $m_P$ = 0.3-1.0 $M_\oplus$ (the more massive, the shorter the timescale). This suggests that the currently observed orbital structure in the classical region was established early in the solar system, before planetary migration. However, a drawback in the scenario was the shortage of bodies with $i$ >15-20 deg. Using a slightly different planetoid-resonance hybrid model, the findings of Yeh & Chang (2009) support the idea that the classical region could reflect the perturbation of a massive planetoid during the early solar system.

Concerning the outer classical region, the planetoid was effective in creating the trans-Neptunian belt outer edge at 48 AU. It is worth noting that the model reproduced, for the first time, both the absence of low-eccentricity bodies and the abrupt decrease of the number density of TNOs with heliocentric distance beyond 45 AU. This finding is connected to planetoids that evolved on pre-migration orbits with $a_P$ = 60-80 AU, $q_P$ = 20-30 AU and $i_P$ < 15 deg. In summary, the excitation in the classical region and the trans-Neptunian belt outer edge were possibly created during the first tens of Myr before planetary migration. Thus, these observational features would tell us about the origin and orbital history of the planetoid, and other relic features during very early stages of the formation of the solar system (see also Fig. 25).

## 9.3 Resonant structure

A large number of objects were identified within various resonances across the trans-Neptunian region after 4 Gyr, namely the 5:4, 4:3, 7:5, 3:2, 8:5, 5:3, 7:4, 9:5, 11:6, 2:1, 13:6, 11:5, 9:4, 7:3, 5:2, 8:3, 11:4, and 3:1 resonances. The great majority of the members of these resonances were captured from the primordial planetesimal disk at locations between 30 and 50 AU. This distance range approximately overlaps with the proposed location of the cold classical region, implying that these resonant populations and cold classical TNOs would share similar physical properties based on their birth places in the disk. It is also worth noting that other models that included a migrating planetoid can reproduce quite well the 3:2, 2:1, and the 5:2 resonances (Yeh & Chang 2009).

In addition, the eccentricities of resonant TNOs were very well reproduced. In particular, the eccentricities of resonant TNOs were better matched for disks truncated at approximately $a$ = 51-54 AU. Concerning inclination distributions, the obtained resonant objects yielded typically $i$ < 20-30 deg, which is reasonably comparable with observations. Furthermore, the distribution of libration amplitudes is also in good agreement with those derived from observations. In particular, stable populations in the 9:4, 5:2, and 8:3 resonances yielded a wide range of libration amplitudes that satisfy the constraints posed by the observed populations. The existence of stable resonant populations in the scattered disk obtained in the simulations agree well with the existence of long-term 9:4, 5:2, and 8:3 resonant TNOs (Lykawka & Mukai 2007a). Lastly, resonant particles experienced the Kozai mechanism inside the following resonances: 5:4, 4:3, 7:5, 3:2, 8:5, 5:3, 7:4, 2:1, 7:3, 5:2, 8:3, and 3:1. The relative fraction of Kozai librators inside these resonances is reasonably compatible with observations. Symmetric and asymmetric 2:1 resonant TNOs were also reproduced. The fact that so many distinct properties of the resonant populations can be satisfied attest that the model is robust.

Concerning the perturbative effects of the planetoid on resonant populations, the maximum eccentricities of 5:2, 8:3, 11:4, and 3:1 resonant bodies were limited by the planetoid's perihelion approximately via the relation $Q < q_P$. That is, objects with sufficiently high eccentricities were removed from resonance by the planetoid's perturbation (see Sections 2.1.3 and 8.3.3). For this reason, no long-term resonant populations were found beyond the 3:1 resonance. In short, if this scenario is correct, we would not expect to find significant stable resonant populations in the 4:1, 5:1, and other strong resonances in the scattered disk.

Table 3 summarizes the main results and various properties of the resonant populations obtained in the model.

### 9.4 Scattered population

A significant population of scattered objects was obtained at the end of the simulations on typically eccentric orbits with a broad range of inclinations (<50 deg). The obtained scattered population also spread across the classical region and the scattered disk (i.e., with no boundary in semimajor axis). The inclusion of a planetoid hardly affected the process of the formation of scattered objects, so the formation of this population proceeded by gravitational encounters with Neptune over billions of years, similar to the standard fashion, as discussed in Section 4.3. A similar result was also obtained in the planetoid-resonance model of Yeh & Chang (2009) (see their Fig. 11). In the end, the survival fraction of scattered bodies after 4 Gyr of evolution was of the order of 1-2% to the original population of objects on Neptune-encountering orbits.

Despite the apparent minor role of the planetoid, its influence during the early stages of excitation of the trans-Neptunian belt was important in allowing the region at $a > 40$ AU to contribute significantly as a source of scattered bodies. This suggests that about half of the scattered TNOs would have originated somewhere around 40-50 AU, and a few tens of percent from the region at 30-40 AU that was sculpted by Neptune over Gyr. The remaining scattered TNOs originated from resonances (dynamical diffusion) and the path transversed by the migrating Neptune.

In addition, scattered bodies evolving on orbits that passed near the planetoid's sphere of influence (within a few AU of its orbit) suffered significant excitation in inclination, leading to $i$ = 40-50 deg, including analogs of Eris. Indeed, in contrast to the standard models (e.g., Duncan & Levison 1997), in which the gravitational perturbation by the giant planets can produce only < 1% of highly-inclined objects ($i > 40$ deg), this model is much more efficient at producing these highly-inclined objects beyond 50 AU. Indeed, the scattered population acquired up to $i < 50$ deg in most runs, while inclinations up to 90 deg are possible for objects with orbits near that of the planetoid and/or for more massive planetoids. Therefore, it is possible that the planetoid perturbation may explain this highly-inclined subpopulation, as apparently evinced by the first discoveries by observations (Section 4.5).

A representative outcome for the scattered disk is illustrated in Fig. 26.

### 9.5 Detached population

A substantial detached population was obtained at the end of 4 Gyr in several simulations. These objects originally evolved on Neptune-scattered orbits, but later acquired their detached orbits via continuous perturbations by the planetoid. In general, this process lasted until these objects acquired perihelion large enough to be detached from the gravitational sphere of influence of Neptune. The bulk of the detached population resulted in $q$ = 40-60 AU and $i < 60$ deg. More massive planetoids were more efficient in creating detached objects with larger semimajor axes and perihelia, because the perturbations of the planetoids were stronger within the same time span (4 Gyr). In addition to the mechanism described above, detached bodies were also obtained from the outer regions of the cold planetesimal disk, namely objects that acquired eccentric orbits and remained with $q > 40$ AU. Thus, this model could explain the existence of detached TNOs with low inclinations, which are not obtained in scenarios based on resonant interactions (Sections 2.1.4 and 4.4). In addition, some extreme bodies with $a$ = 500-800 AU and 40 AU < $q$ < 50 AU, and analogs of typical detached TNOs such as 2004 XR190, 2000 CR105 and Sedna were obtained in several runs.

More importantly, the obtained detached populations are comparable to, or a few times larger than, the final population of scattered objects. This result is in excellent agreement with unbiased observational estimates (Section 4.4). The calculations also reveal that the detached objects were essentially part of the scattered population during the early stages of the system's evolution. That is, scattered and detached TNOs observed today would be indistinguishable because their origins in the primordial disk would span approximately the same wide region of the planetesimal disk at ~20-50 AU. For this reason, they would also show the larger variety of physical properties (colors, spectra, etc.) among the trans-Neptunian populations.

Finally, an intriguing result of the long-term residence of the planetoid is that it can leave observable orbital signatures in the scattered disk. First, distant resonant populations would possess stable members conditioned to $Q < q_P$ (Sections 8.3.3 and 9.3). In addition, scattered and detached objects with orbits near that of the planetoid acquired, in general, the largest inclinations and perihelia in these respective distributions. Thus, a better and more accurate characterization of the orbital structure in the trans-Neptunian region is warranted to test for the existence of such peculiar signatures (see Fig. 27).

Representative outcomes of both scattered and detached populations, according to key parameters of the planetoid, are illustrated in Fig. 27, while Table 4 shows the statistics of scattered and detached populations during their evolution, as compiled from several independent runs.

Finally, it is worth noting that other models that included a scattered massive planetesimal were able to obtain a substantial population of detached bodies (e.g., Gladman & Chan 2006).

## 9.6 The loss of 99% of the ancient trans-Neptunian belt total mass

As discussed in Section 9.1, in this scenario the planetesimal disk was strongly dynamically depleted during the pre-migration stage, and, to a lesser extent, during the remaining period of the system's history (until 4 Gyr). Overall, approximately 60-85% of the planetesimals were removed from the system after 4 Gyr for disks of ~51-54 AU radius. Moreover, with the planetoid's perturbation, a large fraction of planetesimals initially on cold orbits were excited to higher levels of eccentricities and inclinations. Recall that the accretion vs. fragmentation outcome of small bodies is strongly related to their random velocities, which are given by

$$v_{rnd} = v_K \sqrt{e^2 + i^2} \quad , \qquad (27)$$

where $v_K$ is the Keplerian velocity, and orbital elements refer to the planetesimal (e.g., Kokubo & Ida

2000). The Keplerian velocity can be obtained with $v_K = 29.8a^{-½}$ (kms$^{-1}$).

By using the eccentricities and inclinations of excited planetesimals obtained during stages I and II (before and during planetary migration), Eq. 27 yields collisional velocities high enough that the systems perturbed by a planetoid would have easily entered a fragmentation regime during this early period in solar system history (Stern & Colwell 1997; Kenyon & Bromley 2004a). Consequently, the trans-Neptunian belt experienced intense collisional grinding. If we combine the dynamical depletion of 60-85% with the collisional grinding levels of 92-97% found in detailed models (Kenyon & Bromley 2004a; Kenyon et al. 2008), the remainder of the original belt mass would be ~ 0.5-3%. It is worth noting that an overabundance of resonant populations (e.g., Zhou et al. 2002) and the less-excited conditions considered in collisional evolution models imply that the dynamical and collisional grinding levels above are probably overestimates. Therefore, since a higher degree for the loss of the total mass is expected, this model could explain consistently the current small total mass in the trans-Neptunian belt by the loss of 99% or more mass of the ancient belt.

The trans-Neptunian belt can be considered a debris disk that evolved for ~4.5 Gyr, so it is useful to compare the main features of debris disks with those of the belt. A significant fraction of debris disks exhibit dust excess in their spectral energy distributions. Since, in general, the dust lifetime is much shorter than the age of the star, this dust must have been produced by collisions of (presumably) planetesimals present in the system (e.g., Jewitt et al. 2009). This implies that, similarly to the evolution of debris disks in extrasolar systems, it is expected that the trans-Neptunian belt has evolved in a similar way. As such, collisional evolution of disk planetesimals has been likely active throughout the history of the solar system, producing an excess of dust and a gradual loss of the disk total mass via collisional grinding.

## 9.7 Primordial planetesimal disk size and the possible formation of TNOs beyond 48 AU

Taking into account the detailed orbital distributions/properties of distant resonant populations and detached bodies located within approximately 60 AU, the ancient trans-Neptunian belt appears to have had a radius of at least 50 AU. This disk size supports the formation of the observed distant resonant populations beyond 50 AU (Lykawka & Mukai 2007a). It also suggests that the planetesimal disk extended beyond the trans-Neptunian outer edge. Indeed, the two detached TNOs with low inclinations and moderate eccentricities 2003 UY291 ($a$ = 49.5 AU; $q$ = 41.3 AU) and (48639) 1995 $TL_8$ ($a$ = 52.9 AU; $q$ = 40.1 AU) support the hypothesis of a continuous primordial disk extending beyond 48 AU. If this is correct, the minimum size of the primordial planetesimal disk would be about 53 AU. From the point of view of the science of debris disks, this size would be approximately typical, since most of observed debris disks with similar ages as the solar system appear to be 40 AU or larger in radius (Jewitt et al. 2009, and references therein).

The possible growth of planetesimals beyond 48 AU may also provide an intriguing constraint. Recalling that the timescale for accretion growth is proportional to $R^3$, and that the largest cold classical TNO at 45 AU has a diameter of about 400 km (assuming an albedo $p$ = 0.1), the size of this object at 50 and 55 AU would be a factor of 0.73 (300 km) and 0.55 (200 km) for a fixed growth time span. This implies that TNOs formed in situ beyond 48 AU may have suffered a growth cutoff if the planetoid managed to perturb their orbits during the growth period, because their excited eccentricities and inclinations would result in non-accreting collisions (Section 9.6). As a result, these distant local TNOs might be intrinsically smaller, perhaps with maximum diameters not larger than 200 km. It is worth noting that the distribution of the largest classical TNOs shows an apparent preference for $a$ < 45 AU, which appears to support the existence of such an anomalous size distribution among some TNOs beyond 45 AU (Lykawka & Mukai 2005b). Detailed collisional models also suggest that external perturbers operated on the primordial planetesimal disk (Kenyon et al. 2008). Analysis of the classical population also seems to agree that it experienced long-term collisional evolution (Fuentes et al. 2010).

### 9.8 Nature of Neptune's migration

During planetary migration, Neptune's radial displacement is directly proportional to a function describing the total mass of planetesimals in planet-encountering orbits (Levison et al. 2007). Past studies also claimed that Neptune should migrate beyond 30 AU for massive disks with radius larger than 30 or 35 AU (Gomes et al. 2004).

Lykawka & Mukai (2008) investigated whether Neptune could stop at 30 AU after integrating self-consistently the orbits of the four giant planets with or without planetoids in 50-60 AU-sized cold planetesimal disks with disk total masses from 0.1 to 1 in units of the minimum mass solar nebula (MMSN; e.g., Hayashi et al. 1985). The disks were composed of 10,000-20,000 equal-mass bodies with an inner edge at 10-20 AU, and extended to 50-60 AU. It is worth noting that less massive disks (<1 MMSN) are an expected result of collisional grinding during the early stages of solar system history, according to recent collisional models (Kenyon et al. 2008). Irrespective of the presence of planetoids in the systems, the outcomes of these special simulations resulted in Neptune migrating to ~25-30 AU for almost all 0.9-0.5-MMSN disks after a few hundred million years. Therefore, provided that 50-60 AU-sized planetesimal disks acquired total masses <1 MMSN via collisional evolution, or dynamical depletion, before Neptune's migration, or that >1MMSN disks ran-out of feeding planetesimals, a massive planetesimal disk need not be truncated at 30-35 AU in order for Neptune to stop migrating near its current location. Therefore, the planetoid-resonance hybrid scenario is compatible with Neptune's current orbit.

In addition, the migration behavior of a migrating planet can also be significantly modified if it experiences close encounters, or giant collisions, with massive planetesimals (Ida et al. 2000b; Gomes et al. 2004; Murray-Clay & Chiang 2006; Chiang et al. 2007; Levison et al. 2007). Indeed, the obliquities of Neptune and Uranus strongly suggest that both planets experienced such giant impacts (Brunini et al. 2002; Lee et al. 2007). In fact, giant collisions of planetoids with Uranus and Neptune and mutual resonance crossings involving both planets can significantly change the migration behavior of the planets (Lykawka & Mukai 2008).

## 10 Discussions and implications

The planetoid-resonance sweeping model can robustly explain the main properties of all observed trans-Neptunian belt populations and satisfy many other constraints. Nevertheless, is a resident planetoid necessary to explain the observations? How is it possible to detect such a planetoid?

### 10.1 Hints from observations about the existence of a resident trans-Neptunian planetoid

Three pieces of evidence support the current existence of such a planetoid, as described below.

(1) *A prominent population of detached TNOs with peculiar orbits*. The observed detached population is expected to be quite large, with an intrinsic ratio to the scattered population of ≥1.0. In addition, detached TNOs possess a broad range of inclinations from a few degrees to almost 50 deg, and include several peculiar members, such as 2003 UY291, 2004 XR190, 2000 CR105, and Sedna. The sole gravitational and resonant perturbations by the four giant planets are highly unlikely to explain this population consistently (Section 4.4).

(2) *A large fraction of TNOs have inclinations higher than 40 deg*. The existence of a significant population of very high-$i$ TNOs cannot be explained by previously-published models (Section 4.5). In particular, gravitational scattering solely by Neptune is insufficient to explain this population. It is also possible that the high-$i$ components of currently known resonant populations, such as the Neptune Trojans and 3:2 resonants, reflect a primordial planetesimal disk that contained objects with an initial broad inclination distribution, including a fraction of objects with $i$ > 40 deg.

(3) *The lack of long-term resonant populations beyond the 8:3 resonance*. Although 3:1, 7:2, and 4:1 resonant objects could display stability over 4 Gyr in the trans-Neptunian region (Hahn & Malhotra 2005; Lykawka & Mukai 2007a), no evidence exists of such populations (Lykawka & Mukai 2007b).

To consistently explain all the aforementioned observational evidence, long-term perturbations by a massive undiscovered trans-Neptunian body seems highly plausible. Indeed, the planetoid would lift the perihelia and inclinations of a significant fraction of TNOs in the entire trans-Neptunian region and produce several analogs of peculiar TNOs. Finally, the absence of long-term resonant TNOs in the 3:1 resonance and other resonances beyond can be explained by the gravitational perturbation of a planetoid with $q_P$ = 80-90 AU on timescales comparable to the age of the solar system.

## 10.2 Prospects for the existence of a trans-Neptunian planet

According to the hybrid planetoid-resonance sweeping model, the possible capture of the planetoid in a distant resonance suggests $a_P$ ~100-175 AU, near the current locations of 6:1-14:1 resonances. The most successful model runs suggest $m_P$ = 0.3-0.7 M$_\oplus$, $q_P$ > 80 AU and $i_P$ = 20-40 deg. Concerning the physical properties of this massive object, it would be probably a differentiated body with a rocky interior and layers composed of ices, similar to the solar system's dwarf planets, thus suggesting mean densities of ~2-3 g cm$^{-3}$ (de Sanctis et al. 2001; Merk & Prialnik 2006; Stern 1992; Rabinowitz et al. 2006; Brown & Schaller 2007; Lacerda & Jewitt 2007). Given its appreciable mass, the planetoid would have large reservoirs of $CH_4$, $H_2O$, $N_2$ and other icy compounds on its surface (Brown et al. 2005; Barkume et al. 2006; Licandro et al. 2006a; Licandro et al. 2006b; Rabinowitz et al. 2006; Dumas et al. 2007; Tegler et al. 2007; Trujillo et al. 2007; Schaller & Brown 2007). Assuming a spherical shape and mean density of 2-3 g cm$^{-3}$, the planetoid's diameter would be $D_P$ = 10,000-16,000 km.

Although large TNOs tend to possess high albedos (Section 5.2), this observation is restricted to objects relatively close to the Sun (i.e., at $R$ < 100 AU). On the other hand, hypothetical more distant icy objects should be relatively dark (Stern 1991; Thompson et al. 1987; Moroz et al. 1998; Cooper et al. 2003; Brunetto et al. 2006). Sedna is the best-known representative of such distant bodies, and its albedo has been constrained to the range 0.1-0.3 (Emery et al. 2007). Therefore, considering the planetoid's suggested orbital properties ($a_P \geq$ 100 AU; $q_P$ > 80 AU), it is reasonable to assume albedo $p_P$ = 0.1-0.3 for the planetoid, using Eq. 24 with the derived ranges of $D_P$ and $p_P$ above, the planetoid would have $h_P$ ~15-17 mag for $q_P$ = 80-90 AU during a perihelion passage (Fig. 28).

Another important factor is the apparent rate of movement across the sky, which can be determined using Eq. 18. For typical trans-Neptunian belt orbits (30-60 AU), TNOs move at a rate of 2-5 arcsec h$^{-1}$, while the planetoid would show $\theta_P$ = 1.5-1.7 arcsec h$^{-1}$ during perihelion approach ($q_P$ = 80-90 AU). Nevertheless, having a large eccentricity, the planetoid would spend more time near aphelion during its orbit, thus implying apparent rates of around 0.5-1.0 arcsec h$^{-1}$ (Fig. 28). The majority of past dedicated surveys are sensitive, at most, to 1.5 arcsec h$^{-1}$ (Brown et al. 2004; Trujillo & Brown 2003; Brown et al. 2005). So, only a few recent wide-area surveys have probed apparent motion comparable to, or below, these critical values (Larsen et al. 2007; Schwamb et al. 2010; Sheppard et al. 2011).

Several past wide-area surveys have searched for bright bodies in the outer solar system ($h$ < 17 mag), but were unsuccessful in finding TNOs larger than currently known dwarf planets (Tombaugh 1961; Luu & Jewitt 1988; Kowal 1989; Jewitt & Luu 1995; Jewitt et al. 1998; Sheppard et al. 2000; Gladman et al. 2001; Trujillo et al. 2001a; Trujillo et al. 2001b; Trujillo & Brown 2003; Brown et al. 2004; Brown et al. 2005; Brown et al. 2006b; Jones et al. 2006; Larsen et al. 2007; Sheppard et al. 2011). Other wide-area surveys have not found any very bright TNOs within about 10 deg of the ecliptic (Sheppard et al. 2000; Trujillo & Brown 2003; Jones et al. 2006; Larsen et al. 2007). In addition, recalling that the planetoid would have an inclined orbit (20-40 deg), it would be more likely to be discovered at ecliptic latitudes, $\beta \sim i$, than in the ecliptic ($\beta$ = 0 deg), and the fraction of its orbit spent near the ecliptic would be only ~1.5-4% (Trujillo et al. 2001a). More recent wide-area surveys have been probing higher ecliptic latitudes and slow-moving objects, but planetoids or other peculiar distant objects (e.g., Sedna-like objects) have not been found yet (Schwamb et al. 2010; Sheppard et al. 2011). Thus, in general, the discovery of objects with higher inclinations has been less likely in the great majority of such wide-area surveys.

In conclusion, any hypothetical massive and distant trans-Neptunian planet, such as the planetoid described in Section 9, probably escaped detection because it is currently either moving with sky motion below survey sensibility or at an ecliptic latitude far from the ecliptic plane. Independent of the planet's properties, the probability of non-detection should be ~10-15% because surveys avoid the region near the galactic plane (Chiang & Jordan 2002; Trujillo & Brown 2003). Other potential observational biases are discussed elsewhere (Bernstein & Khushalani 2000; Horner & Evans 2002).

Nevertheless, ongoing and future survey programmes (e.g., Pan-STARRS, LSST, WISE) are much less afflicted with the limitations discussed above, so the existence of planets in the trans-Neptunian region will likely be resolved during the next decade (Jewitt 2003; Milani & Trilling 2009; McMillan et al. 2011).

## 11 Conclusions

In order to explain the complex orbital structure of the trans-Neptunian region and satisfy the aforementioned constraints and others detailed in Section 4, I have performed extensive simulations to investigate the origin and evolution of that region with the presence of massive planetoids (0.01-1.0 $M_\oplus$) in distant and varied orbits.

First, a trans-Neptunian planet on a fixed or migrating orbit cannot reproduce the detailed architecture of the trans-Neptunian belt; in particular, the excited orbital structure of classical TNOs, the outer edge of the belt, and the formation of stable distant resonant populations at $a$ > 50 AU. Currently known 3:2, 2:1 and 5:2 resonant TNOs also set a lower limit of $q_P$ > 80 AU for any hypothetical resident distant planetoid. These constraints rule out all models based on the existence of trans-Neptunian planets published before Lykawka & Mukai (2008) (see Section 8.1).

The proposed planetoid-resonance hybrid model could solve these issues, and is based on three main stages: the excitation of the trans-Neptunian belt before planetary migration (over tens of Myr timescales), planetary migration (within 100 Myr), and long-term sculpting (over the last 4 Gyr). I assumed the planetoid formed in the realm of the icy giant planets, and was thus probably a member of a large population of bodies that formed at 10-20 AU during the late stages of planet formation, and that were later scattered by these planets (Stern 1991; Kenyon 2002; Goldreich et al. 2004a;

Goldreich et al. 2004b; Parisi & Del Valle 2011). The overall results showed that planetesimal disks excited by such scattered massive planetesimals were remarkably similar to current observations in the 40-50 AU region, and, at the same time, provided the necessary disk orbital conditions for the formation of distant resonant populations. In the end, the planetoid acquired $a_P$ = 100-175 AU, whilst the Kozai mechanism changed its orbital elements to $q_P$ > 80 AU and $i_P$ = 20-40 deg.

Finally, the complex orbital excitation at 40-50 AU and the abrupt truncation near 48 AU in the trans-Neptunian belt probably represent evidence of the planetoid's perturbation back to the first Myr after planet formation during the early solar system (~ 4.5 Gyr ago), whereas the prominent detached and very high-*i* populations resulted from the planetoid's perturbations over the age of the solar system. In summary, this model reproduces all the main aspects of currently known orbital characteristics of TNOs and offers observationally testable predictions.

In line with the model and results described in Sections 8 and 9, it is interesting to note that, in several other scenarios, the inclusion of additional planets during the early solar system seem to provide better results over standard models (Chambers 2007; Ford & Chiang 2007; Yeh & Chang 2009; Brasser & Morbidelli 2011; Nesvorny 2011; Gomes & Soares 2012). This suggests that a consistent model for the entire solar system should include not only the known terrestrial and giant planets, but also other massive planetary bodies that probably existed in large numbers at the end of planet formation.

### 11.1 Main achievements of the planetoid-resonance hybrid model

- Explains the shortage of TNOs in the region between Neptune and 39 AU.
- Explains the cold and hot populations in the classical region, including their orbital excitation and distinct physical properties.
- Reproduces the properties of resonant populations in the trans-Neptunian region, including eccentricities, inclinations, and libration amplitudes. This also includes Kozai members and their fundamental properties. Analogs of Pluto and Haumea were fully obtained.
- Reproduces the scattered population with their distribution of orbital elements, including analogs of Eris.
- Produces a prominent detached population, including low-*i* objects and analogs of several peculiar members of this population (e.g., 2004 $XR_{190}$, 2000 $CR_{105}$, and Sedna). The obtained ratio of the detached population to the scattered population is in excellent agreement with the intrinsic estimates.
- Produces a very high-*i* population in agreement with the intrinsic fraction estimates in the trans-Neptunian belt.
- Reproduces the trans-Neptunian belt outer edge at 48 AU without threatening the stable 3:2, 2:1, and 5:2 resonant populations.
- Reproduces the dearth of TNOs beyond 45 AU, including the very abrupt decrease as a function of heliocentric distance.
- Possibly explains the current small total mass of the trans-Neptunian belt by the simultaneous action of enhanced collisional grinding induced by the planetoid and dynamical depletion.
- Explains Neptune's current orbit at 30.1 AU.

## 12 Summary of results

I summarize below the main results obtained in several investigations conducted with collaborators over the last few years in an attempt to better understand the physics and dynamics of TNOs. These results can also provide new constraints and hints for future studies of the trans-Neptunian region.

- Large TNOs have albedos intrinsically higher than the canonical value of 0.04.
- All TNOs larger than approximately 800 km would be able to sustain thin atmospheres/icy frosts composed of $CH_4$, CO or $N_2$. This would be the main reason for a systematic increase of albedos for larger TNOs.
- TNOs brighter than ~21 mag would have larger sky densities than the other TNOs. Following this trend, trans-Neptunian dwarf planets would appear two orders of magnitude more common in the sky.
- Small bodies in the outer solar system can be classified in five main known classes: classical, resonant, scattered, detached TNOs and the Centaurs.
- Resonant TNOs were found in the following resonances: 1:1 (Neptune Trojans), 5:4, 4:3, 11:8, 3:2, 18:11, 5:3, 12:7, 19:11, 7:4, 9:5, 11:6, 2:1, 9:4, 16:7, 7:3, 12:5, 5:2, 8:3, 3:1, 4:1, 11:2, and 27:4. In particular, the 3:2, 7:4, 2:1 and 5:2 appear to be the most populated resonances. Kozai TNOs were also found inside the 3:2, 5:3, 7:4 and 2:1 resonances.
- Scattered and detached TNOs (non-resonant) have $q < 37$ AU and $q > 40$ AU, respectively. TNOs with 37 AU $< q <$ 40 AU occupy an intermediate region where both classes coexist.
- Classical objects are non-resonant TNOs, usually divided into cold and hot populations. Their boundaries are as follows: cold classical TNOs ($i < 10$ deg) are located at 37 AU $< a <$ 45(50) AU ($q > 37$ AU), and hot classical TNOs occupy similar orbits with $i > 10$ deg.
- The orbital structure of the classical region ($a$ = 37-45(50) AU) strongly depends on inclination and the location of resonances. In particular, classical TNOs evolve differently in the inner (<45 AU) and outer regions (>45 AU) due to concentration of stronger resonances and dynamical effects associated with the proximity of Neptune in the inner region.
- There is an erosion of low-$i$ TNOs in the inner classical region (<45 AU), except for those objects in stable resonant orbits.
- The stability and number density of theoretical classical TNOs increases with simultaneous larger semimajor axis and higher inclinations. An optimal region was found at 45 AU $< a <$ 47 AU, $q > 40$ AU and $i > 5$ deg. Conversely, regions of instability and paucity of bodies were identified near the 5:3 and 2:1 resonances.
- TNOs in the 7:4 and other high order resonances located in the classical region usually show irregular eccentricity and inclination evolution over billions of years.
- TNOs in the 40-42 AU region are likely hot classical TNOs. Thus, they probably did not originate locally.
- The lack of low-$e$ classical TNOs beyond $a \sim 45$ AU and the trans-Neptunian outer edge at $a \sim$48 AU cannot be explained solely by the dynamical influence of the 2:1 resonance or the perturbation by the four giant planets. An additional perturbation is needed.
- The origin of detached and very high-i TNOs cannot be explained by a purely dynamical influence of the four giant planets, nor the resonant effects associated with them. An additional perturbation is needed.
- The evolution of scattered TNOs is determined by multiple temporary resonance trapping (resonance sticking) and scattering by Neptune.
- Each scattered TNO experiences tens, to hundreds, of resonance captures over 4 Gyr, representing about 36% of the object's lifetime (median value).
- Timescales of temporary resonance captures are proportional to resonance strength. In particular, resonances described by the ratio $r$:$s$ with the lowest argument $s$ are the strongest in the scattered disk.
- Resonance sticking is important mostly at $a < 250$ AU. This region concentrates all sufficiently strong resonances.
- Trapping in $r$:1 or $r$:2 resonances coupled with the Kozai mechanism in the current solar system can lead to the increase of perihelion (40 AU $< q <$ 60 AU), thus contributing to the population of detached TNOs. In addition, inclinations reach up to ~50 deg.

- Some scattered disk resonant TNOs have been residing in the 9:4, 5:2, and 8:3 resonances over the age of the solar system.
- The origin of Gyr-resident 9:4, 5:2, and 8:3 resonant TNOs is better explained by resonance sweeping over a pre-excited primordial planetesimal disk (ancient trans-Neptunian belt) of at least 45-50 AU radius with 0.1-0.3 or greater eccentricities, and a range of inclinations up to ~20 deg.
- The trans-Neptunian belt must have suffered a dynamical perturbation during the early stages of the solar system's existence, before planetary migration, to satisfy the point above.
- A model to explain the origin and evolution of the entire trans-Neptunian belt is detailed in Sections 8 and 9 of this paper. The orbital history of a massive planetoid can explain all main characteristics of the belt architecture with unprecedented detail and offers insightful, observationally testable predictions. *This is currently the most comprehensive scenario for explaining the orbital structure beyond Neptune.*
- The Neptunian Trojans were likely captured from the primordial planetesimal disk during a slow and extended migration of Neptune.
- The four giant planets were able to capture and retain a significant population of Trojans from the primordial planetesimal disk after planetary migration. The capture efficiencies of Trojans on tadpole orbits were ~$10^{-6}$-$10^{-5}$ (Jupiter and Saturn), ~$10^{-5}$-$10^{-4}$ (Uranus) and ~$10^{-4}$-$10^{-3}$ (Neptune).
- The bulk of captured Trojans acquired unstable orbits. When evolved for 4 Gyr, the survival fractions for the captured Trojan populations of Jupiter and Neptune Trojans were approximately 25% and 1-5%.
- Taken together, the lost Trojans of Jupiter and Saturn probably contained 3-10 times the current mass of observed Jovian Trojans, while the lost Trojan populations of Uranus and Neptune amounted to tens, or even hundreds, of times that mass. This implies that the Trojan populations have been providing an important source of Centaurs and SPCs throughout the entire history of the solar system.
- The intrinsic Haumea family probably occupies wide ranges in semimajor axes and eccentricities, and its members populate the four main TNO classes. When evolved over 4 Gyr, the non-diffusing character of theoretical family fragments implies that the observed properties can be used to draw conclusions on the nature of the creation of the family, billions of years ago.

## Acknowledgements

P. S. Lykawka is grateful to Prof. Tadashi Mukai for fruitful discussions and support over the last eleven years. Many thanks also to both the referees, who made several specific comments that improved significantly the presentation of this paper.

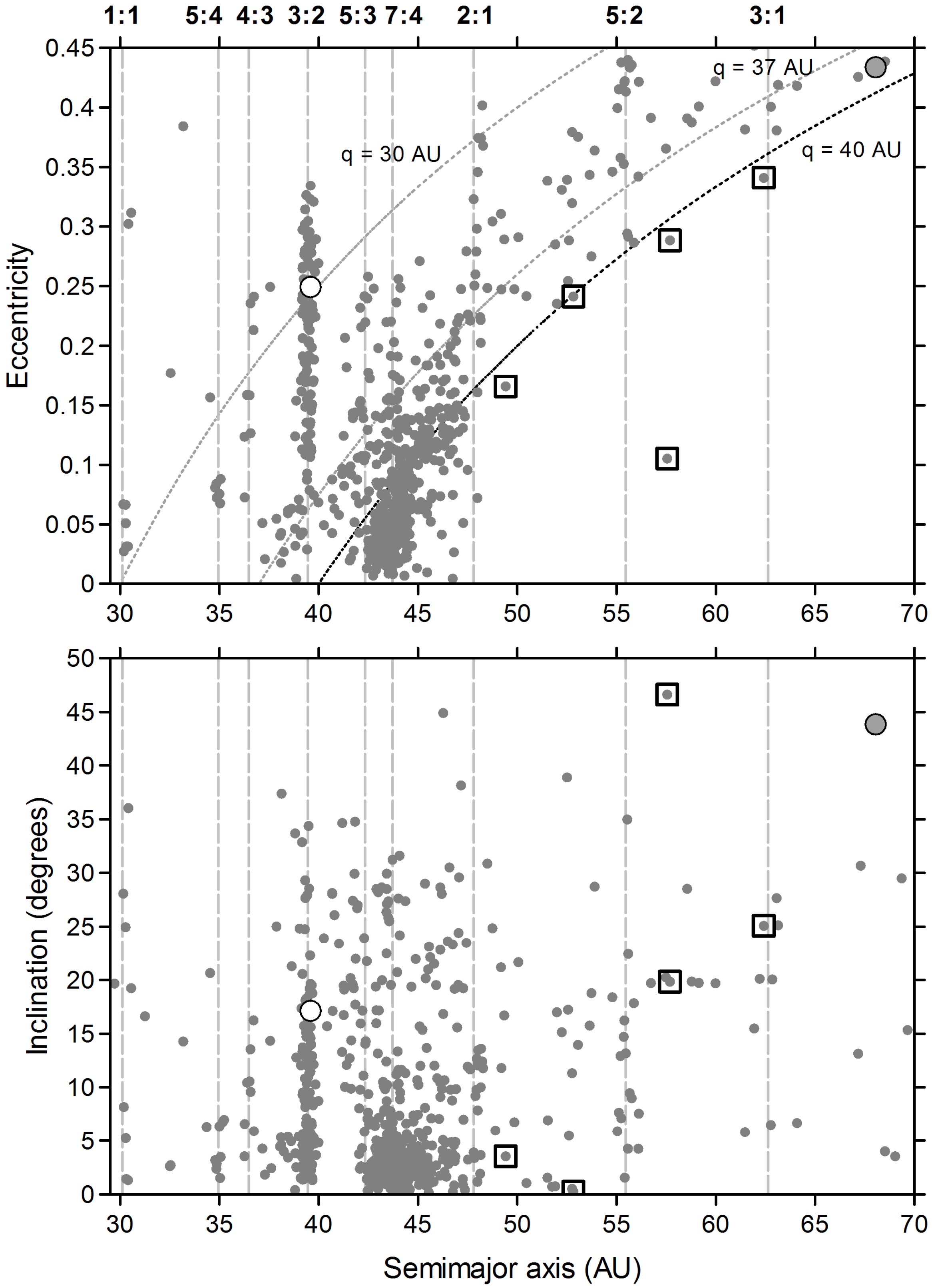

Figure 1: The orbits of 759 TNOs (gray circles) taken from the Asteroids Dynamic Site, AstDyS, in October 2010. For clarity, only those objects with orbital uncertainties of ($a_{uncertainty}$ / $a$) < 1% (1σ) are shown. Typical detached TNOs are denoted by squares. Perihelion distances of 30, 37 and 40 AU are illustrated by dotted lines (upper panel). The locations of Neptunian mean motion resonances are indicated by vertical dashed lines. Pluto and Eris are shown as white and gray large circles, respectively.

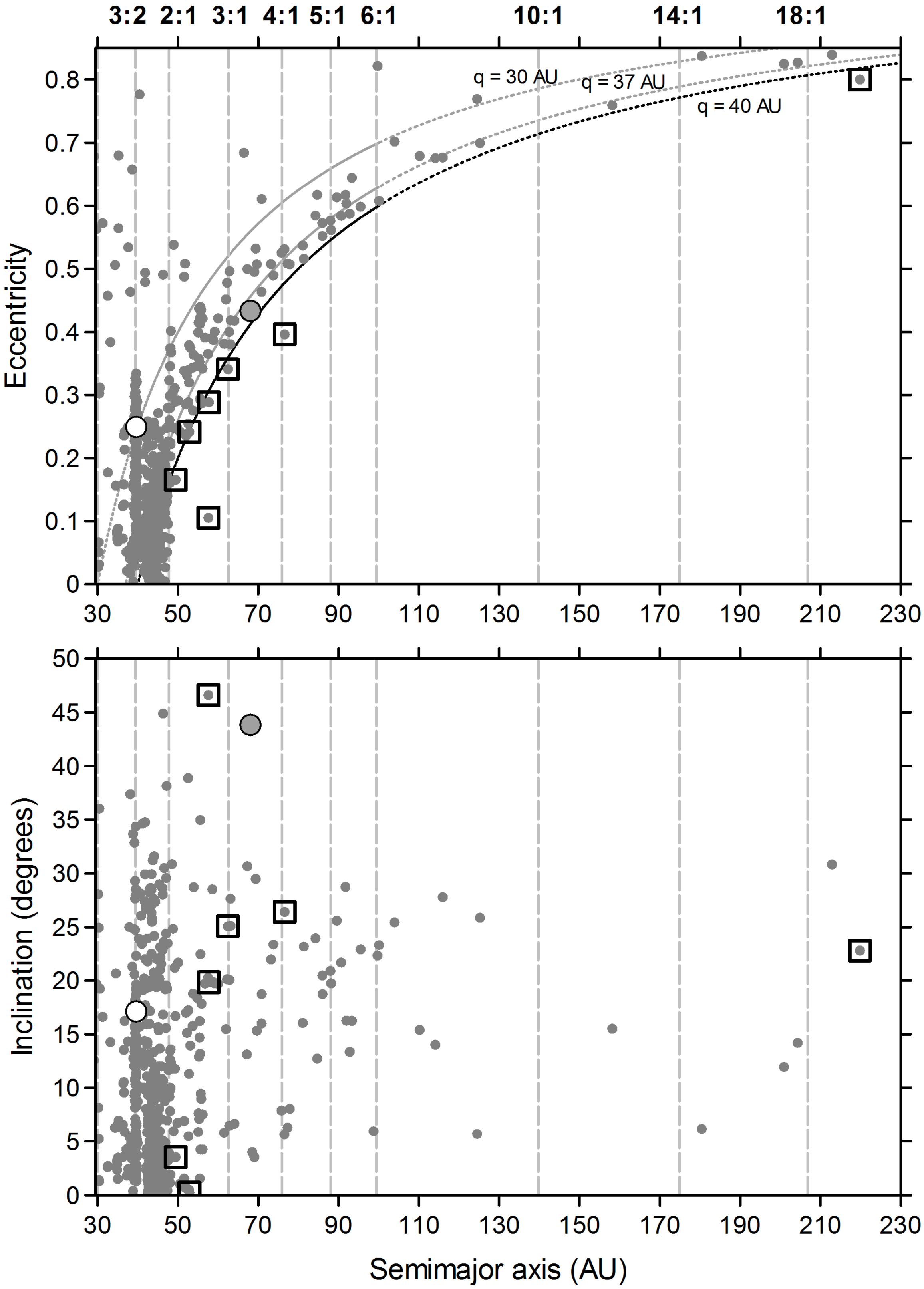

Figure 2: The orbits of 759 TNOs (gray circles) taken from the Asteroids Dynamic Site, AstDyS, in October 2010. The meaning of the symbols and other details are the same as shown in the caption of Fig. 1. Here, we focus on the trans-Neptunian region extended to $a$ = 230 AU. The strongest Neptunian mean motion resonances beyond the classical region ($a$ > 45-50 AU) are those of the $r$:1 type. For clarity, only a few of them are shown. Sedna is out of the range of this figure ($a$ = 535.5 AU; $q$ = 76.4 AU; $i$ = 11.9 deg).

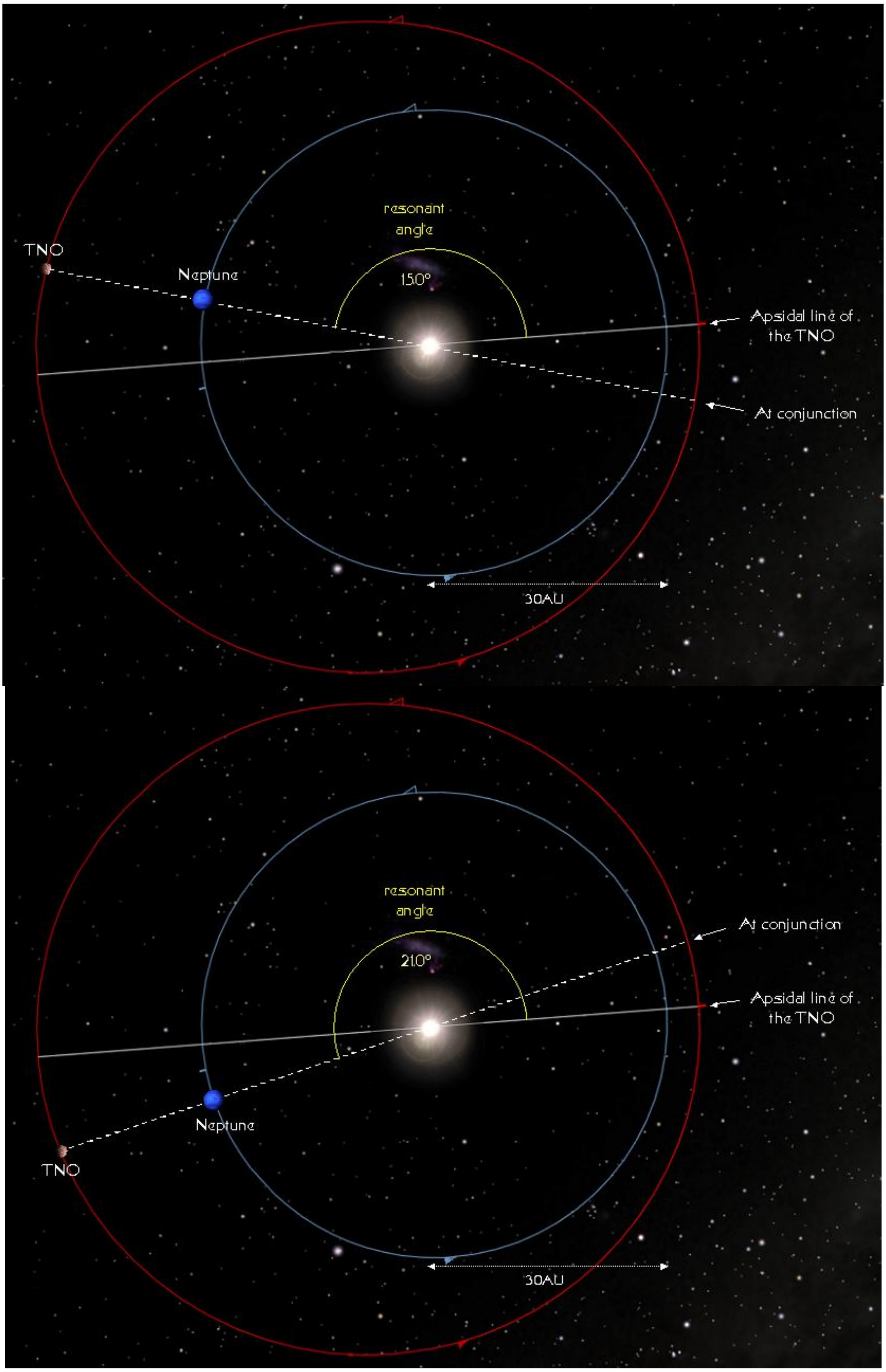

Figure 3: Geometry of conjunctions between a TNO in a mean motion resonance and Neptune. The resonant angle is defined as the angle formed between the perihelion line of the TNO and the longitude when at conjunction with Neptune. In this example, due to resonant motion, the object librates about 180 deg (the resonant angle oscillates periodically between 150 and 210 deg) with a full width amplitude of 60 deg.

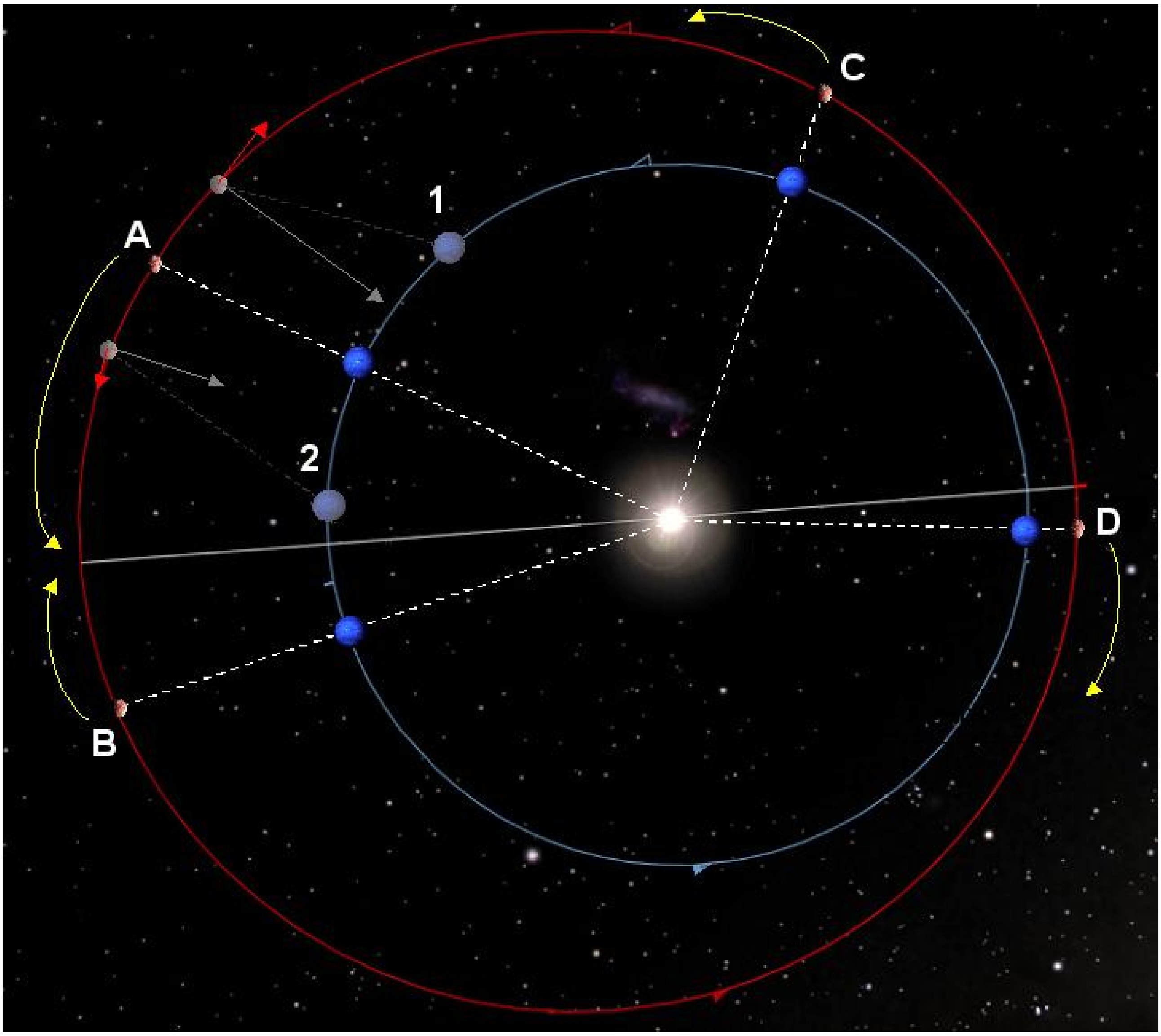

Figure 4: A simplistic illustration of the physics of a mean motion resonance, where a TNO is locked in a mean motion resonance with Neptune librating about 180 deg libration center. Points A, B, C and D represent distinct geometries of conjunctions for the TNO and Neptune. Numbers 1 and 2 indicate the moments before and after conjunction at A, respectively. Before the conjunction, there is a net loss of angular momentum and respective gain of angular velocity. Conversely, in the case of conjunction at point B the body experiences a net gain of angular momentum and loss of angular velocity. Thus, the next conjunctions will tend to occur near the aphelion.

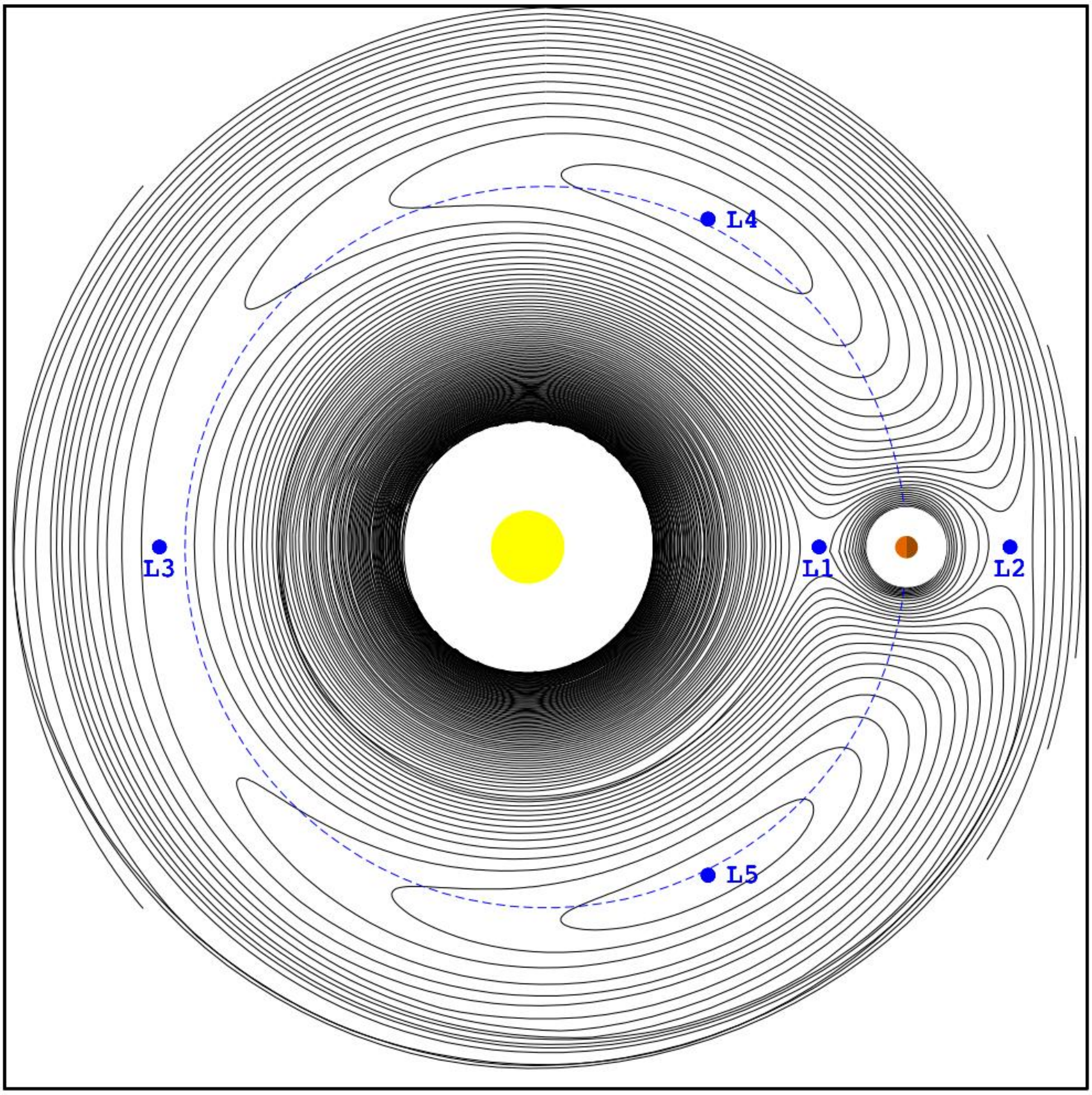

Figure 5: The locations of the five Lagrange points in the circular restricted three-body problem. The brownish planet represents Neptune, while the yellow body at the center is the Sun. The solid lines give areas of equal gravitational potential. The regions around the L1, L2 and L3 points can be considered regions of instability (or short-term stability), while the regions around the L4 and L5 points can provide long-term stability for objects moving on resonant orbits about these points. As can be seen in Table 1, the population of Neptunian Trojans occupies the L4 and L5 Lagrange regions. (Figure created by Jonathan Horner using a modified version of the gnuplot code detailed at http://commons.wikimedia.org/wiki/File:Lagrange_points.jpg)

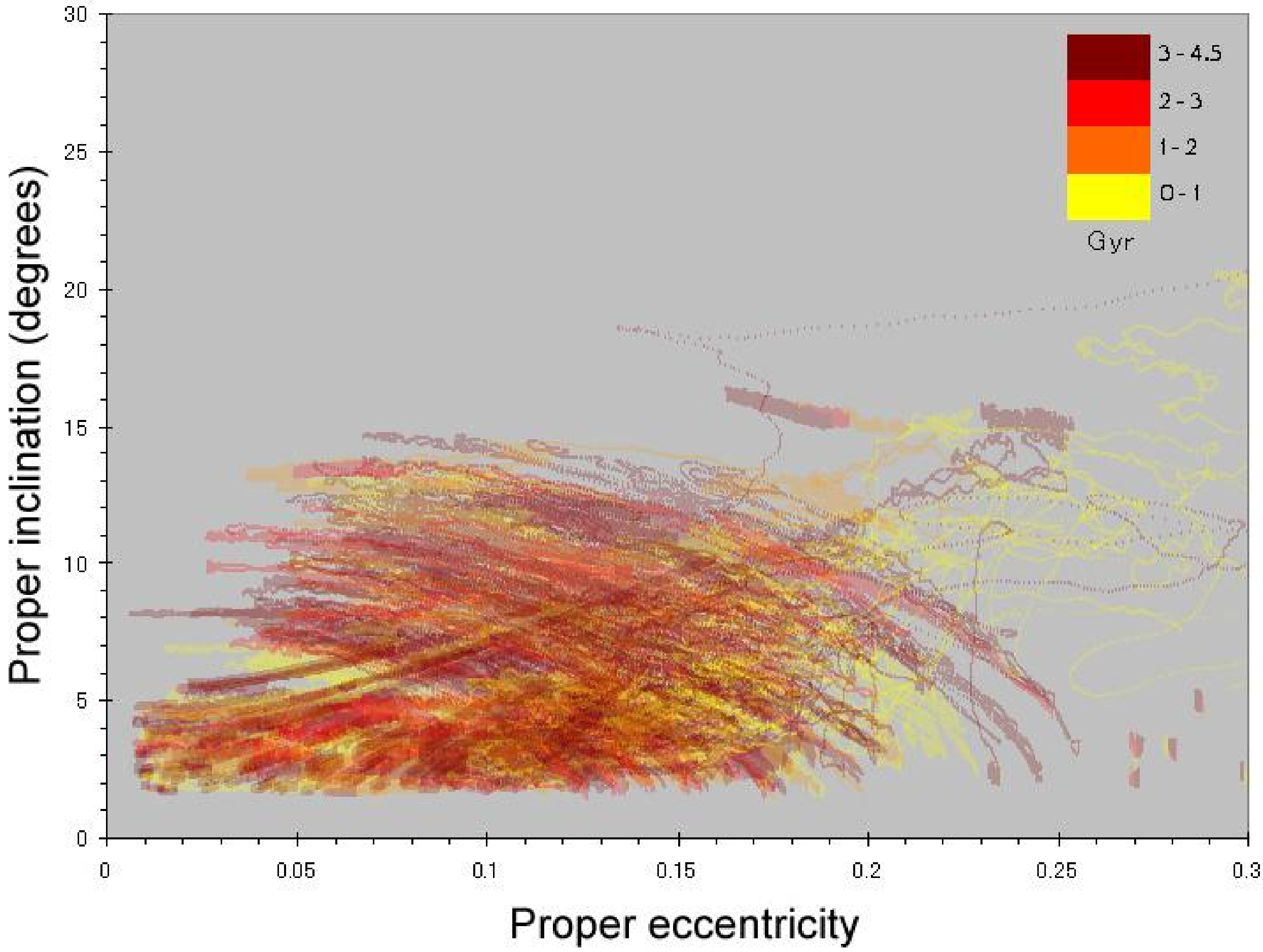

Figure 6: The mobility of 25 particles locked in the 7:4 mean motion resonance ($a$ = 43.7 AU) in proper elements over 4.5 Gyr of orbital evolution (based on Lykawka & Mukai 2005a). All particles had initial inclinations less than 5 degrees, and varied eccentricities at the start of the simulations (the initial $e$, $i$ of the 25 particles are indicated by the open circles shown in Fig. 11). The diffusion in element space of each survivor was computed by the evolution in proper elements every 0.1 Myr following Eq. 1 (Section 2).

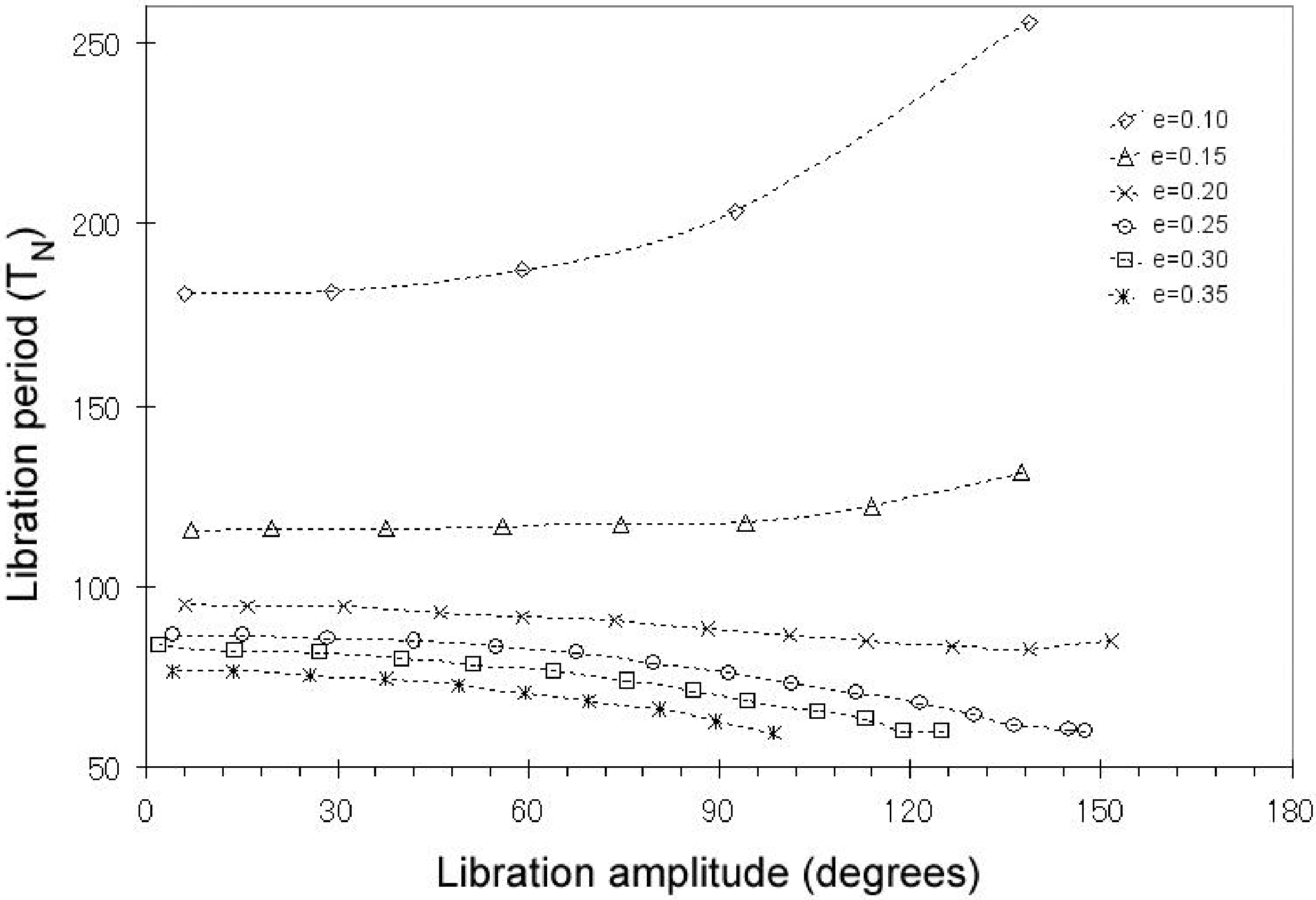

Figure 7: Libration amplitude and libration period for objects trapped in the 7:4 mean motion resonance ($a$ = 43.7 AU) as a function of eccentricity: 0.10, 0.15, 0.20, 0.25, 0.30 and 0.35. The libration periods are shown in units of Neptune orbital period (~165yr) and range within 10-100 kyr time scales (Figure from Lykawka & Mukai 2005a).

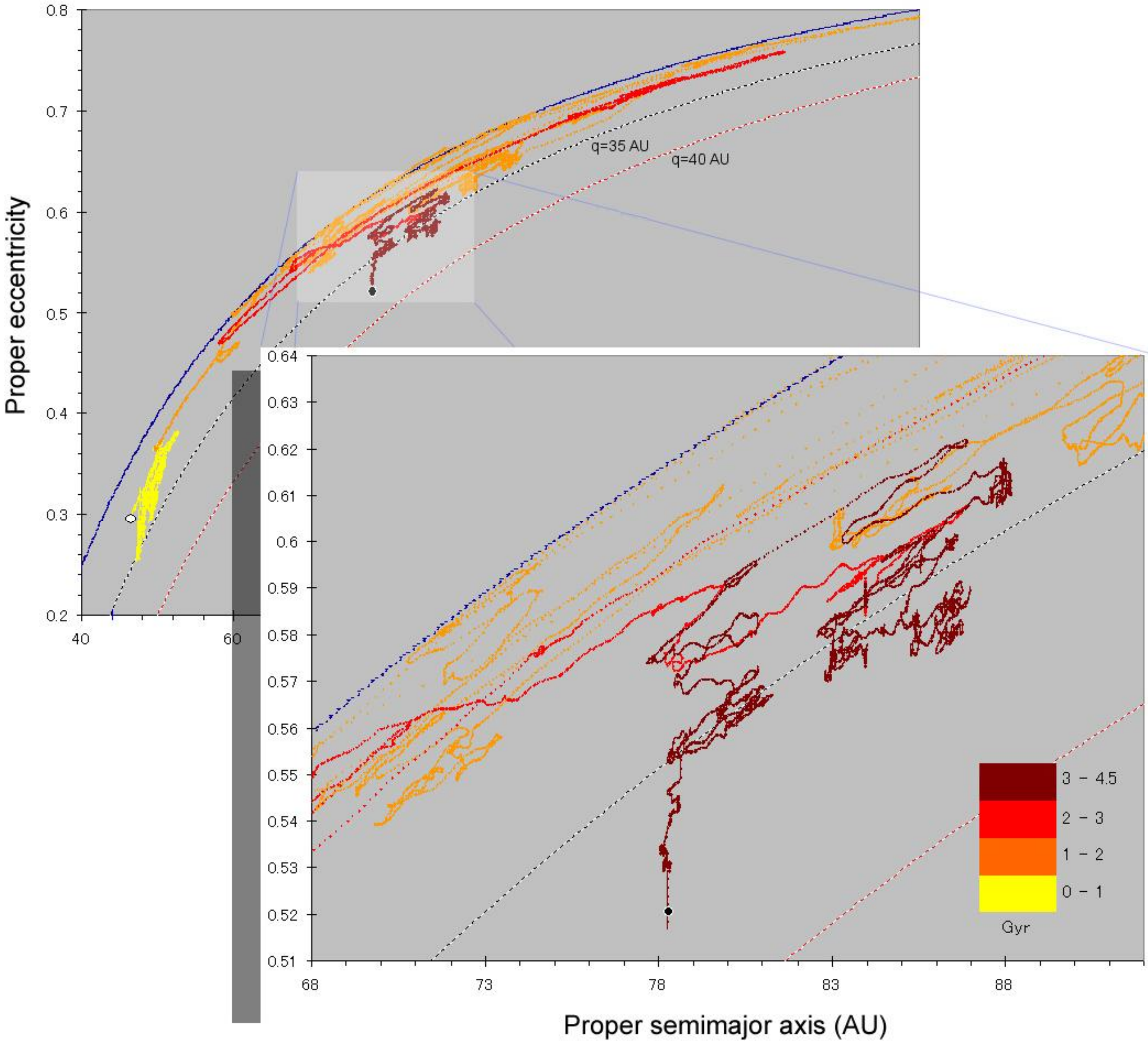

Figure 8: A typical example of an object experiencing gravitational scattering by Neptune (regions of large mobility in *a*-*e* element space) and multiple temporary captures in mean motion resonances with the same planet (the resonance sticking phenomenon; regions of small or negligible variation in semimajor axis) over 4.5 Gyr of orbital evolution (based on Lykawka & Mukai 2006). The perihelion distances of Neptune, 35 AU and 40 AU are shown by dashed curves. The diffusion in element space of each survivor was computed by the evolution in proper elements every 0.1 Myr following Eq. 1 (Section 2). Initial and final proper elements are shown by the big white and black circles, respectively. This single object ended the simulation locked in the 21:5 resonance ($a$ = 78.2 AU), after evolving for almost 1 Gyr in a narrow region at 78-88 AU.

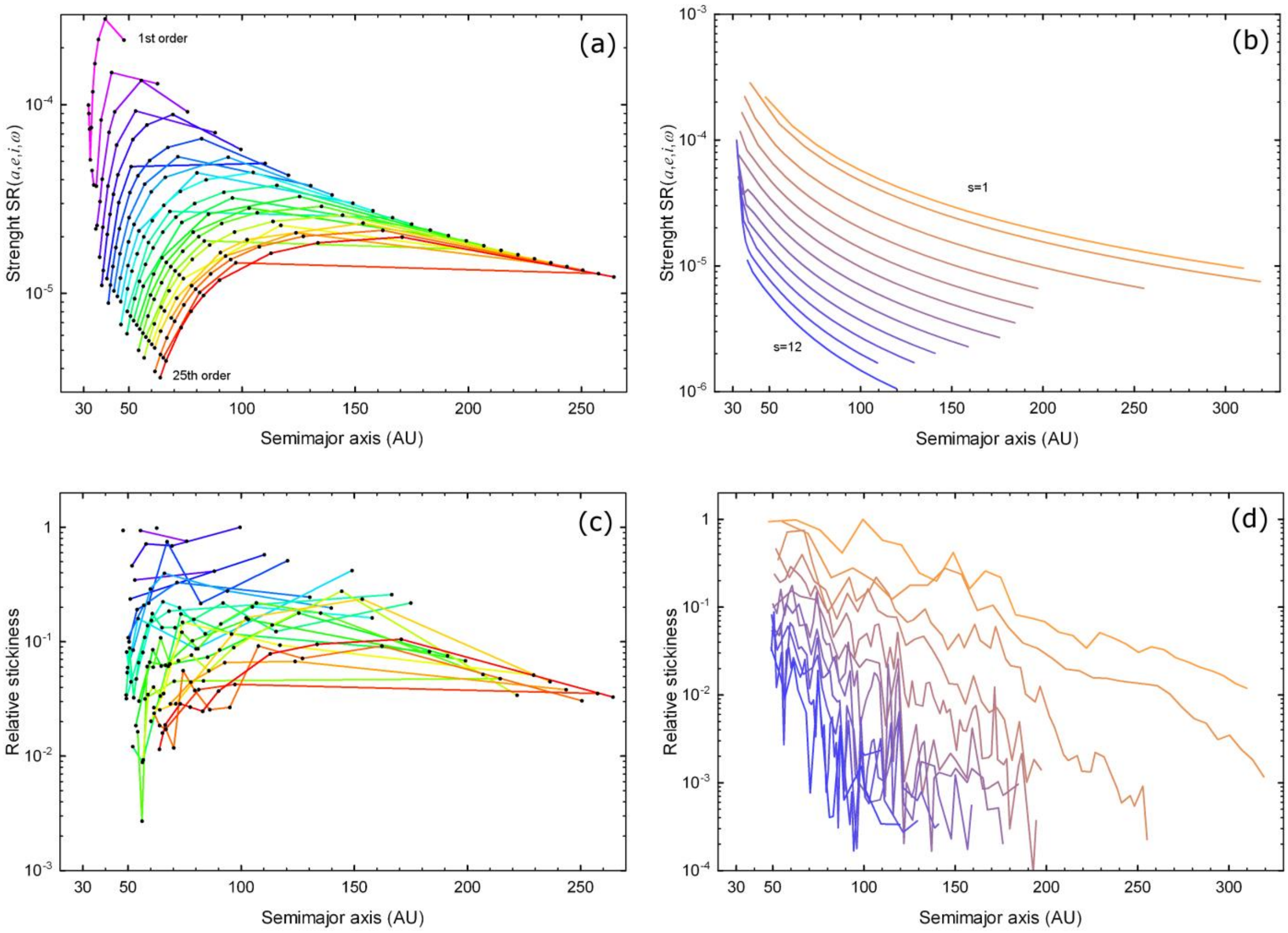

Figure 9: Resonance strength (panels **a** and **b**) and relative resonance stickiness (panels **c** and **d**) of relevant mean motion resonances (described as *r:s*) beyond 45 AU. Resonance stickiness was determined from the behavior of 255 particles that evolved over 4 Gyr, while resonance strength was calculated by solving the strength function (see Section 2.1.4 for details). Resonance stickiness illustrates the likelihood of capture into a resonance, and the ability of that resonance to retain a captured object (i.e., timescale). Resonance stickiness was normalized to the largest value, found at the 6:1 resonance ($a$ = 99.4 AU). Relative resonance stickiness and resonance strength are given as a function of resonance order (panels **a** and **c**) and argument *s* (panels **b** and **d**). Resonances indicated in this figure range from 1st to 25th order and argument *s* from 1 to 12, with values increasing from purple to red (left panels) and orange to blue (right panels) (Figure adapted from Lykawka & Mukai 2007c).

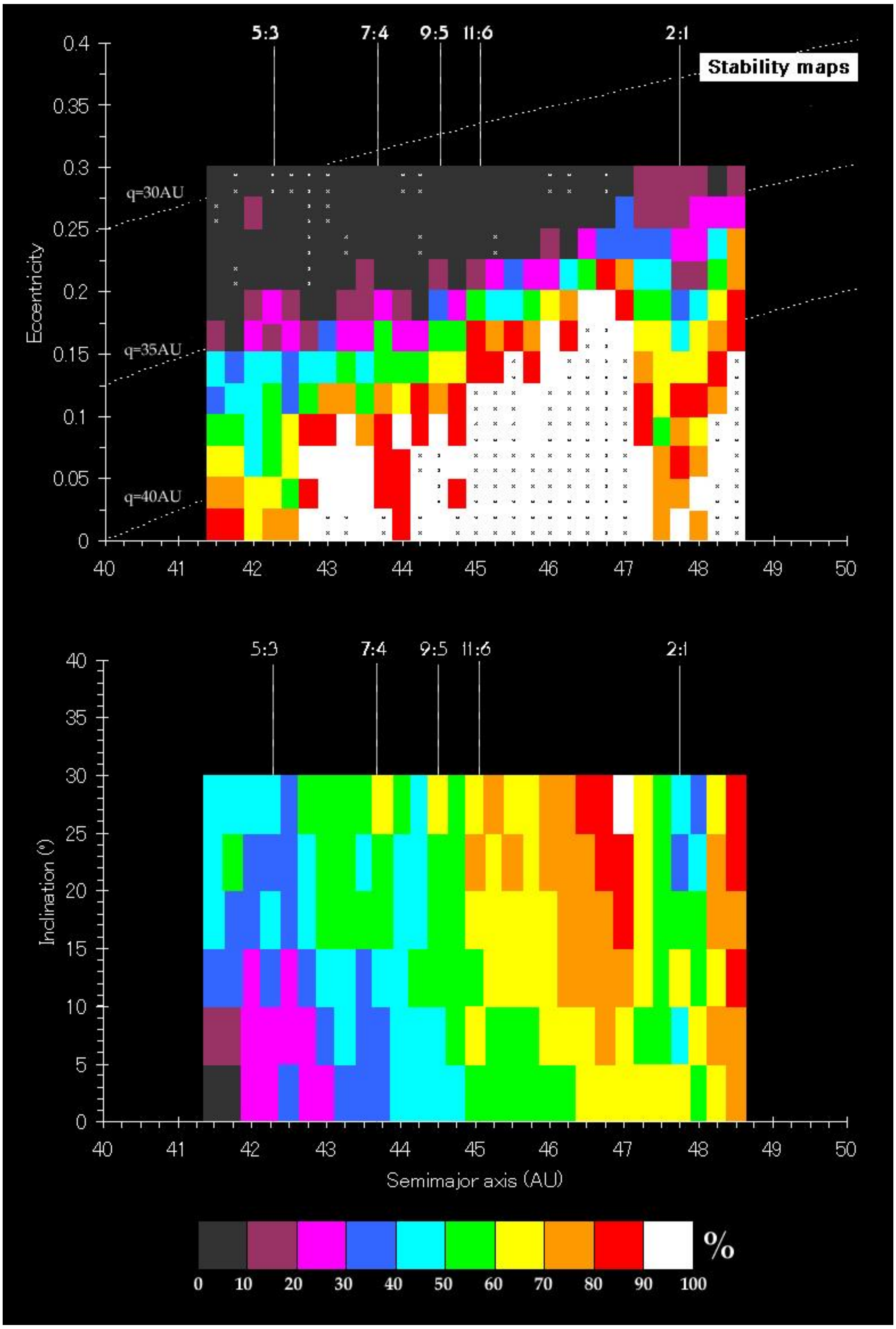

Figure 10: Stability maps for the classical region of the trans-Neptunian belt based simulations of classical bodies over 4 Gyr. Vertical lines indicate the locations of the main mean motion resonances with Neptune. In the upper panel, perihelion distances of 30 AU, 35 AU and 40 AU are indicated by dashed curves. The white regions with small marks indicate areas with 100% survival rates, while the marked grey regions indicate total elimination. For example, a region with 90% stability means that 10% of the particles that started in that region were ejected from the solar system during the simulation (Figure from Lykawka & Mukai 2005c).

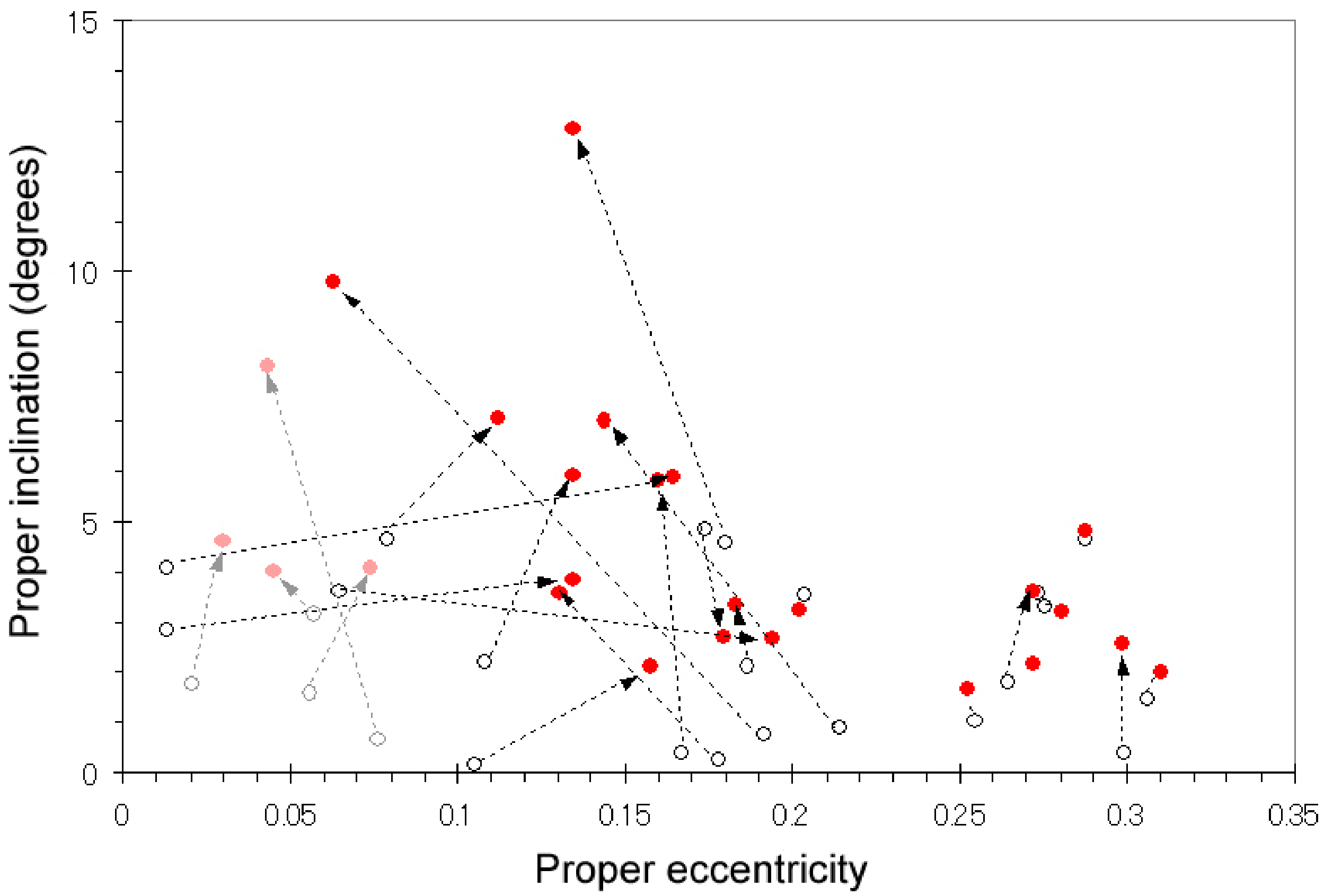

Figure 11: The initial and final positions of 25 bodies located in the 7:4 mean motion resonance ($a$ = 43.7 AU) after dynamically evolving over 4.5 Gyr. Open circles and red circles indicate the initial and final orbital states. Objects in light shaded colors displayed irregular libration behavior. Two groups are apparent. The first at $0.1 < e < 0.2$ and the other at $0.25 < e < 0.3$, represented by higher and lower mobility in element space, respectively (Figure from Lykawka & Mukai 2005a).

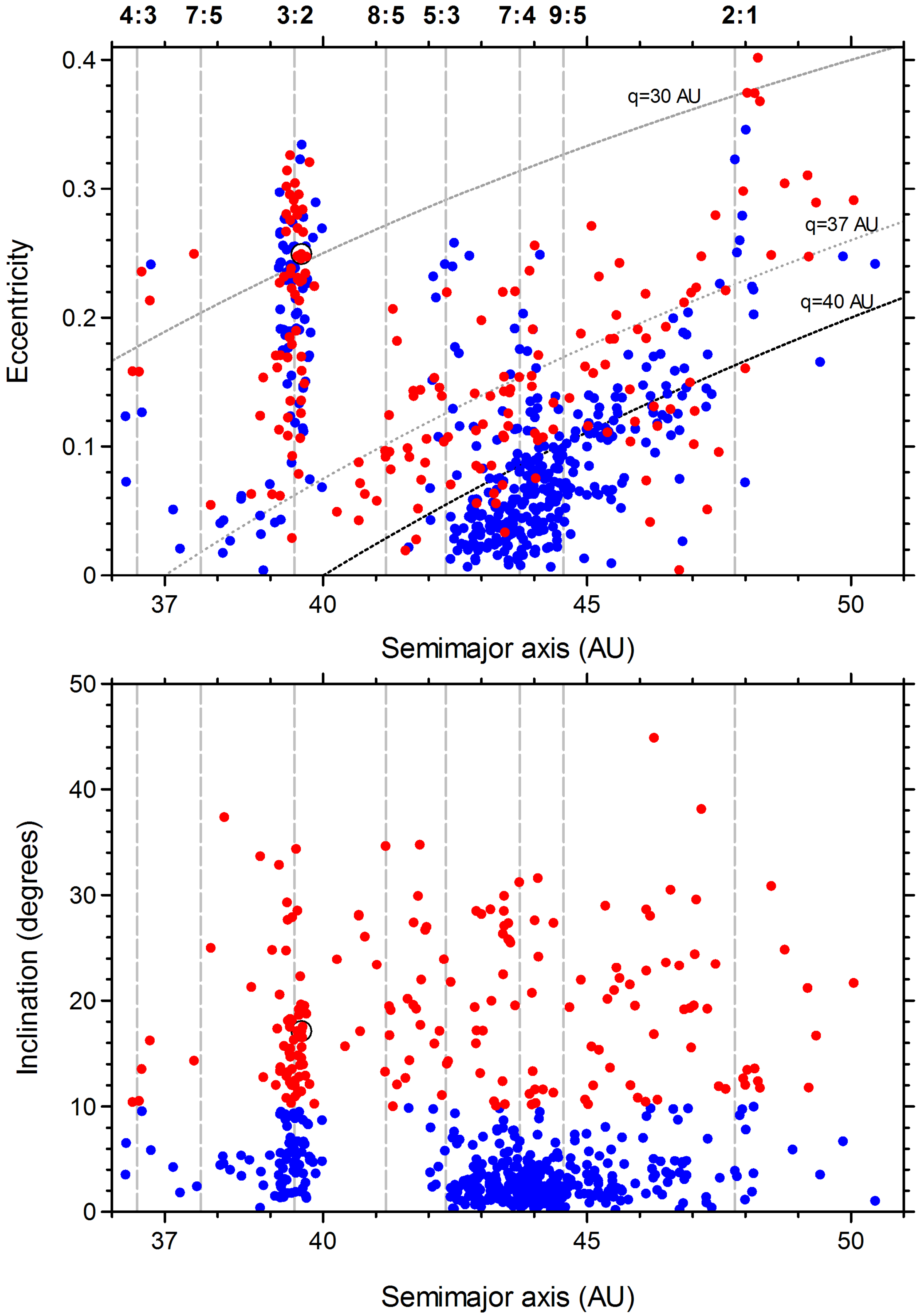

Figure 12: Orbital distribution of cold (*i* < 10 deg; blue circles) and hot classical TNOs (*i* > 10 deg; red circles) with elements taken from the Asteroids Dynamic Site, AstDyS, in October 2010. Perihelion distances of 30, 37 and 40 AU are illustrated by dotted lines in the upper panel. The locations of Neptunian mean motion resonances are indicated by vertical dashed lines.

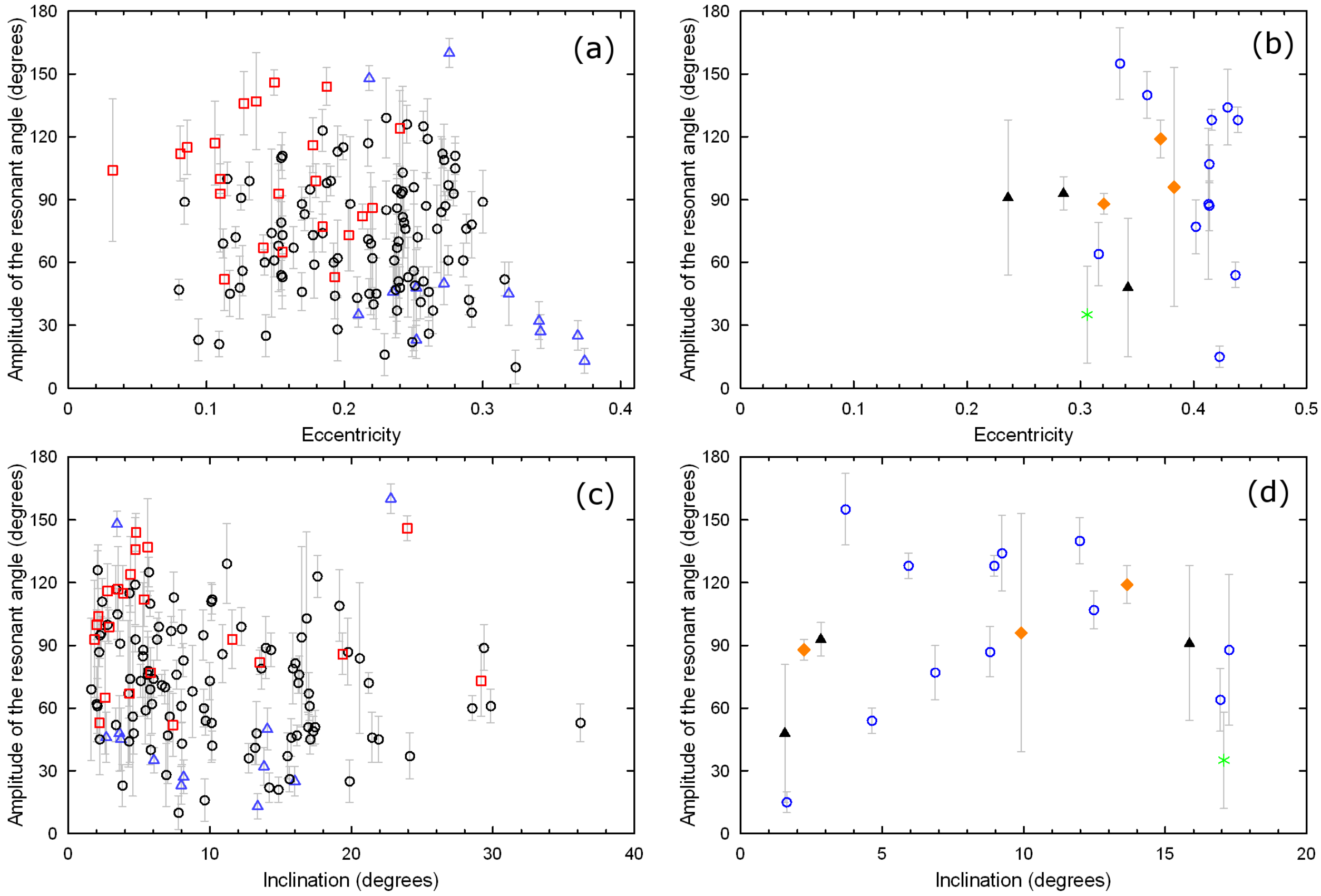

Figure 13: Relation between averaged eccentricity (panels **a** and **b**) and averaged inclination (panels **c** and **d**) with amplitude of the resonant angle, *A*, (libration amplitude) of TNOs for selected mean motion resonances in the trans-Neptunian belt (based on Lykawka & Mukai 2007b). All 2:1 resonant TNOs with *A* < 60 deg are asymmetric librators. In panels **a** and **c**, 3:2 (circles), 7:4 (squares), and 2:1 (triangles) resonant TNOs are shown, while panels **b** and **d** display the 9:4 (triangles), 7:3 (diamonds), 5:2 (circles), and 8:3 (stars) resonant bodies. The symbol used for each resonant population is given within the parenthesis.

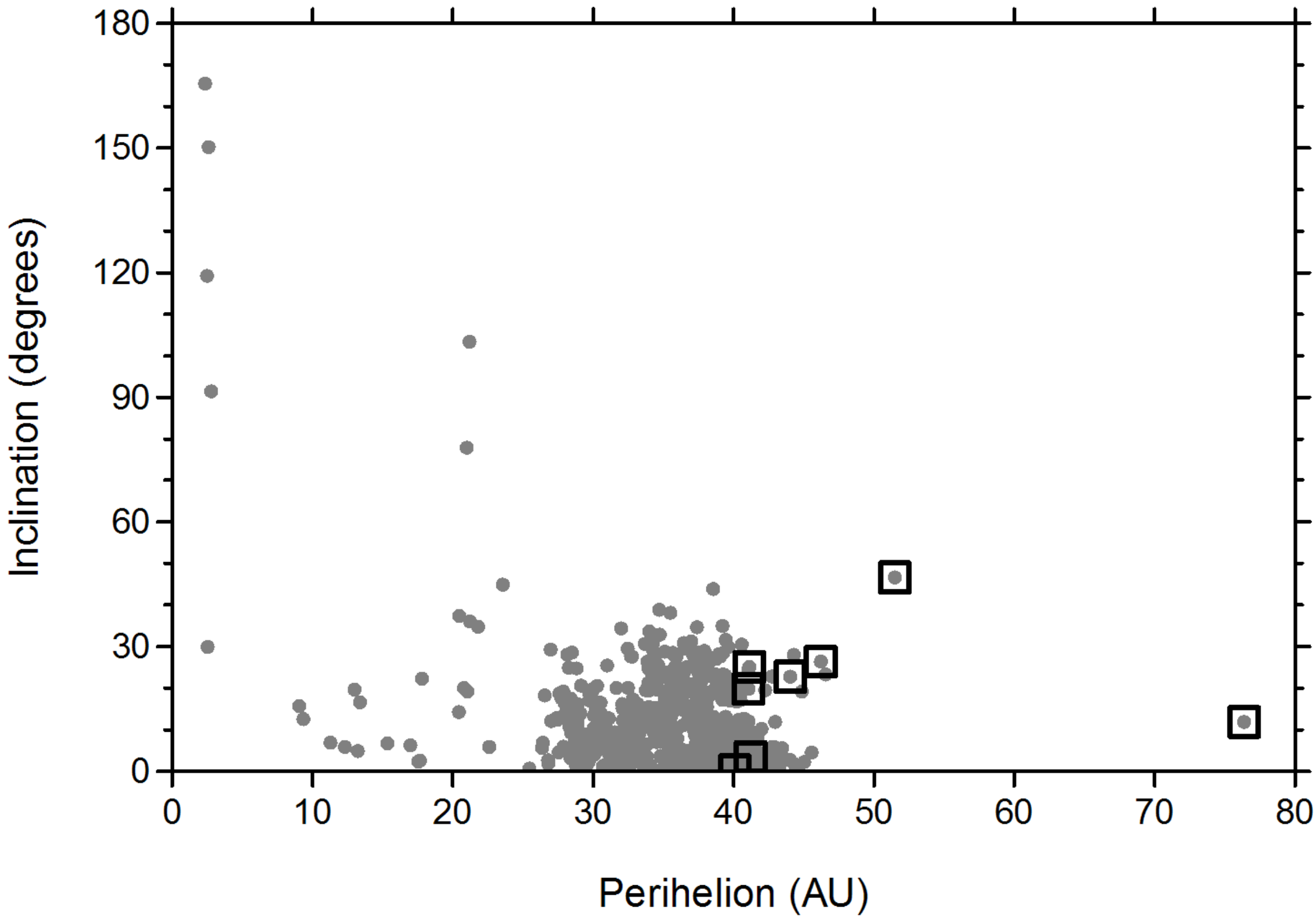

Figure 14: The orbits of 759 TNOs (gray circles) taken from the Asteroids Dynamic Site, AstDyS, in October 2010. The meaning of the symbols and other details are the same as shown in the caption of Fig. 1. Here, we emphasize the existence of extreme TNOs possessing either large perihelia (in particular, the detached TNOs; squares), or very high inclinations (>40 deg). TNOs with approximately $q$ < 25 AU are evolving on unstable orbits (the Centaurs).

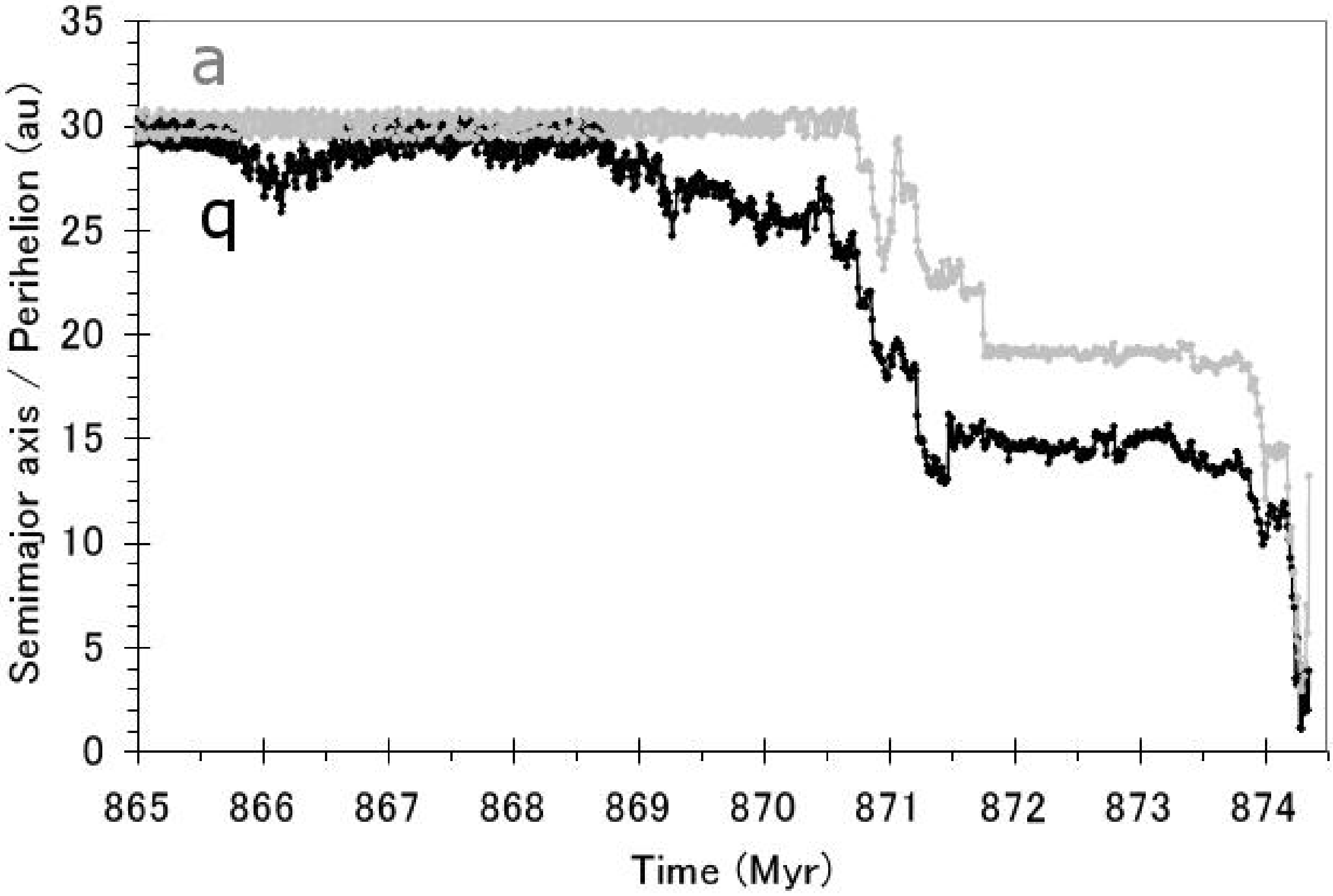

Figure 15: The dynamical evolution of a Neptune Trojan during the few million years around its escape from the Neptunian Trojan cloud at ~870.5 Myr for the Centaur population. The grey line shows the evolution of the object's semimajor axis, *a*, and the black line its perihelion distance, *q*. The object rapidly moves inwards, when a series of close encounters with Uranus, Saturn and Jupiter inject it to the inner solar system as a short-period comet, where it remains for the last hundred kyr of its life before being ejected from the solar system by a close encounter with Jupiter.

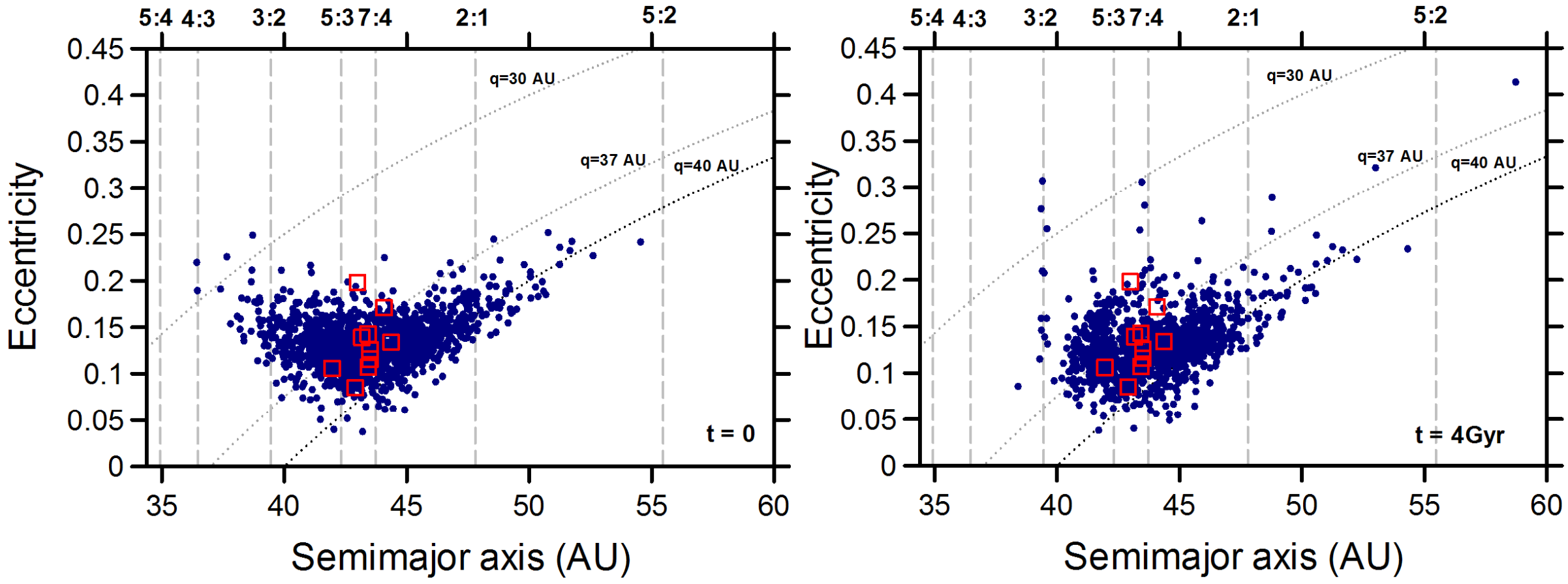

Figure 16: Orbital distribution of a representative Haumea family after 4 Gyr of orbital evolution. Currently known members of the Haumean family are shown by red squares. Perihelion distances of 30, 37 and 40 AU are illustrated by dotted lines, while relevant Neptunian mean motion resonances are indicated by vertical dashed lines (Figure adapted from Lykawka et al. 2012).

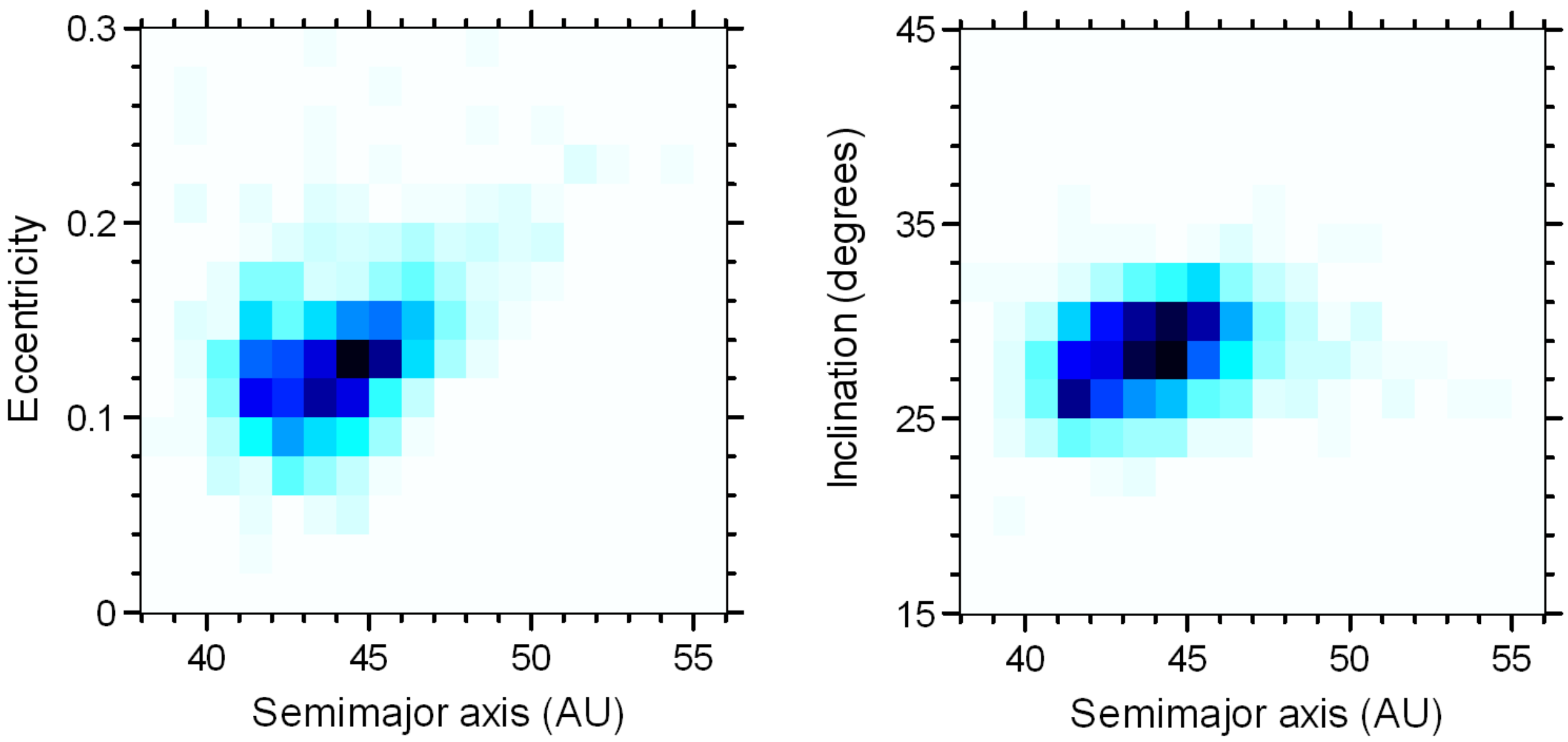

Figure 17: Number densities of representative Haumea family members in *a*-*e* and *a*-*i* element space after 4 Gyr of dynamical evolution. Regions containing different concentrations of objects are indicated by distinct blue scale shaded regions. The densest region was normalized by the highest number of objects in a single region for each panel. The darkest and lightest shaded regions represent roughly an order of magnitude difference (Figure adapted from Lykawka et al. 2012).

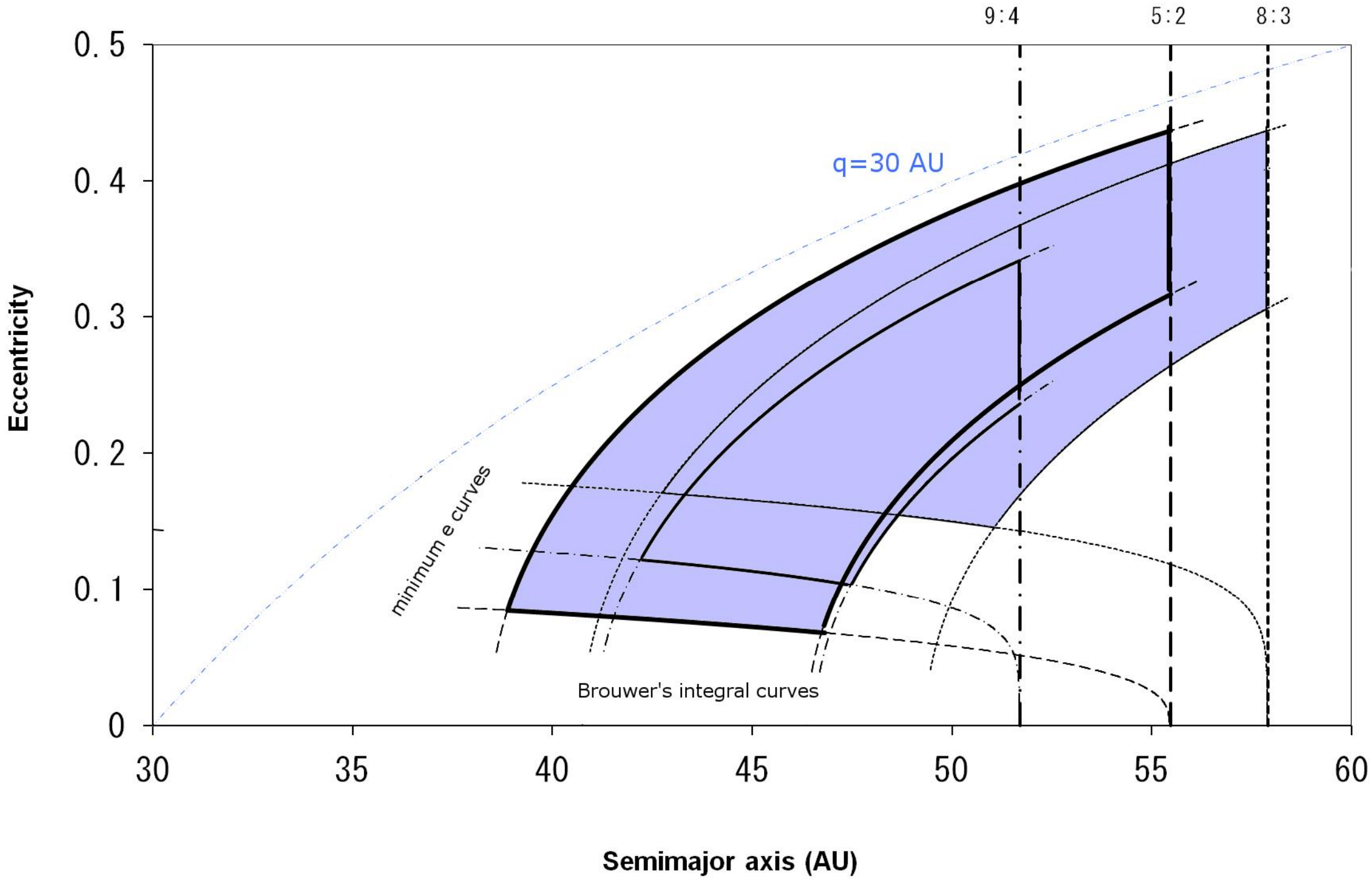

Figure 18: Initial orbital elements needed for the capture of currently observed 9:4, 5:2, and 8:3 resonant TNOs according to the resonance sweeping model of Lykawka & Mukai (2007a) (represented by bluish regions) (see also Section 7.2.1). The perihelion of 30 AU is shown by the two dotted-dashed blue curve. Lower curves set the minimum eccentricities for capture in mean motion resonance (Eq. 17). For each resonance, two curves represent Brouwer's integral conservation for minimum and maximum eccentricities constrained by 9:4, 5:2, and 8:3 resonant TNOs (Eq. 16).

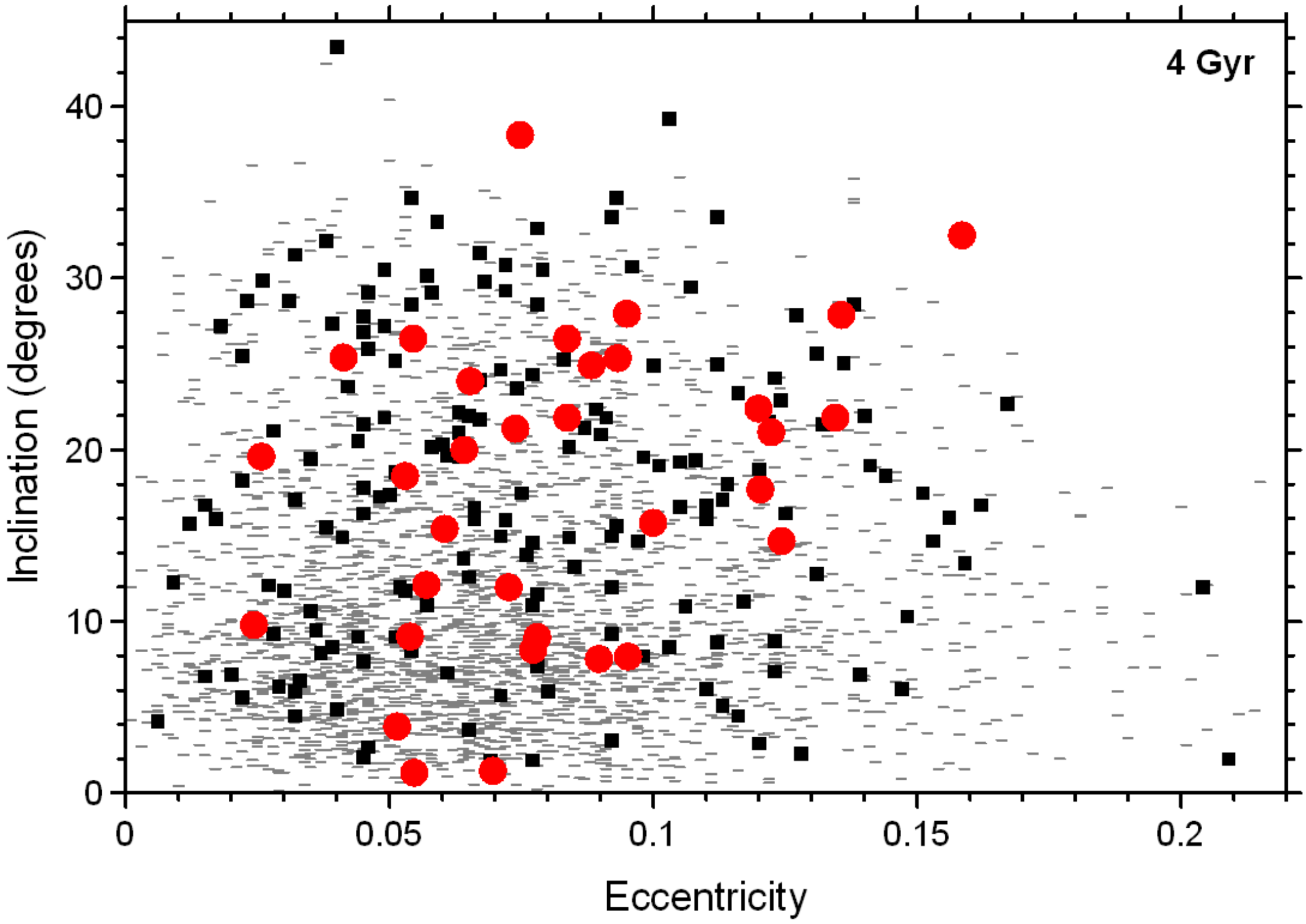

Figure 19: The orbital distribution of objects captured as Jovian Trojans during planetary migration and after evolving them over 4 Gyr (red circles). The objects plotted were all found to have been moving on tadpole orbits around the L4 and L5 Jovian Lagrange points. Currently known Trojans with more accurate orbits (i.e., those with two or more opposition observations) are shown for comparison, taken from the IAU Minor Planet Center* in January 2010. Large Trojans with absolute magnitudes, $H$, less than 10.5 are represented by squares, while small Trojans ($H > 10.5$) are shown as minus signs (Figure adapted from Lykawka & Horner 2010).

* http://www.cfa.harvard.edu/iau/lists/JupiterTrojans.html

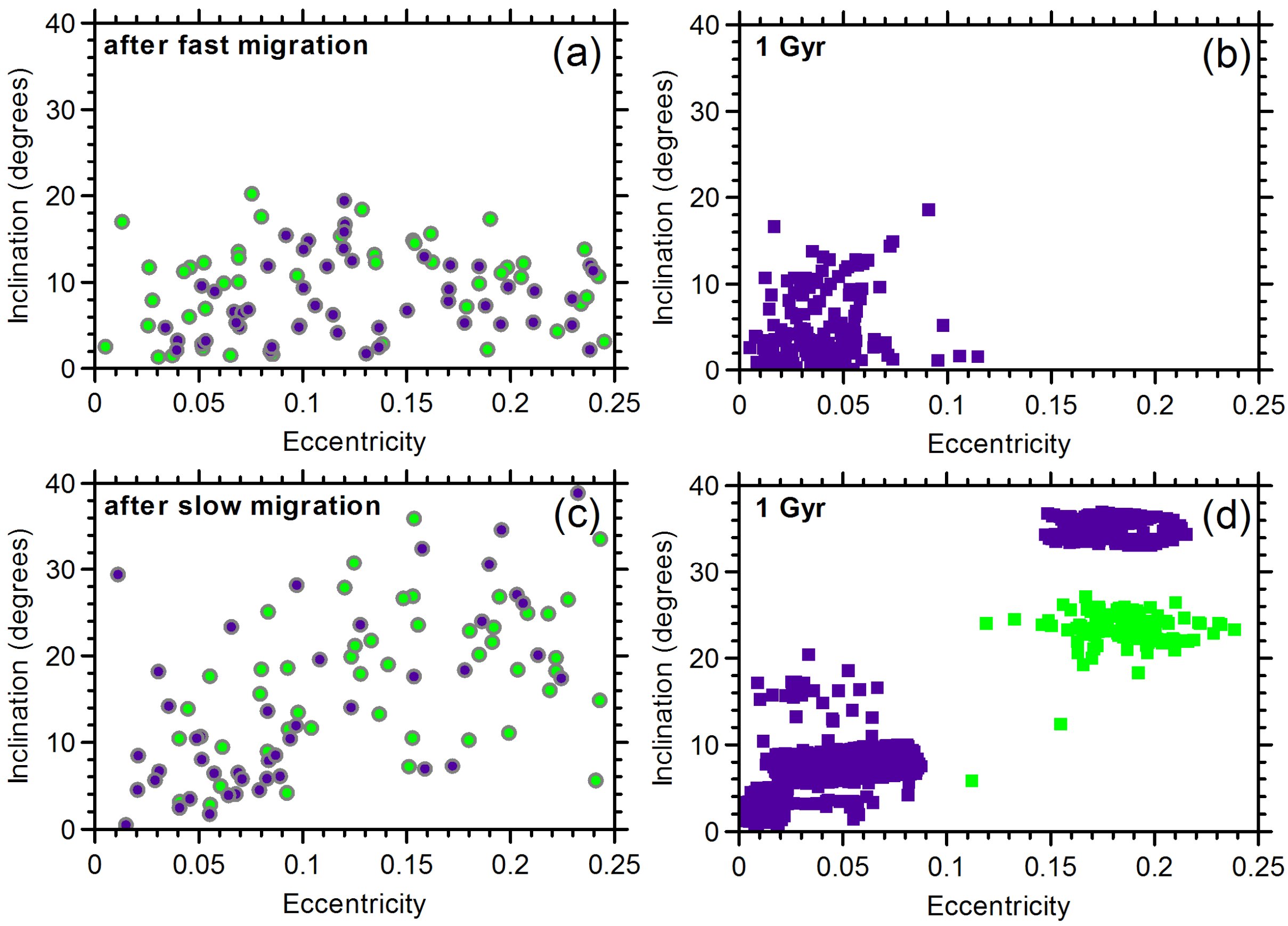

Figure 20: The distribution of Neptunian Trojans obtained at the end of planetary migration, in which Neptune slowly migrated outward from 18.1 AU to its current location at 30.1 AU. The left panels show the distribution of objects for both fast (total time = 5 Myr; panel **a**) and slow (total time = 50 Myr; panel **b**) migrations, while the distribution of those objects after 1 Gyr of dynamical evolution under the gravitational influence of the four giant planets can be seen in panels **c** and **d**, respectively. The points in violet represent Trojan objects moving about the L4 and L5 Lagrange points, while those in green denote objects moving on horseshoe and similar orbits (Figure adapted from Lykawka et al. 2011)

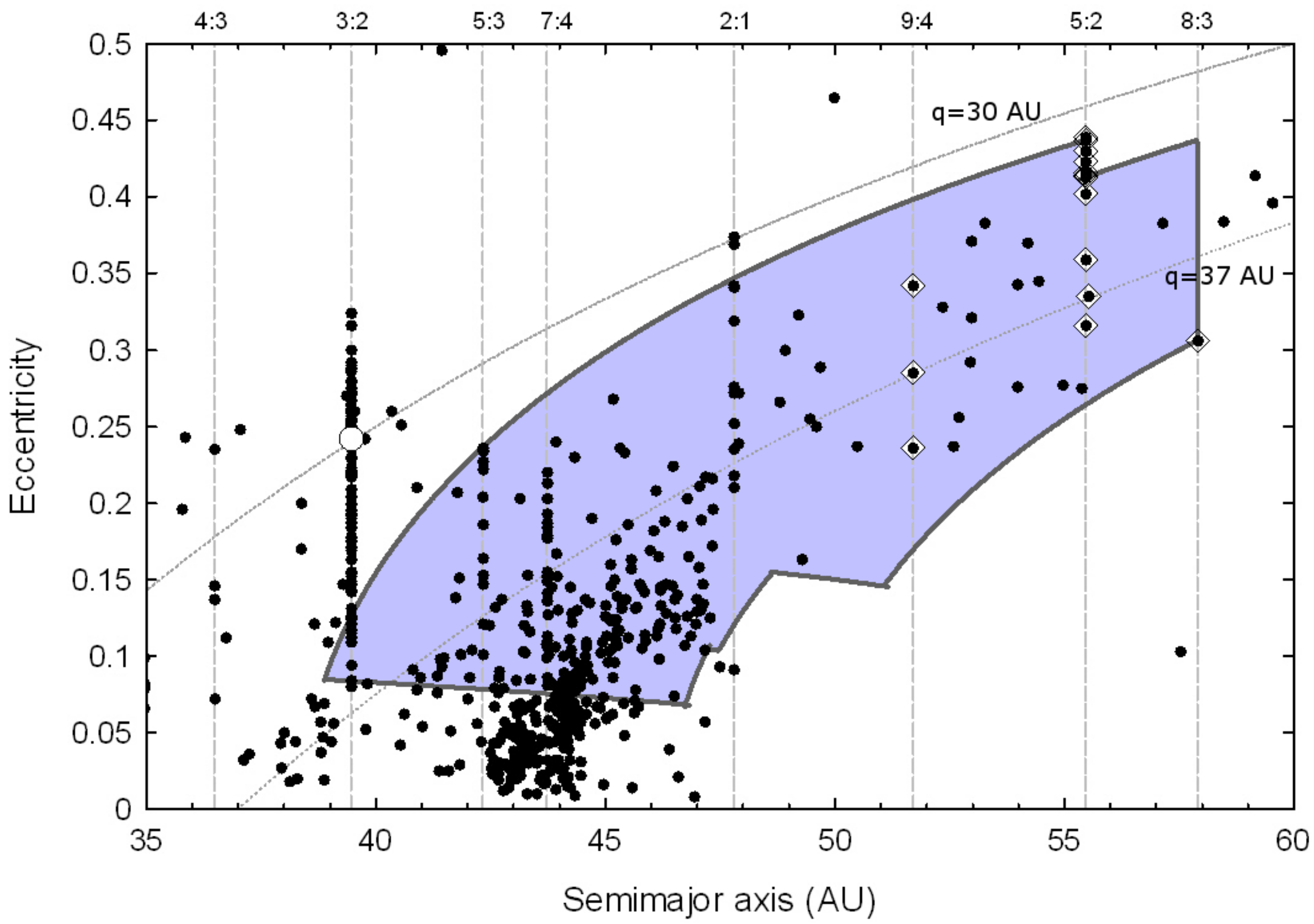

Figure 21: Orbital distribution of TNOs with long arc observations averaged over 10 Myr of dynamical evolution. Vertical dashed lines indicate mean motion resonances with Neptune. Dotted curves represent the perihelia of 30 and 37 AU. Pluto is shown as a white large circle. The 9:4, 5:2, and 8:3 resonant TNOs are enclosed with diamonds, and the bluish region defines the eccentricities needed in an excited planetesimal disk to reproduce long-term members in the latter resonances, as derived in Lykawka & Mukai (2007a) (Figure adapted from Lykawka & Mukai 2008).

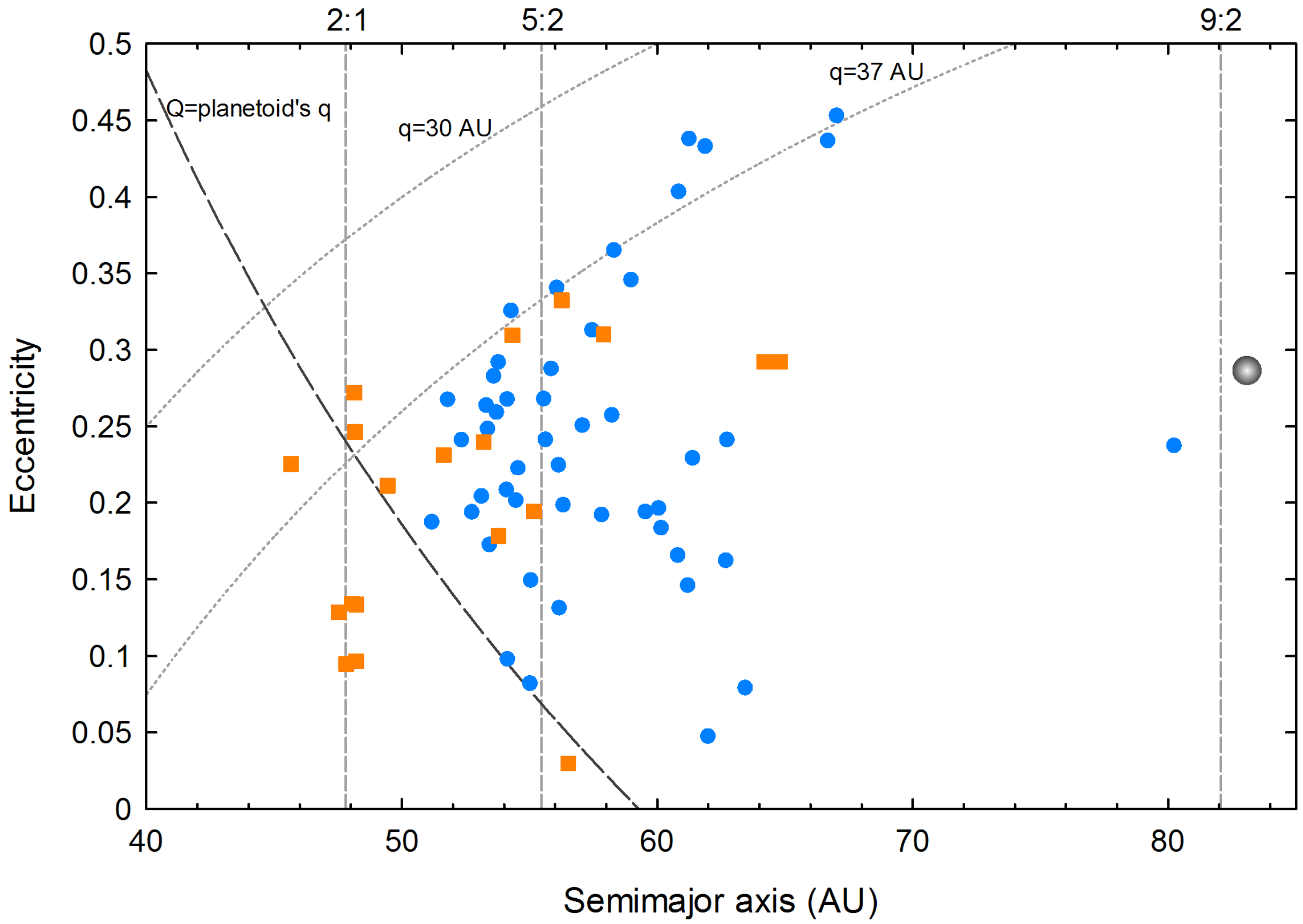

Figure 22: Orbital distribution of particles in the 2:1 and 5:2 mean motion resonances with a resident massive planetoid after 4 Gyr (represented by squares and circles, respectively). The initial orbital elements of 2:1 and 5:2 resonant populations were $0.07 < e < 0.4$ and $0.35 < e < 0.45$ ($i < 25$ deg). Dotted curves represent the perihelia of 30 and 37 AU. Dashed vertical lines indicate the locations of the 2:1, 5:2, and 9:2 resonances. The planetoid has 0.3 $M_{\oplus}$ and $i_P \sim 11$ deg (gray sphere), and objects above the long-dashed curves could encounter it. (Figure adapted from Lykawka & Mukai 2008).

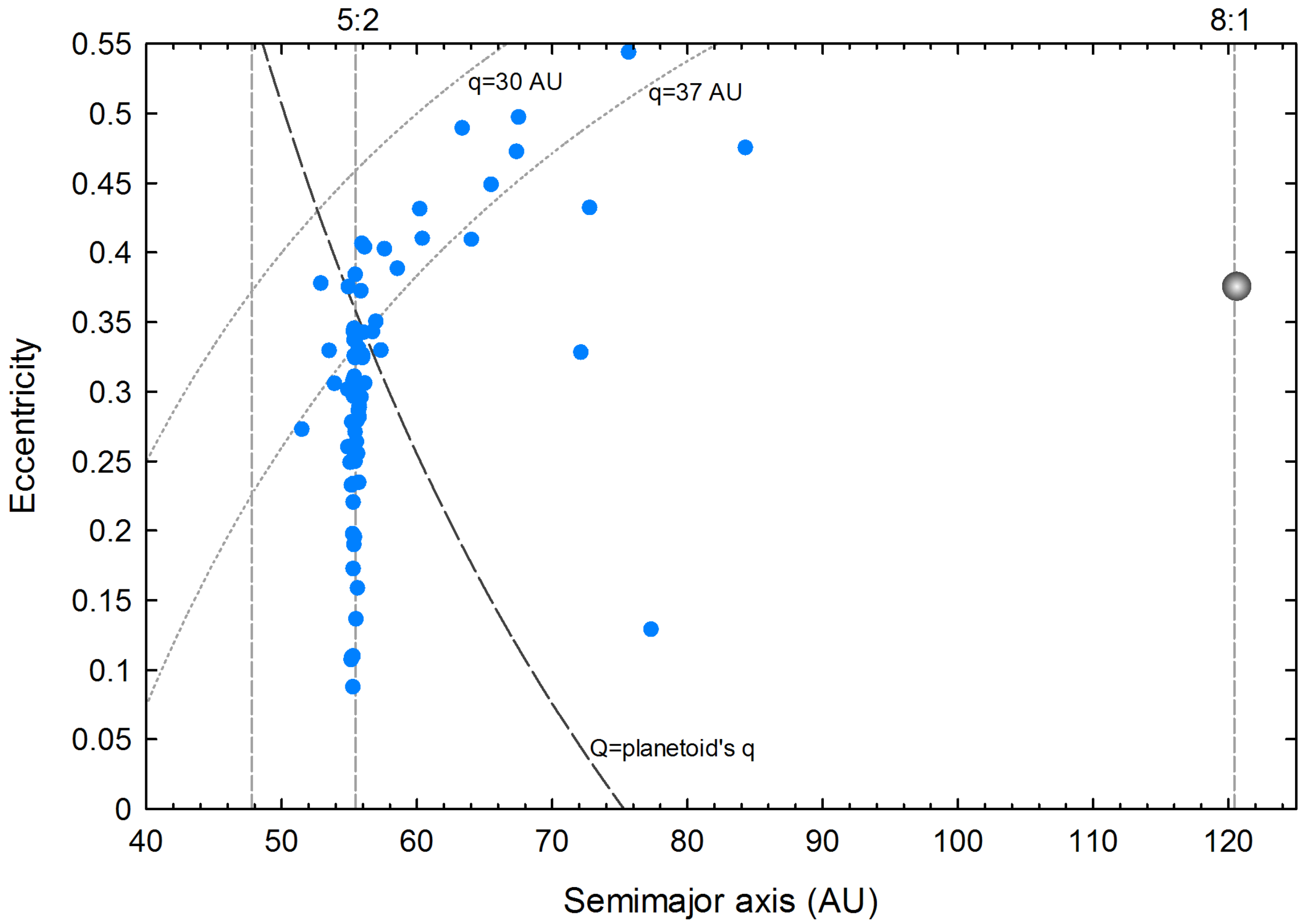

Figure 23: Orbital distribution of particles in the 5:2 mean motion resonance with a resident planetoid after 4 Gyr (represented by circles). The initial orbital elements of 5:2 resonant populations were 0.35 < $e$ < 0.45 ($i$ < 25 deg). Dotted curves represent the perihelia of 30 and 37 AU. Dashed vertical lines indicate the locations of the 5:2 and 8:1 resonances. The planetoid has 0.4 $M_\oplus$ and $i_P$ ~ 25 (gray sphere), and objects above the long-dashed curves could encounter it. The 5:2 resonant population was quite depleted over 4 Gyr (80%) (Figure adapted from Lykawka & Mukai 2008).

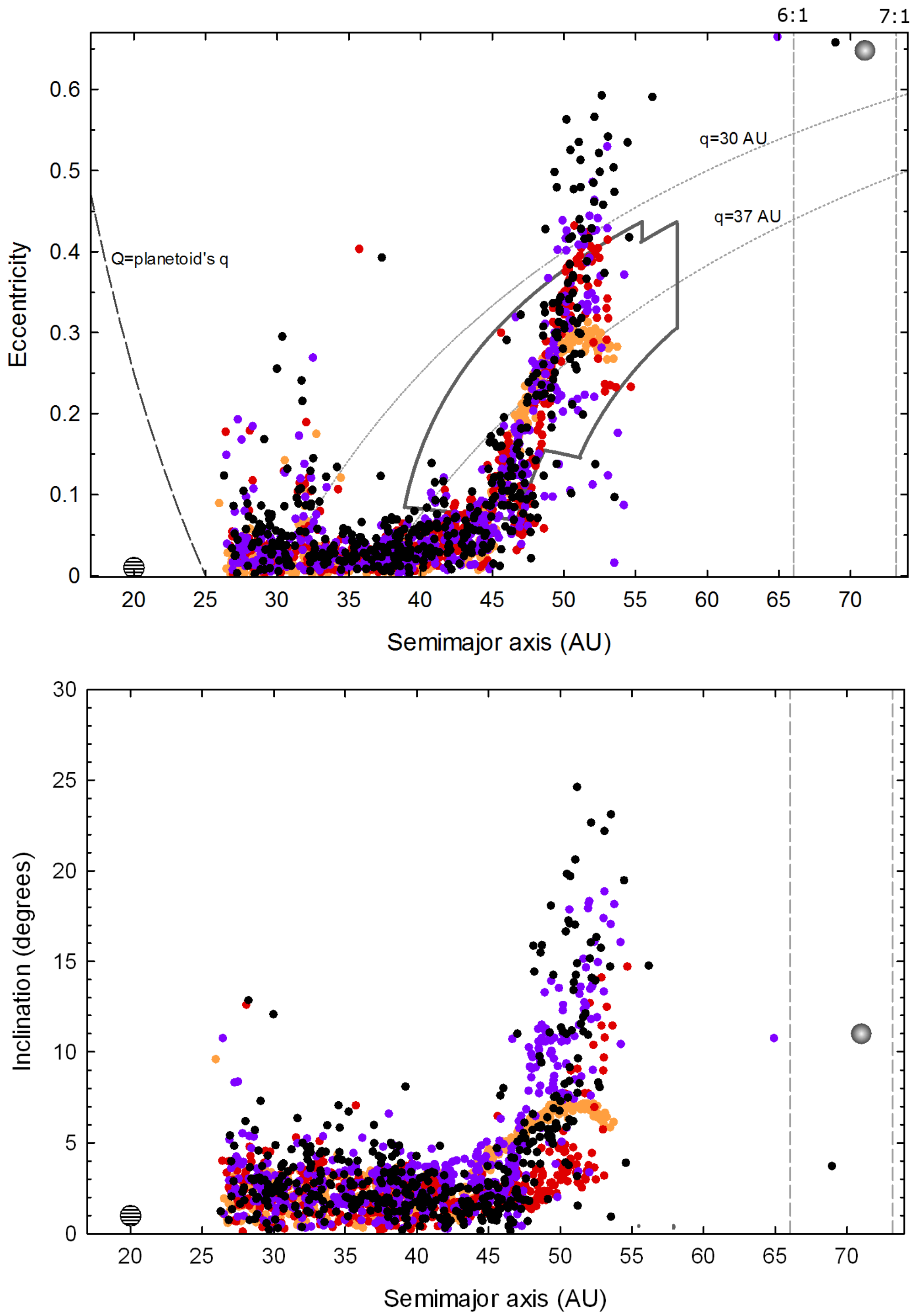

Figure 24: Orbital excitation of planetesimal disks with the presence of a scattered planetoid (gray sphere), before planetary migration. The disks were initially in cold orbital conditions ($e = i \sim 0$). Dotted curves represent the perihelia of 30 and 37 AU (upper panel). Neptune is at 20 AU and is indicated by a hatched circle. The dashed vertical line indicates the primordial locations of the 6:1 and 7:1 mean motion resonances. The enclosed region defines the conditions needed in an excited planetesimal disk to reproduce long-term TNOs in the 9:4, 5:2, and 8:3 resonances, according to Lykawka & Mukai (2007a). In both panels, the perturbation of a planetoid with $a_P = 71$ AU and 0.7 $M_{\oplus}$ was obtained at four distinct timescales $t$ = 20, 50, 80, and 150 Myr, represented by orange, red, violet and black circles, respectively (Figure adapted from Lykawka & Mukai 2008).

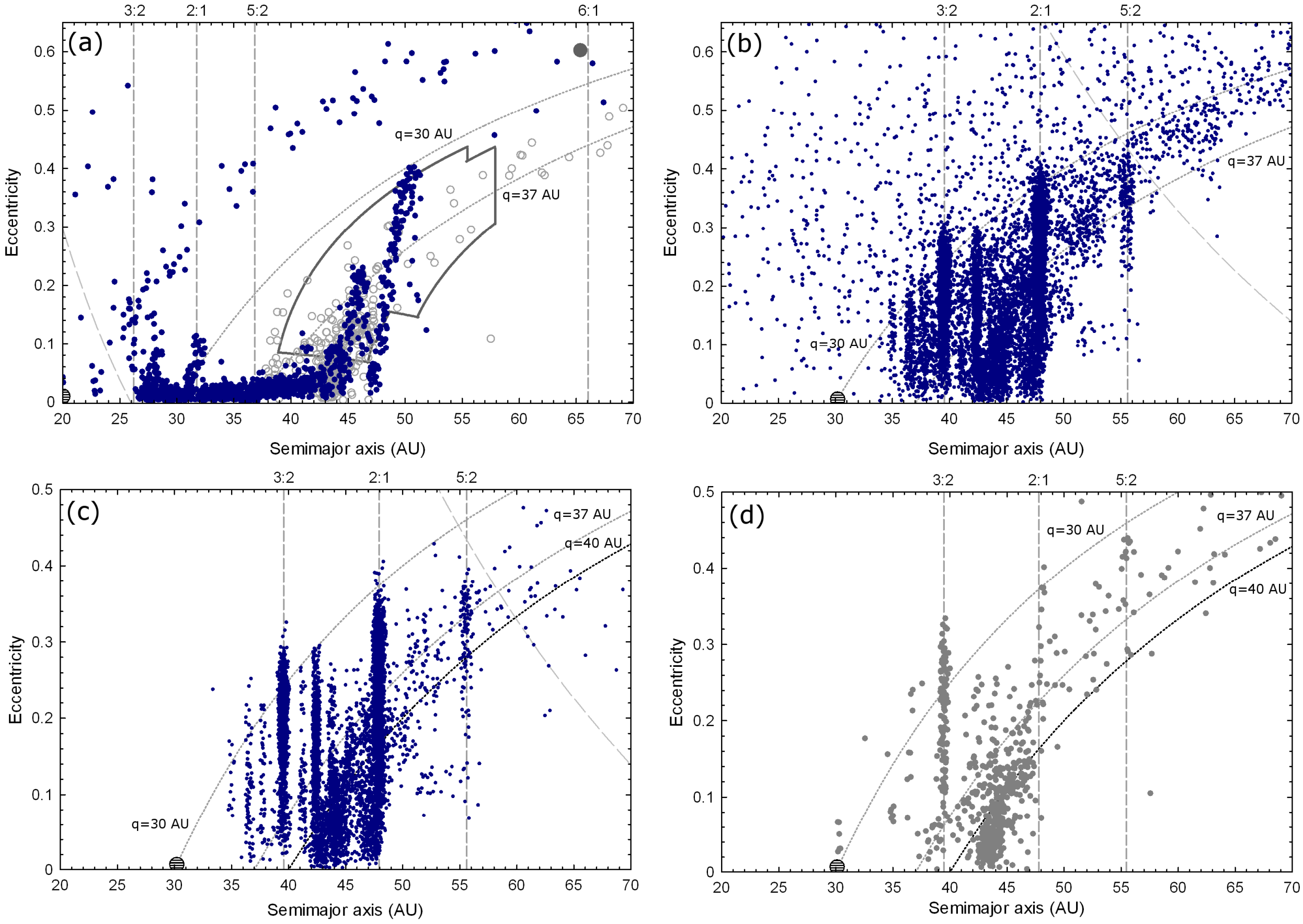

Figure 25: Orbital evolution of a planetesimal disk with a planetoid, as illustrated by theoretical disk objects in panels **a**, **b**, and **c**, obtained from a reference simulation of Lykawka & Mukai (2008) (blue circles). Panel **d** shows observational data for comparison (as of 2007). The planetesimal disk was initially cold ($e = i \sim 0$), and extended from 17 to 51 AU. Jupiter, Saturn, Uranus, and Neptune were located at 5.4, 8.7, 15, and 20 AU, respectively, and a 0.4 $M_{\oplus}$ scattered planetoid at ~65 AU with $i_P \sim$ 10 deg (big circle). The dotted curves represent the perihelion distances of 30, 37, and 40 AU. Vertical lines indicate mean motion resonances with Neptune. Panel **a**: After 60 Myr, before planetary migration. Non-resonant TNOs are illustrated for reference (open circles). The enclosed region defines the eccentricities needed in a stirred planetesimal disk to reproduce long-term TNOs in resonances beyond 50 AU, according to Lykawka & Mukai (2007a); Panel **b**: After planetary migration, performed within 100 Myr. The planetoid acquired semimajor axis $a_P \sim$ 100 AU and $i_P \sim$ 30 deg, following the location of the 6:1 resonance; Panel **c**: After evolving the system over 4 Gyr; Panel **d**: Orbital distribution of TNOs with more reliable orbits (with long-arc).

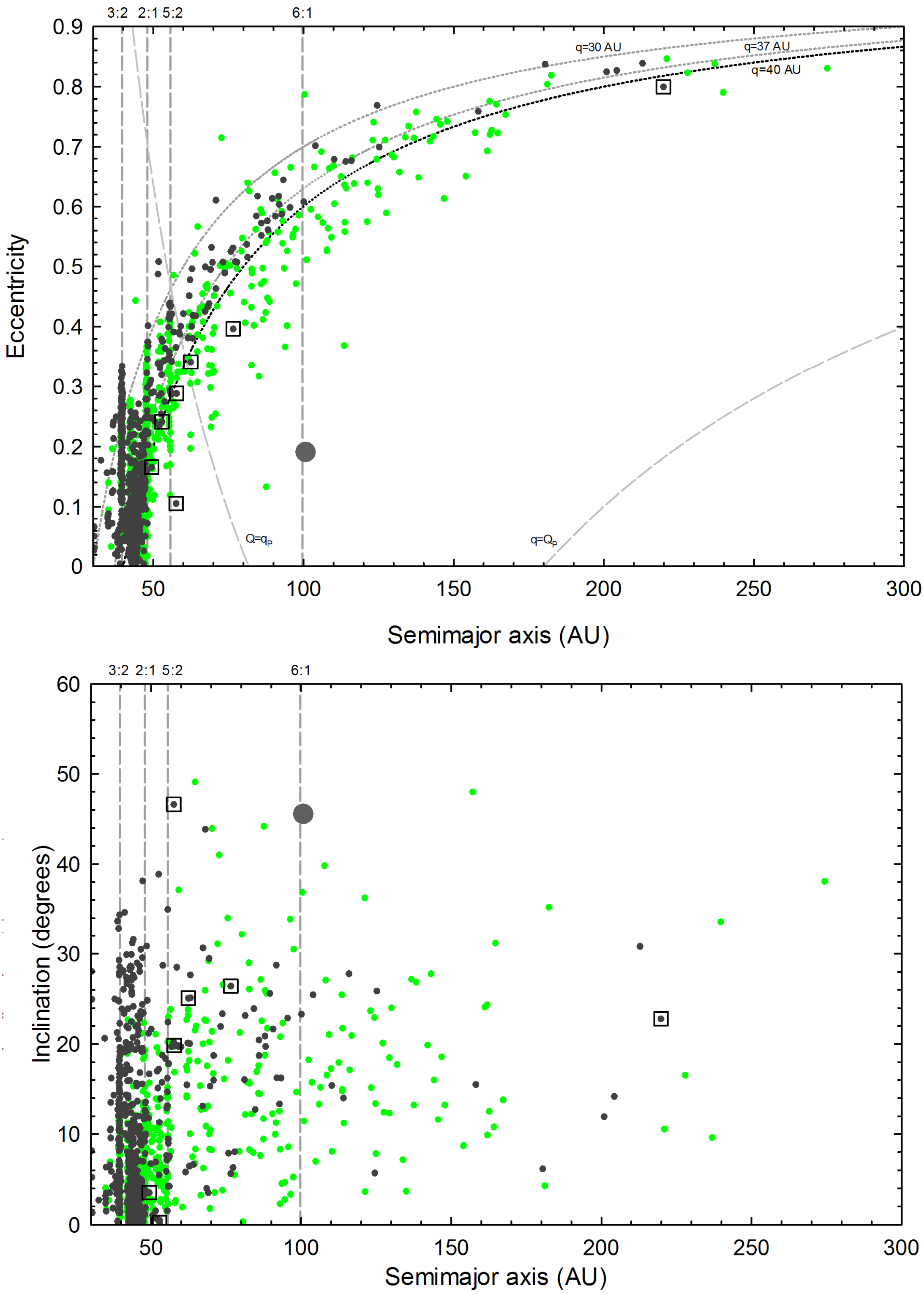

Figure 26: Comparison of orbital distributions between the planetoid-resonance sweeping hybrid model and observations obtained from a reference simulation of Lykawka & Mukai (2008). In this reference simulation, initial conditions were very similar to those shown in Fig. 25. Vertical lines indicate mean motion resonances with Neptune. Dotted curves represent the perihelia of 30, 37, and 40 AU. The results represent outcomes after 4 Gyr (green circles). The outer planet (0.4 $M_\oplus$) acquired semimajor axis $a_P$ ~ 100 AU, eccentricity $e_P$ ~ 0.2, and inclination $i_P$ ~ 45 deg (large circle). Only TNOs with more reliable orbits are plotted (with long-arc; grey circles). Squares represent detached TNOs.

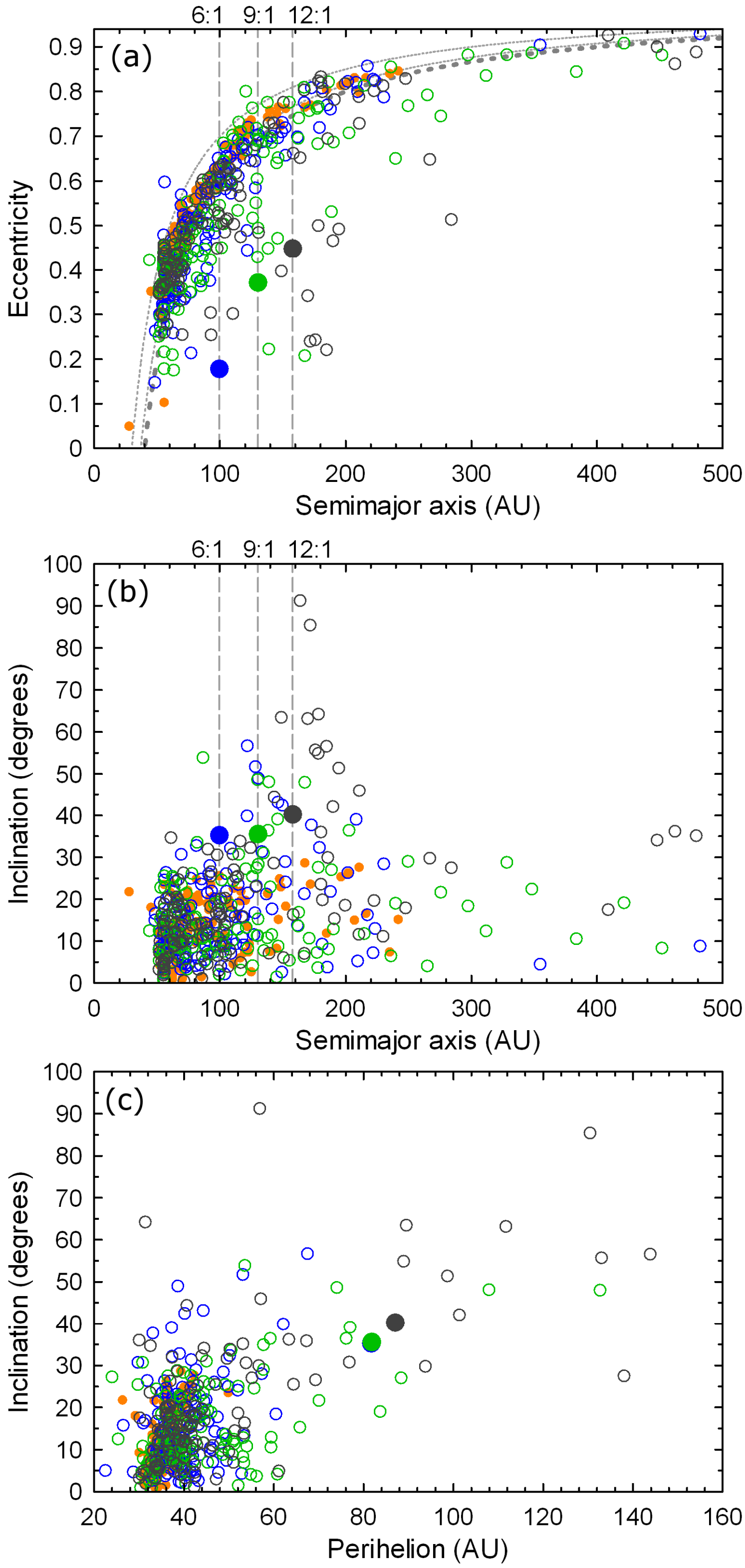

Figure 27: Orbital distributions of objects in the scattered disk after 4 Gyr (circles). The plot shows the superposition of three simulations of Lykawka & Mukai (2008) that included the perturbation of the giant planets and a resident planetoid located in distinct mean motion resonances. The three planetoids were located near the 6:1 (0.3 $M_\oplus$), 9:1 (0.5 $M_\oplus$), and 12:1 (0.7 $M_\oplus$) resonances (big filled circles), represented by blue, green, and black symbols, respectively. Dotted curves represent the perihelia of 30, 37, and 40 AU (panel **a**). Dashed vertical lines indicate the locations of the 6:1, 9:1, and 12:1 resonances (panels **a** and **b**). Finally, the outcomes of scattered objects after 4 Gyr without a planetoid are indicated by filled orange circles.

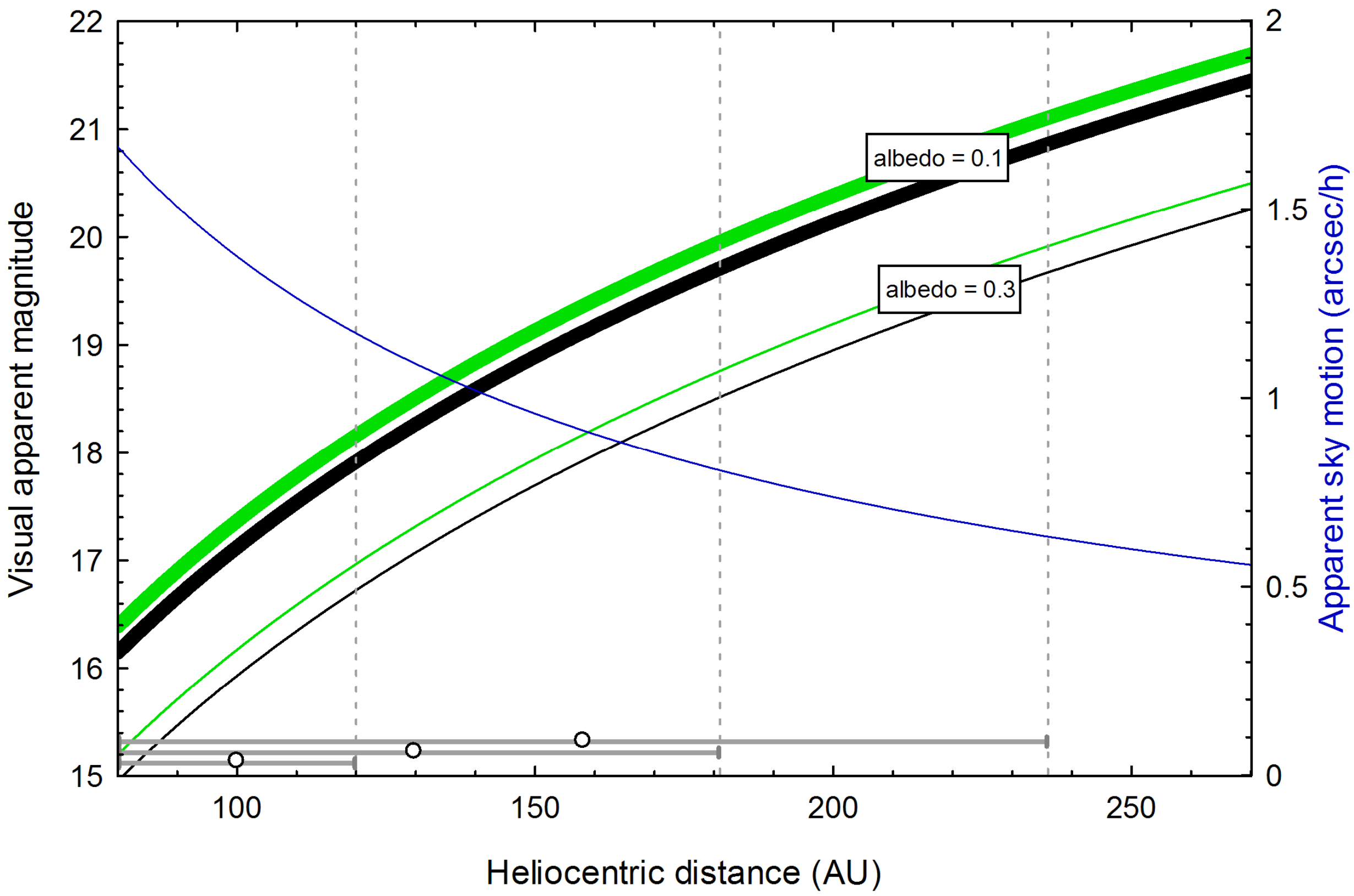

Figure 28: Apparent magnitudes of trans-Neptunian planets in distant orbits. Green and black curves represent planetoids with 0.5 and 0.7 $M_\oplus$ (mean density ρ = 2 g $cm^{-3}$) for two assumed albedos of 0.1 and 0.3. A decreasing curve in blue represents the apparent sky motion of a planetoid as a function of heliocentric distance. Three hypothetical orbits for the planetoid are shown assuming $q_P$ = 80 AU, $a_P$ ~ 100 AU (near the 6:1 resonance), $a_P$ ~ 130 AU (near the 9:1 resonance), and $a_P$ ~ 158 AU (near the 12:1 resonance). (Figure adapted from Lykawka & Mukai 2008).

Table 1
List of the currently known Neptunian Trojans.

| Prov. Des. | L*x* [a] | *a* [b] (AU) | *e* [b] | *i* [b] (deg) | *D* [c] (km) | *A* [d] (deg) | $T_L$ [d] (yr) |
|---|---|---|---|---|---|---|---|
| 2001 QR322 | 4 | 30.37 | 0.032 | 1.3 | 100-200 | 25 ± 2 | 9200 |
| 2004 KV18 | 5 | 30.13 | 0.184 | 13.6 | 50-100 | ~70-100 | ~10000 |
| 2004 UP10 | 4 | 30.28 | 0.031 | 1.4 | 50-100 | 12 ± 2 | 8900 |
| 2005 TN53 | 4 | 30.24 | 0.066 | 25.0 | 50-100 | 8 ± 2 | 9400 |
| 2005 TO74 | 4 | 30.25 | 0.050 | 5.2 | 50-100 | 9 ± 2 | 8500 |
| 2006 RJ103 | 4 | 30.15 | 0.027 | 8.2 | 100-200 | 7 ± 2 | 8600 |
| 2007 VL305 | 4 | 30.12 | 0.066 | 28.1 | 80-150 | 14 ± 1 | 9600 |
| 2008 LC18 | 5 | 30.01 | 0.082 | 27.5 | 80-150 | 15 ± 8 | 9500 |

The orbital elements and observational properties were taken from the Asteroids Dynamic Site - AstDyS, while the resonant properties were obtained from calculations using RESTICK (Lykawka & Mukai 2007b). 2004 KV18 was added to this table during the review of this paper.

[a] The Neptunian Lagrange point about which the object librates.
[b] *i* gives the inclination of the orbit with respect to the ecliptic plane, *e* the eccentricity, and *a* the semimajor axis.
[c] Estimated diameter of the object assuming the objects have albedos of 0.05 (upper estimate) or 0.20 (lower estimate).
[d] The values of mean libration amplitude (*A*, the time-averaged maximum displacement of the object from the center of libration) and median libration period ($T_L$). The error bars show the statistical errors (at the 1$\sigma$ level) resulting from averaging the libration amplitudes.

Table 2
Capture of Trojans by the giant planets at the end of planetary migration

| Planet | $\varepsilon_{min}$ [a] | $\varepsilon_{max}$ [a] | $M_{min}$ ($M_\oplus$) [b] | $M_{max}$ ($M_\oplus$) [b] |
|---|---|---|---|---|
| Jupiter | $5 \cdot 10^{-6}$ | $5 \cdot 10^{-5}$ | $3 \cdot 10^{-5}$ | $2 \cdot 10^{-4}$ |
| Saturn | $<10^{-6}$ | $10^{-5}$ | $<8 \cdot 10^{-6}$ | $6 \cdot 10^{-5}$ |
| Uranus | $5 \cdot 10^{-5}$ | $5 \cdot 10^{-4}$ | $6 \cdot 10^{-4}$ | $7 \cdot 10^{-3}$ |
| Neptune | $3 \cdot 10^{-4}$ | $10^{-3}$ | $4 \cdot 10^{-3}$ | $2 \cdot 10^{-2}$ |

[a] Minimum and maximum capture efficiencies ($\varepsilon_{min}$, $\varepsilon_{max}$). These calculations assume that a mass of between 13 and 25 $M_\oplus$ of material was initially present in the planetesimal disk through which the giant planets migrated (based on Lykawka & Horner 2010).
[b] Estimated minimum and maximum masses of the captured Trojan populations for each of the giant planets at the end of their migration.

Table 3
Resonant populations using ~51-54 AU sized planetesimal disks after 4 Gyr (based on Lykawka & Mukai 2008)

| Resonant population | $a_{res}$ [a] (AU) | $N$ [b] | $e$ [c] | $i$ [c] (deg) | $N_{KM}$ [b] | $f_{KM}$ [b] (%) | *Max A* [d] (deg) |
|---|---|---|---|---|---|---|---|
| 5:4 | 34.9 | 20 | 0.11 | 5.6 | 1 | 5 | 125 |
| 4:3 | 36.5 | 152 | 0.11 | 9.7 | 18 | 12 | 140 |
| 7:5 | 37.7 | 48 | 0.13 | 12.4 | 15 | 30 | 115 |
| 3:2 | 39.4 | 2157 | 0.22 | 5.3 | 535 | 24 | 145 |
| 8:5 | 41.2 | 34 | 0.15 | 7.9 | 9 | 17 | 110 |
| 5:3 | 42.3 | 1163 | 0.18 | 2.8 | 69 | 6 | 165 |
| 7:4 | 43.7 | 134 | 0.14 | 5.7 | 59 | 35 | 130 |
| 9:5 | 44.5 | 12 | 0.10 | 2.8 | 0 | 0 | 115 |
| 11:6 | 45.1 | 11) | 0.16 | 4.4 | 0 | 0 | 130 |
| 2:1 | 47.8 | 4088 | 0.26 | 4.1 | 223 | 5 | 175 |
| | | | | | | | 60 |
| 13:6 | 50.4 | 5 | 0.26 | 5.2 | 0 | 0 | 75 |
| 11:5 | 50.9 | 22 | 0.26 | 6.8 | 0 | 0 | 130 |
| 9:4 | 51.7 | 44 | 0.23 | 4.5 | 0 | 0 | 135 |
| 7:3 | 53.0 | 64 | 0.24 | 5.8 | 11 | 15 | 130 |
| 5:2 | 55.4 | 264 | 0.30 | 8.2 | 118 | 43 | 160 |
| 8:3 | 57.9 | 5 | 0.33 | 5.8 | 1 | 11 | 130 |
| 3:1 | 62.6 | 8 | 0.41 | 17.8 | 1 | 5 | 175 |

[a] Resonance semimajor axis (Eq. 6).
[b] $N$ = number of resonant particles. $N_{KM}$ = number of objects that experienced the Kozai mechanism and their fraction of the total resonant population, $f_{KM}$.
[c] Orbital elements represent the median of each population, where $e$ = eccentricity and $i$ = inclination.
[d] *Max A* gives the maximum values of libration amplitudes (amplitudes of the resonant angle). The error is approximately ±5 deg. In the case of 2:1 resonant objects, the second line represents asymmetric librators.

Table 4
Scattered and detached populations with resident planetoids in the scattered disk after 4 Gyr (based on Lykawka & Mukai 2008)

| Planetoid location | $m_P$ [a] ($M_\oplus$) | $i_P$ [a] (deg) | $q_P$ [a] (AU) | $P_{scat}$ [b] (%) | $P_{det}$ [b] (%) | $Ratio_{SD}$ [c] | Median $q_{det}$ [d] (AU) |
|---|---|---|---|---|---|---|---|
| $a$ ~ 100AU (6:1) | 0.1 | 36 | 82 | 83.9 | 16.1 | 5.2 | 42.5 |
| | 0.2 | 36 | 82 | 68.3 | 31.7 | 2.2 | 42.5 |
| | 0.3 | 36 | 82 | 58.0 | 42.0 | 1.4 | 43.3 |
| | 0.4 | 36 | 82 | 47.2 | 52.8 | 0.9 | 45.7 |
| | 0.5 | 36 | 82 | 44.2 | 55.8 | 0.8 | 44.8 |
| | 0.5 | 41 | 91 | 46.2 | 53.8 | 0.9 | 44.2 |
| | 0.5 | 46 | 81 | 45.4 | 54.6 | 0.8 | 44.8 |
| $a$ ~ 130AU (9:1) | 0.4 | 36 | 82 | 47.6 | 52.4 | 0.9 | 45.4 |
| | 0.5 | 36 | 82 | 50.0 | 50.0 | 1.0 | 49.2 |
| | 0.5 | 41 | 83 | 59.8 | 40.2 | 1.5 | 48.1 |
| | 0.5 | 41 | 91 | 54.2 | 45.8 | 1.2 | 43.5 |
| | 0.7 | 20 | 87 | 34.8 | 65.2 | 0.5 | 51.4 |
| | 0.7 | 40 | 87 | 39.7 | 60.3 | 0.7 | 47.4 |
| $a$ ~ 158AU (12:1) | 0.5 | 21 | 83 | 37.8 | 62.2 | 0.6 | 54.6 |
| | 0.5 | 40 | 82 | 52.0 | 48.0 | 1.1 | 47.4 |
| | 0.7 | 40 | 87 | 51.2 | 48.8 | 1.0 | 48.3 |
| | 1.0 | 35 | 92 | 30.5 | 69.5 | 0.4 | 51.5 |
| | 1.0 | 50 | 92 | 46.0 | 54.0 | 0.9 | 47.4 |
| No planetoid [e] | - | - | - | 93.2 | 6.8 | 13.7 | 40.6 |
| No planetoid(2) [f] | - | - | - | ~84-88 | ~12-16 | 5.25-7.33 | 41.5 |
| Observations apparent [g] | - | - | - | ~90 | ~10 | ~9.0 | >41.3 |

[a] $m_P$ = mass in Earth masses, $i_P$ = inclination, and $q_P$ = perihelion distance of the planetoid.
[b] $P_{scat}$ and $P_{det}$ represent the proportion of scattered and detached particles ($q$ > 40 AU) to the total population beyond 48 AU after evolving the orbits of populations of Neptune-encountering objects ($q$ < 35 AU).
[c] Ratio of scattered to detached populations.
[d] Median perihelion distance of the detached population.
[e] Results of simulations in which only the four giant planets were included (without planetary migration).
[f] Results of extra simulations in which only the four giant planets were included for cases with, and without, planetary migration.
[g] Computed with the identification of 72 scattered and 9 detached TNOs (Lykawka & Mukai 2007b). Because of severe observational biases, the intrinsic ratio of scattered to detached TNOs is expected to be approximately ≤1.0 (Gladman et al. 2002), and the median of this population should be >41.3.